\documentclass[twocolumn,preprintnumbers,amsmath,amssymb]{revtex4}

\usepackage{graphicx}
\usepackage{dcolumn}
\usepackage{bm}

\usepackage{amsmath,amsfonts}
\usepackage{amssymb}
\usepackage{float}
\usepackage[squaren]{SIunits}
\usepackage[T1]{fontenc}
\usepackage[english]{babel}
\usepackage[export]{adjustbox}
\title{Mechanothermics}

\begin{document}


\title{ On the link between mechanics and thermal properties: mechanothermics\\}
\author{Eddy Collin$^{1,*}$, Ilya Golokolenov$^1$, Olivier Maillet$^2$, Laurent Saminadayar$^1$
\& Olivier Bourgeois$^1$  }

\address{$^1$Univ. Grenoble Alpes, Institut N\'eel - CNRS,  38042 Grenoble, France\\ 
$^2$Univ. Paris-Saclay, CEA, CNRS, SPEC, 91191 Gif-sur-Yvette, France\\ 
* Corresponding Author: {\it eddy.collin$@$neel.cnrs.fr}.}

\date{\today} 
\begin{abstract} 

We report on the theoretical derivation of macroscopic thermal properties (specific heat, thermal conductivity) of an electrically insulating rod connected to two reservoirs, from the linear superposition of its mechanical mode Brownian motions. 
The calculation is performed for a weak thermal gradient, in the classical limit (high temperature).
The development is kept basic as far as geometry and experimental conditions are concerned, enabling an almost fully analytic treatment.
In the modeling, each of the modes is subject to a specific {\it Langevin force}, which enables to produce the required temperature profile along the rod.
The theory is predictive: the temperature gradient (and therefore energy transport) is linked to motion amplitude {\it cross-correlations} between nearby mechanical modes. 
This arises because energy transport is actually mediated by {\it mixing 
between the modal waves}, and not by the modes themselves. 
This result can be tested on experiments, and shall extend the concepts underlying equipartition and fluctuation-dissipation theorems. 
The theory links intimately the macroscopic size of the clamping region 
where the mixing occurs 
to the microscopic lengthscale of the problem at hand: the phonon mean-free-path. 
This clamping region, which is key, has received recently a renewed attention in the field of nanomechanics with topical works on "phonon shields" and "soft clamping".
We believe that our work should impact the domain of thermal transport in nanostructures,
 with future developments of the theory toward the quantum regime.

\end{abstract}

\maketitle




\section{Introduction}
\label{intro}

{\it Context } - Solid matter is constituted of atoms located at specific places in space. 
These positions are defined by the interactions among these atoms, whatever they might be (e.g. covalent, ionic, van der Waals). While these bonding forces are actually rather difficult to describe from first principles (especially at defects like e.g. grain boundaries in crystals, surfaces, voids) \cite{volker91}, they result in an effective potential $V(\vec{r}_i)$ experienced by each atom (located at $\vec{r}_i=\{ x_i,y_i,z_i \}$) that keeps the piece of matter together: which means $\partial V /\partial x_i =0 $ and similarly for $y_i,z_i$ at the rest positions $\vec{r}_{i,0}$ \cite{clelandBk}.

At equilibrium, atoms oscillate weakly around their rest positions: this is {\it Brownian motion}, understood as the response of the solid to the energy supplied randomly by the thermal baths to which it is inevitably connected (let it be the surrounding air and mostly the support that holds the object, at temperature $T$) \cite{ziman}. This agitation is conveniently described in terms of {\it phonons}, the collective mode excitations arising from the coupling between atoms (the proper eigenstates of the system). At lowest order, this coupling can be seen as a collection of restoring forces with effective spring constants $\partial^2 V /\partial x_i\partial x_j $ (with all permutations $x_i \longleftrightarrow y_i,z_i$ and $x_j \longleftrightarrow y_j,z_j$); at higher orders, the anharmonicity of the potential $V$ leads to the finite lifetime of these excitations \cite{clelandBk,ziman}.
Modern numerical methods which take into account interactions over a large number of atoms, namely Molecular Dynamics Simulations (MDS), can be used to calculate thermal properties of widely used materials, like e.g. crystalline Silicon \cite{siliconMD}.
Considering solids with no free electrons (insulators or intrinsic semiconductors at low enough $T$)  
and with no extra 
degrees of freedom (like e.g. a magnetic moment),
phonons lead at the macroscopic scale to both the description of thermal equilibrium, with the definition of the {\it specific heat} $C_v$ (at constant volume), and to the description of heat transport with the definition \linebreak
of the {\it thermal conductivity} $\kappa$ \cite{clelandBk,ziman}. 

		\begin{figure}[t!]
		\centering
	\includegraphics[width=12.5cm]{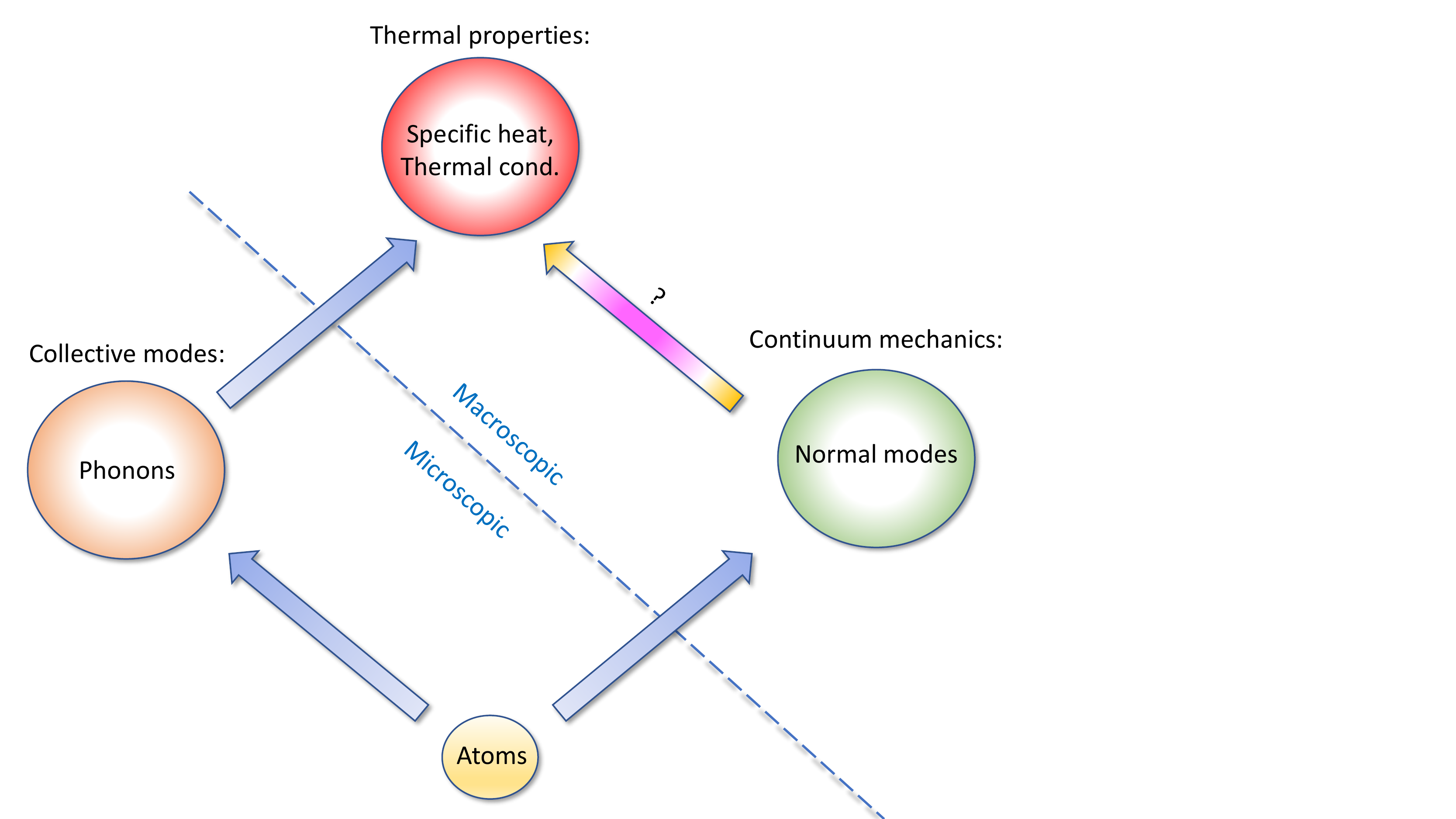}
			\caption{\small{ 
			The microscopic atomistic theory enables to built on one side elementary excitations (phonons) and on the other one elasticity theory (stress vs. strain relations, i.e. continuum mechanics). The former leads at the macroscopic scale to the bulk thermal properties: specific heat and thermal conductivity. 
It should therefore be possible to derive these properties from the latter. This is the subject of the manuscript (the `` ? '' marked arrow above). }}
			\label{fig_1}
		\end{figure}
		
Exerting a small force onto a solid results into a small reversible deformation of this body, of amplitude proportional to the force (the linear limit). Again, this can be seen as a consequence of the effective springs 
 that link all the atoms together. 
One can then derive {\it elasticity theory} by averaging together these interactions over a small volume $\delta \tau$, and transforming sums over atoms into integrals over space. The procedure is exact, provided this volume is small enough from a macroscopic point of view (which requires the inter-atomic interactions to be short-ranged), but still large enough to host a decent number of atoms; besides, deformations have then to occur on lengthscales larger than this volume, thus much larger than the inter-atomic distance (i.e. long wavelengths).
This leads to the definition of the stiffness tensor $[c_{ijkl}]$ that relates stresses to strains (with each index $i,j,k,l$ referring to the 3 directions in space $x,y,z$) \cite{clelandBk}.
As for thermal properties, MDS can be used to evaluate the stiffness coefficients $c_{ijkl}$ of relevant materials, e.g. Silicon \cite{siliconMD2}.
On the macroscopic scale, elasticity theory enables to describe how an object will distort when driven by an arbitrary DC or AC force applied onto a region of it. In the framework of the linear theory discussed here, this is performed by defining the {\it normal modes} of the structure, i.e. the standing wave motional states that the object can sustain, and then performing the proper linear superposition \cite{clelandBk}.
		
The atomistic description of solids thus leads both to the definition of macroscopic thermal {\it and} mechanical properties, Fig. \ref{fig_1};
these are just two facets of the same underlying physics (i.e. the inter-atomic potential $V$).
And indeed, thermodynamics can be probed at the level of an {\it individual} mechanical normal mode (as opposed to the whole object). 
Brownian motion of micromechanical beams has been reported using various systems and techniques \cite{OptoTh,VinantePRL,sansa,regal2008}, from room temperature down to millikekvin temperatures. 
It is detected as peaks in the motion spectrum at the frequencies of the mechanical modes $\omega_n$, with widths $\Gamma_n$ the mechanical damping rates, and  areas proportional to the stored elastic energies $\frac{1}{2} k_n \!\!\!< x_n^2 > = \frac{1}{2} k_B T$, with $x_n(t)$ the amplitude of motion of mode $n$ having an effective spring constant $k_n$. $k_B$ is Boltzmann's constant and $< \cdots >$ the ensemble average. This result is known as the {\it equipartition theorem}, and applies when in-equilibrium \cite{statbook}.
The mathematical formalism describes Brownian motion as due to a stochastic force $F_n$, the so-called {\it Langevin force}, that acts upon mode $n$. It is due to the thermodynamic bath (at temperature $T$) that generates the damping $\Gamma_n$; there is therefore a link between $F_n$ and $\Gamma_n$, which is called {\it the fluctuation-dissipation theorem} \cite{statbook}.

Experiments on single modes are also performed out-of-equilibrium \cite{bellon1,bellon2}. In this case, measured properties (local temperature along the beam, mode spectrum) do depend on the heat flow imposed in the structure. 
The situation is thus much more complex than for the in-equilibrium case (where everything is defined through the single parameter $T$), and heat flows are usually extremely large in the experimental realizations, which adds another degree of complexity \cite{bellon3}.
New thermodynamics concepts that describe the steady-state have to be developed {\it beyond} standard equipartition and fluctuation-dissipation theorems \cite{bellon1,bellon3}.

Developing these new concepts is a very difficult task, and the present manuscript proposes an original theoretical approach to these questions, drastically different from existing literature.
From Fig. \ref{fig_1} and what has been stated above about single-mode thermodynamics, it appears that {\it there should be a method to derive the macroscopic thermal properties from a pure mechanistic approach}.
In order to do so, we shall consider a rather simplified textbook problem which is essentially analytically solvable: a cylindrical rod made of a single isotropic, homogeneous and simple material (e.g. non-magnetic) with no conduction electrons, connected to two reservoirs in vacuum (see Fig. \ref{fig_2} below). 
For the sake of concreteness, we shall present numerical estimates using properties of polycrystalline Silicon.
The dimensions will be assumed large enough to be fairly macroscopic, with respect to the atomic size and lengthscales related to phonons (see discussion below).
We will limit ourselves to the treatment of linear response for both thermal and mechanical properties. Finally, the average temperature $T$ will be assumed high enough so that everything can be treated in the framework of classical physics. 

{\it Structure of the paper } - In Section \ref{phonons}, we will first remind the reader about fundamental theoretical aspects related to phonons, thermal equilibrium and thermal transport. 
Basic parameters will be introduced, and the temperature profile $T(z)$ along the rod presented.
Section \ref{meca} gives the mathematical formalism that defines energies (kinetic, potential, friction) in the rod in terms of its normal modes. These are obtained from the {\it Pochhammer-Chree} waves \cite{poche}, which are described in the Appendix \ref{pochhammer} for the sake of completeness.
The simple case of in-equilibrium conditions is treated first in Section \ref{ineq}, introducing the standard equipartition and fluctuation-dissipation 
 theorems.
Section \ref{gradient} then treats the case where a small thermal gradient is imposed, with a weak heat flow.
The new implications for equipartition and fluctuation-dissipation theorems are then described in Section \ref{FDT}.
In Conclusion, we discuss these findings and how they could be confronted to experiments.

{\it Fundamental outcome of the theory} - 
The modeling {\it predicts nonzero amplitude cross-correlations between the modes}, proportional to the heat flow, with a specific dependence to mode number. 
This actually means that energy transport is not carried by the modes themselves, but by their 2-wave mixing. 
The theory leads to a link between a microscopic lengthscale, namely the phononic mean-free-path, and a macroscopic one which characterizes the size of the clamping zone.
This clamping zone is a key element of the modeling: it is where correlations between modes of nearby 
wavevectors 
are created, leading to nonzero 
mixing amplitudes. 
The full model is a linear theory, even though nonlinear processes within the clamp, which are responsible for correlations, are key.

		\begin{figure*}[t!]
		\center
	\hspace*{1cm} \includegraphics[width=24.cm]{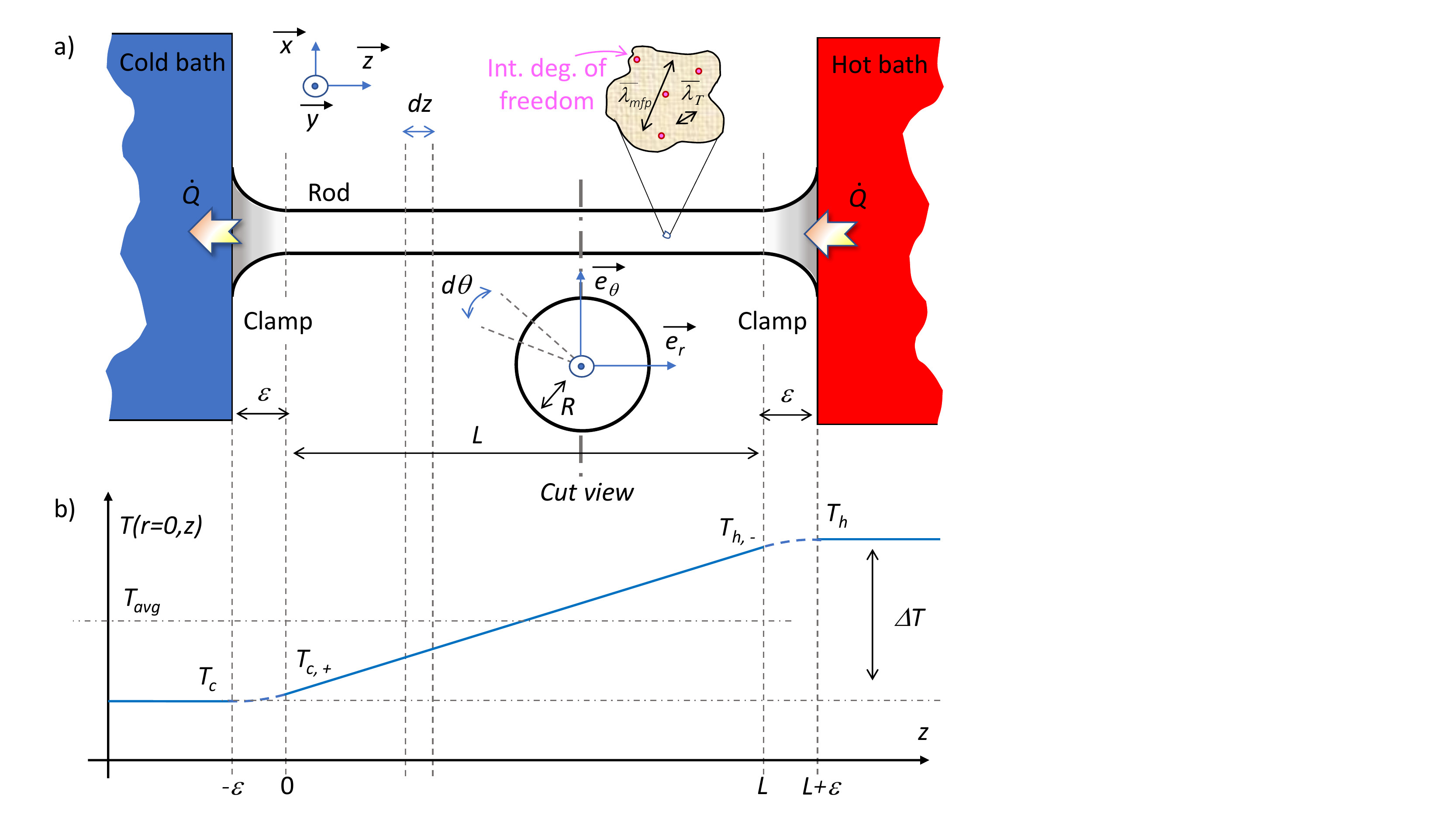}     
			\caption{\small{a) Schematics of the problem at hand (not to scale). A cylindrical rod is clamped between two thermal reservoirs, in vacuum; in inset, the relevant lengthscales $\bar{\lambda}_{mfp}, \bar{\lambda}_T$ are depicted qualitatively, together with microscopic {\it defects} (magenta dots, internal degrees of freedom). Geometrical parameters are discussed in the text; body forces in the rod due to in-built stress (appearing e.g. if one pulls on the clamps) are considered only in Appendix \ref{pochhammer}. b) Theoretical temperature profile along the rod axis (under a heat flow $\dot{Q}$); $T$ is uniform within a section of the length $L$, but has a more complex pattern within the clamps (in dashed, see text).}}
			\label{fig_2}
		\end{figure*}

\section{Phonons and thermal properties}
\label{phonons}

\subsection{Collective modes: phonons}

Let us consider a bulk piece of a homogeneous material, at temperature $T$. 
We shall restrict this introductory discussion to the simplest types of solids: those with crystalline order.
The situation of amorphous matter is much more involved and shall not be described here \cite{hunklinger}; we will just comment on it at the end of this Subsection when addressing crystalline defects. \\

The elementary excitations related to motion, namely the phonons, are waves that propagate in the crystal \cite{clelandBk,ziman,ashcroft}. They have dispersion relations $\omega(\vec{q})$
that depend on the direction of propagation $\hat{q}=\vec{q}/\left|\vec{q}\right|$. For small enough wavevectors $\left|\vec{q}\right|$ compared to the size of the first Brillouin zone (i.e. long wavelengths), these are linear: $\omega(\vec{q}) = c_{n,\hat{q}} \left|\vec{q}\right|$ with $c_{n,\hat{q}}$ the wave (phase) velocity in the direction $\hat{q}$, for mode $n$. In each direction, the motion sustains 3 types of waves: one longitudinal ($n=l$, the vibration is along $\hat{q}$), and two transverse ones ($n=t_1, t_2$ with vibration perpendicular to $\hat{q}$).
The velocities of $t_1$ and $t_2$ can be degenerate, but are in principle distinct.
When $\left|\vec{q}\right|$ gets close to the border of the first Brillouin zone (i.e. small wavelengths), the dispersion relations start to flatten out.
Besides, if the crystallographic structure contains more than one atom per unit cell, phonons also support traveling modes at high frequencies, which frequencies do not extrapolate at zero for $\left|\vec{q}\right| \rightarrow 0$: these are the optical branches. 

These concepts can be illustrated with Silicon, see e.g. Refs. \cite{SiPh1,SiPh2,SiPh3} providing comparison between theory and experimental data. 
Monocrystalline Silicon has a {\it fcc} structure with 2 Si atoms in the cell; the lattice parameter $a$ is about 0.54$~$nm,
which is the {\it fundamental lengthscale} that fixes strict bounds to the modeling (see discussion at the end of this Subsection).
Below about 4$~$THz (i.e. corresponding to 200$~$K which is ``almost'' room temperature), the dispersion relations are reasonably linear. This is the regime we shall concentrate on for the work presented in this manuscript.

\begin{figure}[t!]
		\centering
	\includegraphics[width=11cm]{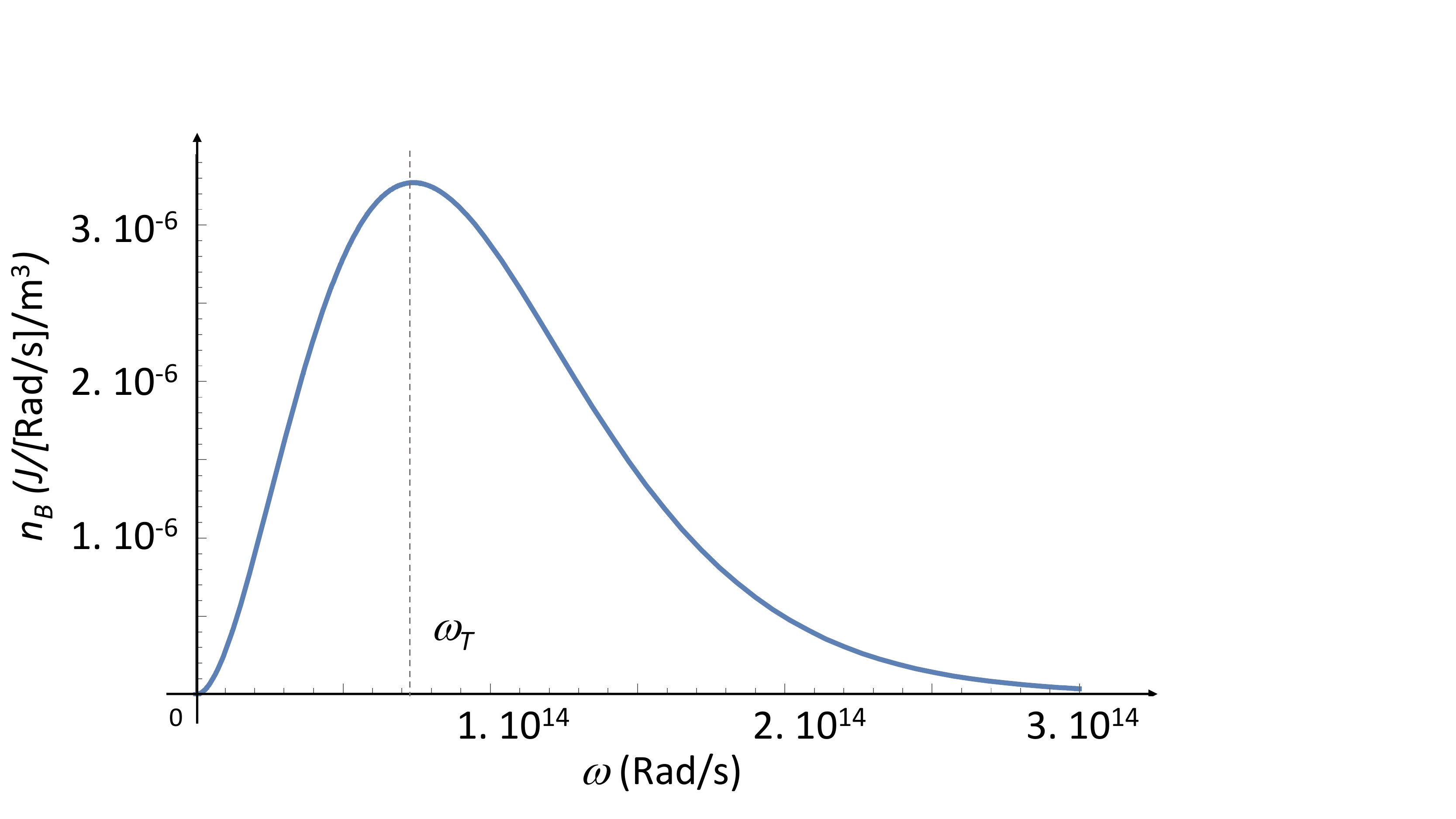}
			\caption{\small{ 
			Energy density distribution of phonons $n_B(\omega)$ at a temperature $T$, Eq. (\ref{distrib}). The position of the maximum is $\omega_T$. Plot realised for Silicon at 200$~$K (see text).}}
			\label{fig_3}
		\end{figure}

In thermal equilibrium, phonons distribute in energy according to Planck's black body law \cite{clelandBk,ziman}:
\begin{equation}
n_B(\omega) = \frac{3 \hbar}{2 \pi^2 \, \bar{c}^3} \frac{\omega^3}{e^{\hbar \omega/(k_B T)}-1}, \label{distrib}
\end{equation}
expressed in (Joules/m$^3$)/(Rad/s), with $\hbar$ Plank's (reduced) constant.
$\bar{c}$ is the average speed of sound in the solid, averaged over all $c_{n,\hat{q}}$ values.
 This function is plotted on Fig. \ref{fig_3}, with $\bar{c} \approx 5\,000~$m/s for Silicon and $T=200~$K.
It is strongly peaked around $\omega_T \approx 2.821 \, (k_B T)/\hbar$, which corresponds to an average wavelength of $\bar{\lambda}_T \approx 2.227 \, \hbar \bar{c}/(k_B T)$ if we consider a strictly linear dispersion law.
This is our first relevant lengthscale known as the {\it dominant  phonon wavelength} \cite{clelandBk,ziman}. At $T=200~$K, it essentially means that motional energy stored in phonons corresponds to distortions occurring over lengthcales of the order of $\bar{\lambda}_T  \sim 0.4~$nm, i.e. about the interatomic distance.
		
\begin{figure}[t!]
		\centering
	\includegraphics[width=11cm]{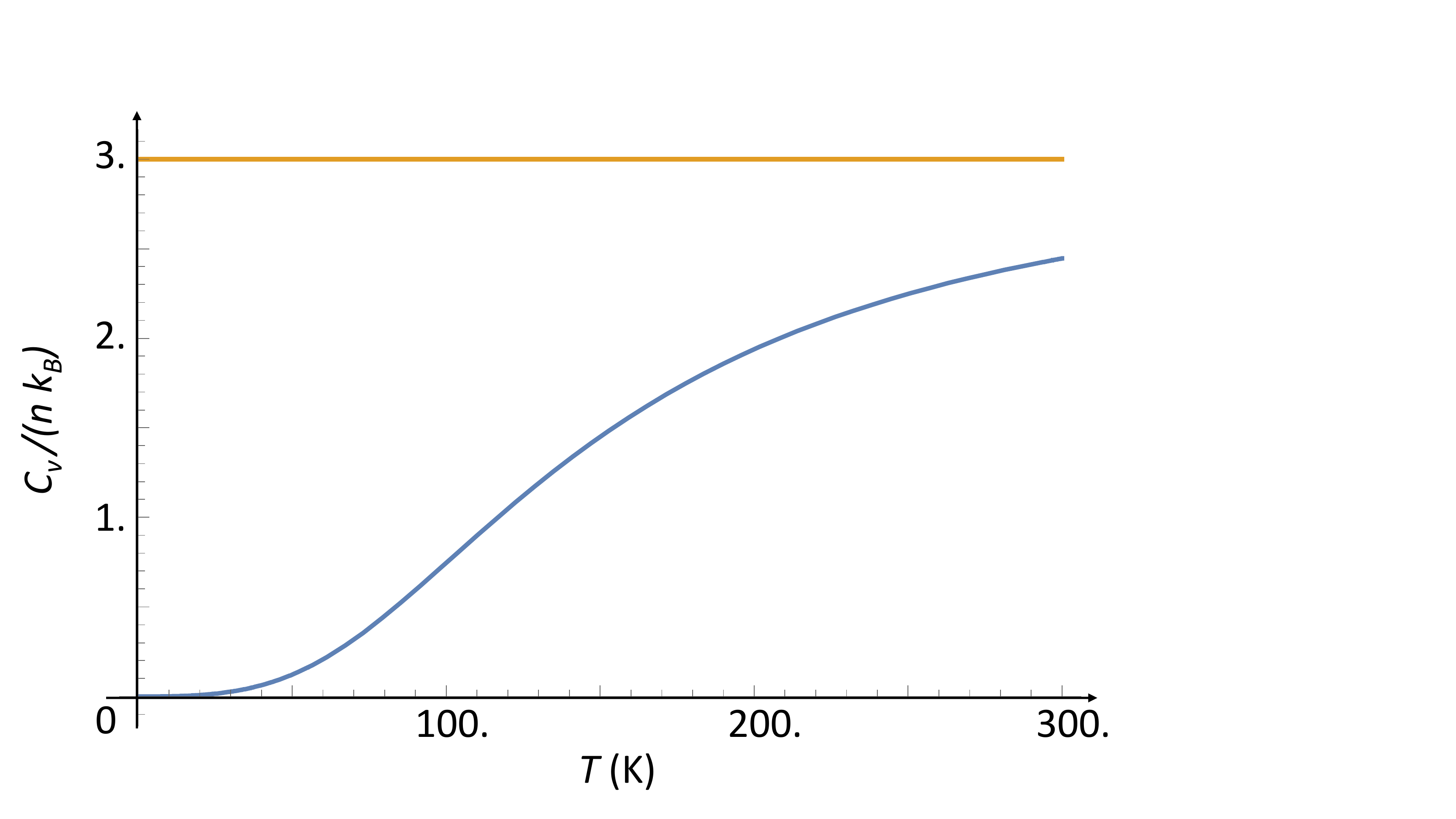}
			\caption{\small{ 
			Debye calculation of the Silicon specific heat $C_v$, with $T_D = \hbar \omega_D/k_B = 620~$K. At high temperature, it saturates at $3 n k_B$ with $n$ the number volumic density (Dulong-Petit law, horizontal line).}}
			\label{fig_4}
		\end{figure}

Summing up the energy stored in the whole phonon bath, and deriving it by temperature, one defines the specific heat $C_v$ (at fixed volume).
It can be written \cite{clelandBk,ziman,ashcroft}:
\begin{equation}
C_v(T) = k_B \int_0^\infty \left( \frac{\hbar \omega}{k_B T}\right)^{\!\!2} \frac{e^{\hbar \omega/(k_B T)}}{\left( e^{\hbar \omega/(k_B T)}-1 \right)^2} {\cal D}(\omega)\, d\omega , \label{capaCv}
\end{equation}
where ${\cal D}(\omega)$ is the density of states, i.e. the number of phononic states available between $\omega$ and $\omega+ d \omega$ per unit volume [in m$^{-3}$(Rad/s)$^{-1}$]. The function ${\cal D}(\omega)$ is directly obtained from the dispersion relations $\omega(\vec{q})$; in practice this displays a rather complex structure with strong peaks. 
A particularly useful simplified treatment has been proposed by Debye \cite{ashcroft}, considering a strictly linear dispersion law with average velocity $\bar{c}$ and replacing the first Brillouin zone by a sphere of radius $q_D$. This radius is chosen such that the total number of degrees of freedom is preserved: having $N$ atoms moving in 3D space, this makes a total of $3N$ modes. One obtains:
\begin{equation}
 {\cal D}(\omega) = \frac{3}{2 \pi^2 \bar{c}^3}\, \omega^2 \, \Theta_{\omega_D}(\omega) ,
\end{equation}
with $\Theta_{\omega_D}(\omega)$ the step function (1 for $\omega < \omega_D$ and zero above; $\omega_D = \bar{c} \, q_D$).
Eq. (\ref{capaCv}) is plotted in Fig. \ref{fig_4} in units of $n k_B$ ($n$ being the number volumic density), with $T_D = \hbar \omega_D/k_B = 620~$K ($1/q_D \approx 0.62~$\AA{}), the value corresponding to Silicon \cite{ashcroft}. The calculated curve is particularly close to experimental data \cite{abe}, despite the simplifications used.
At high temperatures, it saturates at 3: this is known as the {\it Dulong-Petit law}, a limit reached when all modes are equally populated \cite{clelandBk,ziman,ashcroft}. This is the ``classical limit'' which is relevant to the present paper. Around $200~$K, in Si $C_v$ is already of the order of 65$~\%$ of this value. \\

Phonons are collective excitations that behave in the harmonic approximation as a gas of free particles.
But even in an absolutely ideal crystal (free of any defects), they experience some interactions among themselves.  
These interactions are due to the anharmonicity of the actual potential $V$ that describes the solid.
Third order terms of the type $\partial^3 V /\partial x_i\partial x_j \partial x_k$ (with all permutations $x_i \longleftrightarrow y_i,z_i$ ; $x_j \longleftrightarrow y_j,z_j$ ;  $x_k \longleftrightarrow y_k,z_k$) in the Taylor expansion of $V$ lead to 3-phonon processes. Similarly, the fourth order terms lead to 4-phonon processes, and so on. Because of the conservation of energy and momentum, selection rules also apply to these processes.
The scattering rates can be calculated using perturbation theory and the Fermi Golden rule, see e.g. Refs. \cite{clelandBk,ziman} for detailed discussions.
These mechanisms are known as {\it inelastic scattering}, since they redistribute energy among the phonon gas. They are strongly temperature-dependent \cite{clelandBk,ziman}. 

Considering a more realistic situation, phonons also scatter from  {\it structural defects} of the crystal.
These are crystallographic mis-stackings, like vacancies (point-like defects), dislocations (line defects) or grain boundaries (surface defects). 
Even with a chemically pure solid, there is a natural isotopic variation within the material. This is the isotopic replacement of $^{28}$Si (most abundant) with $^{29}$Si and $^{30}$Si in our example, at random places of the lattice. It creates {\it mass point-like defects} off which phonons can scatter \cite{clelandBk}.
These defects do not possess any internal degrees of freedom to which energy can be transferred: thus these processes are {\it elastic}, and only redistribute momentum. They are in principle temperature-independent, because they depend only on geometric parameters \cite{clelandBk,ziman}. 

All these (independent) mechanisms sum up to define a total phonon scattering rate $\Gamma_{\hat{q},n}$, for a given dispersion branch $\hat{q},n$.
Averaging over all $ n,\hat{q} $  one can define $\bar{\Gamma}$, and within 
 the linear dispersion limit, an {\it average mean-free-path} $\bar{\lambda}_{mfp} = \bar{c} / \bar{\Gamma}$ using the average velocity $\bar{c}$. 
This is our second relevant lengthscale: it represents the typical distance over which motional energy is homogenized (in both $\omega$ and $\vec{q}$) over the phonon bath.  

Let us consider the high-temperature limit where $\bar{\lambda}_T <\bar{\lambda}_{mfp}$ 
and both are much smaller than the size of the solid. In this case, at any point $\vec{r}$ within the object we can define a phonon number distribution $n_T(\vec{r})$ which depends on a local temperature $T(\vec{r})$: this is known as the {\it diffusive limit} \cite{clelandBk,ziman}.
There is local equilibrium, while the global solid might be out-of-equilibrium if a heat flow is imposed. This heat flow can be described by the Boltzmann transport equation \cite{ashcroft}. After some manipulations, the heat current $\vec{j}_E$ (energy flow density carried by phonons, in W/m$^2$) can be written in the form \cite{clelandBk,ziman,ashcroft}:
\begin{equation}
\vec{j}_E = - \kappa \, \vec{grad} \, T, \label{fourier}
\end{equation}
with $\kappa = \frac{1}{3} C_v \bar{c} \, \bar{\lambda}_{mfp}$ the thermal conductivity of the material.
For Silicon, $\kappa \approx 200~$W.m$^{-1}$.K$^{-1}$ around 200$~$K \cite{kappaSi}, which leads to $\bar{\lambda}_{mfp} \approx 60~$nm (the typical size at which local equilibrium is realized). 
Eq. (\ref{fourier}) is the well-known {\it Fourier's law}, describing thermal transport at the macroscopic scale, which will be used in the next Subsection.
Since equilibrium is reached at a local scale, the thermal properties appear {\it isotropic} on the scale of the whole solid. \\

The whole point of this Subsection is to identify the conditions which are targeted by our modeling, and what {\it underlying limitations} have to be considered.
For the sake of simplicity, we shall consider in the following a {\it polycrystalline material} with a grain size large enough such that thermal and mechanical properties are well defined over it. 
Assuming the grains small enough compared to the size of the solid, and randomly oriented, 
the macroscopic mechanical properties 
reduce to averaged, and thus {\it isotropic}, quantities: the stiffness tensor $[c_{ijkl}]$ is defined by introducing only 2 independent parameters, see Section \ref{meca}. 
The size of grains is thus our last relevant lengthscale.

In polycrystalline Aluminum thin films used for micro-mechanics or superconducting quantum bits, {\it defects} are known to impact strongly device properties \cite{TLSus,ustinov}. These are {\it Two-Level-Systems} (TLS) that reside in the amorphous Aluminum oxide: atoms, or groups of atoms that can occupy different positions within the solid which are nearly equivalent energetically. The switching between these positions can be induced by tunneling or thermal activation, and the overall contribution of such defects generates a rather flat in energy distribution of low-energy states to which strain can couple \cite{phillips}. 
This means that in a glass, phonons will behave essentially like in a crystal at long wavelengths (where the actual atomic order does not really matter), but they will be much less well-defined: their scattering rate $\bar{\Gamma}$ is strongly impacted by their coupling to TLSs \cite{hunklinger,phillips}.
Thus for a non-perfect material, there are internal degrees of freedom which can absorb or emit energy from/to the phonon gas (Fig. \ref{fig_2} inset). 
This dissipative channel is clearly distinct from the phonon-phonon interaction already introduced. 
However, since defects interact (and thus thermalise) through phonons, and since these phonons obey local equilibrium, 
they shall be described by the same $T(\vec{r})$ profile. 

For intrinsic Silicon, even at room temperature the thermal population of charge carriers remains low enough to be completely ignored \cite{kappaSi}.
This is obviously not the case for a metal, like e.g. Aluminum, were electron-phonon coupling impacts $\bar{\Gamma}$. Besides, thermal conductivity is then dominated by electronic transport; this is a completely different regime than the one discussed here \cite{clelandBk, ashcroft}.

But there is a principle limitation in the description we do of collective modes: we assumed the dispersion relations to be {\it strictly linear} (i.e. the Debye approach). 
Obviously, the simple picture we have developed fails to represent reality at small (i.e. atomic) wavelengths and high energies.
At that scale, even the hypothesis of local equilibrium fails;
phonon frequencies are higher than $\bar{\Gamma}$, meaning that 
collisions are not frequent enough to thermalise over a typical period of the propagating mode.
However, 
numerical estimates (i.e. the Debye specific heat) are in rather good agreement with experiments.
For this reason, we shall use in the following the same pragmatic, and simplified approach for mechanical properties: elasticity theory will be assumed exact down to the smallest scales, and we shall bound the number of normal modes by the same argument as for the specific heat (imposing the total number of degrees of freedom to be $3N$). 
The relevance of the modal description also depends on the {\it damping} each mode experiences, a point discussed explicitly in Subection \ref{clamps} and Section \ref{ineq}.
As a matter of fact, local thermal equilibrium will be assumed within this extrapolation.
This is a sort of ``mechanical adaptation'' of Debye's hypothesis.

\subsection{Macroscopic description: Fourier's law}

The simple geometry we consider is depicted in Fig. \ref{fig_2}. 
A cylindrical rod of a homogeneous and isotropic material is connecting two thermal baths in vacuum. Its length is $L$ and radius $R$. 
On each side, a small region of length $\varepsilon \ll L$ represents the clamping zone where the radius monotonically increases in order to adapt to the bulk heat reservoirs.  
These small portions of the rod will receive particular attention in Subsection \ref{clamps} below. For the time being, 
we shall treat them on the same footing as the bulk of the rod. 
Indeed, everywhere in the bar the hypotheses described in the previous Subsection do apply: there is thus a well defined temperature at every point $T(r,\theta,z)$, and a well defined thermal conductivity $\kappa(T)$. For symmetry reasons, $T$ does actually not depend on $\theta$. \\

Equilibrium writes, in steady state with no internal heat load $div(\vec{j}_E)=0$. Using Fourier's law Eq. (\ref{fourier}) we obtain:
\begin{equation}
\frac{1}{r} \frac{\partial }{\partial r}\left( r \frac{\partial T}{\partial r}\right) + \frac{\partial^2 T }{\partial z^2} =0 .
\end{equation}
In the above, we assumed the thermal gradient along the rod to be very small such that $\kappa$ can be treated as a constant.
For the section of the rod of length $L$ and constant radius $R$, the problem is fairly easy to solve: the temperature is homogeneous across the rod (no radial heat flow), and depends linearly on the $z$ abscissa. However, within the clamps it is much more complex (and outside of the scope of the paper): as the beam flares-out on the right of Fig. \ref{fig_2} a), the energy flow spreads over a larger cross-section while the total flux is preserved. This means that the temperature gradient along $z$ should be reduced, and a local $r$-dependent gradient will appear; the reverse is true for the left side clamp.
The profile $T(0,z)$ on the axis of the rod is qualitatively drawn in Fig. \ref{fig_2} b).

We need now to specify concrete boundary conditions.
The thermal bath on the left is maintained at $T_c$, homogeneously over its entire (and very large) volume. The right bath is at $T_h$, and we arbitrarily chose $T_h>T_c$. On the inner-sides of the clamps, the temperatures at each extremity of the homogeneous section of the rod are defined as $T_{h,-}$ and $T_{c,+}$ (see Fig. \ref{fig_2}).
One of the key assumptions of the modeling is that the clamps are almost ideal, which implies $T_{h,-} \approx T_h $ and $T_{c,+} \approx T_c $.
The temperature profile thus reads (for $0<z<L$):
\begin{equation}
T(z) = \Delta T \left( \frac{z}{L} - \frac{1}{2} \right) + T_{avg} , \label{profile}
\end{equation}
with $\Delta T = T_h - T_c \ll T_{avg}$, and  $T_{avg}=(T_h+T_c)/2$, see Fig. \ref{fig_2} b). The heat flow $\dot{Q} = \int \!\!\! \int \vec{j}_E \, d\vec{S}$
across a section of the rod is then defined as:
\begin{equation}
\dot{Q} = -  \frac{\kappa(T_{avg}) \,\pi R^2}{L} \Delta T , \label{heat}
\end{equation}
the $-$ sign reminding that energy flows from the hot side to the cold one.
We recognize the standard expression for thermal conductance $G = \kappa \,\pi R^2/L$.
In the high-temperature limit where the specific heat is constant $C_v = 3 n k_B$, the {\it total motional energy density} stored in the beam (defined per unit length) can be written as:
\begin{equation}
\frac{d E_{tot}(z)}{d z} =  \frac{3 N}{L} k_B \,  T(z) ,    \label{energydens}
\end{equation}
having made use of $n= N/(\pi R^2 L)$.
This last expression is a straightforward implication of the local equilibrium assumption.

 Eqs. (\ref{profile},\ref{heat},\ref{energydens}) completely describe thermal properties at the macroscopic level, and are the {\it  basic inputs} for the mechanics-based theory we develop below.
	
\section{Continuum mechanics description}
\label{meca}

\subsection{Propagating acoustic waves}

We shall start the Section by recalling basics of continuum mechanics.
The distortion of a solid body is described by the displacement field $\vec{u}(\vec{r},t)$, which represents by how much any point $\vec{r}$ moves under a given solicitation. 
This solicitation is due to forces acting upon the object, which propagate inside it as stresses described by the tensor $[\sigma_{ij}(\vec{r},t)]$.
The strain tensor $[\epsilon_{kl}(\vec{r},t)]$ is constructed from the first derivatives of the displacement vector $\vec{u}$ with respect to space coordinates, keeping only the symmetric components.
Elasticity theory simply relates the two tensors through the stiffness $[c_{ijkl}]$:
\begin{equation}
[\sigma_{ij}(\vec{r},t)] = [c_{ijkl}][\epsilon_{kl}(\vec{r},t)] , \label{stiff}
\end{equation}
where the indexes $ijkl$ run each through the 3 directions of space $x,y,z$ \cite{clelandBk}.
$[\sigma_{ij}(\vec{r},t)]$ and $[\epsilon_{kl}(\vec{r},t)]$ are thus $3 \times 3$, while $[c_{ijkl}]$ contains 81 coefficients. 
But this complexity can be simplified drastically by taking into account the symmetries of the problem at hand.
For an isotropic material, only 2 components will be distinct. In the following, we chose to describe mechanical properties with the {\it Young's modulus} $E_Y$ and the {\it Poisson ratio} $\nu$. The former represents the elasticity of the rod when subject to a force along its axis, while the latter quantifies by how much it expands in the radial direction when it is pressed upon.
With a homogeneous solid, $[c_{ijkl}]$ is also $\vec{r}$ independent. 
Within the ``mechanical adaptation'' of Debye's hypothesis mentioned in the previous Section, we shall also consider that $[c_{ijkl}]$ is strictly constant up to the highest frequencies $\omega$: 
this means, equivalently, that it is also $t$-independent. 
Note that by construction this modeling fails to reproduce the high frequency optical branches. These modes are phenomenologically accounted for in the prolongation of the elasticity law {\it beyond} the atomic size, similarly to Debye's phononic treatment. \\

Newtonian dynamics brings the fundamental equation of motion for the field $\vec{u}(\vec{r},t)$ in the absence of body forces \cite{clelandBk}:
\begin{equation}
\rho \frac{\partial^2 \vec{u} }{\partial t^2} = \frac{E_Y}{2 (1 + \nu) (1 - 2 \nu)} \, \vec{grad} ( div[\vec{u}] )+ \frac{E_Y}{2(1+\nu)}  \, {\mathbf \Delta}  \vec{u}, \label{newton}
\end{equation}
which is the basis for {\it acoustics}, with $\rho $ the mass density of the material, and ${\mathbf \Delta}$ the vector Laplacian.
We assume $R \ll L$ so that the rod can be treated as infinite: then the displacement profile should display the same shape for all $z, t$.
This means that we can separate the variables:
\begin{eqnarray}
u_r(r,\theta,z,t) & = & \phi_{r \left\{\eta\right\}}(r,\theta) \, U_{\left\{\eta\right\}}(z,t) , \label{equr} \\
u_\theta(r,\theta,z,t) & = & \phi_{\theta \left\{\eta\right\}}(r,\theta)\, U_{\left\{\eta\right\}}(z,t), \label{equq}\\
u_z(r,\theta,z,t) & = & \phi_{z \left\{\eta\right\}}(r,\theta) R\, \frac{\partial U_{\left\{\eta\right\}}}{\partial z}(z,t), \label{equz}
\end{eqnarray}
written in cylindrical coordinates. $U_{\left\{\eta\right\}}$ represents a propagating solution along $z$, while the $\phi_{i \left\{\eta\right\}}$ (adimensional, with $i=r, \theta,z$) describe the shape of the distortion within a section. The set $\left\{\eta\right\}$ corresponds to the parameters that will distinguish the different existing solutions (see below). Note the $z$-derivative in Eq. (\ref{equz}) [see Appendix \ref{pochhammer}].

The mathematical formalism of the next Subsections is built on the assumption that {\it all the existing normal modes} within the rod are represented, at least within our ``Debye'' simplification. We shall therefore spend some time on describing how these modes are constructed.
The solutions are found by injecting Eqs. (\ref{equr} - \ref{equz}) into Eq. (\ref{newton}). 
The fundamental equation of motion is satisfied with:
\begin{eqnarray}
\frac{\partial^2 U_{\left\{\eta\right\}}}{\partial z^2}(z,t) & = & {\mathcal K}\, U_{\left\{\eta\right\}}(z,t) , \label{zderive} \\
\frac{\partial^2 U_{\left\{\eta\right\}}}{\partial t^2}(z,t) & = & {\mathcal W}\, U_{\left\{\eta\right\}}(z,t),  \label{tderive}
\end{eqnarray}
with ${\mathcal K}$ and ${\mathcal W}$ two constants. 
Propagating waves require ${\mathcal W}=- \omega^2$ (which we chose $\omega >0$ with no loss of generality), and can be classified in 3 categories (solving generically for ${\mathcal K} \in \mathbb{C}$) \cite{poche}:
\begin{itemize}
\item Traveling, ${\mathcal K}=-k_p^2$, $k_p \in  \mathbb{R}$,
\item Evanescent,  ${\mathcal K}=+k_e^2$, $k_e \in  \mathbb{R}$,
\item Mixed, ${\mathcal K}=(k_e \pm \mathrm{i} \, k_p)^2 , \left\{k_e,k_p\right\} \in  \mathbb{R}^2$.
\end{itemize}
The real-valued solutions for the two first ones are:
\begin{eqnarray}
\mbox{Travel.     } \,\, U_{\left\{\eta\right\},0}     (z,t) & = & U_0 \, \cos (\pm k_p z- \omega t) , \label{travel1} \\
   \mbox{or}        \,\, U_{\left\{\eta\right\},\pi/2} (z,t) & = & U_0 \, \sin (\pm k_p z- \omega t) , \label{travel2} \\
\mbox{Evan.     }   \,\, U_{\left\{\eta\right\},0}     (z,t) & = & U_0 \, \exp (\pm k_e z)  \, \cos (\omega t) , \label{eva1} \\
    \mbox{or}       \,\, U_{\left\{\eta\right\},\pi/2} (z,t) & = & U_0 \, \exp (\pm k_e z)  \, \sin (\omega t) , \label{eva2}
\end{eqnarray}
having defined explicitly the two equivalent time-dependencies with orthogonal phases $0,\pi/2$.
$U_0$ is a (yet arbitrary) amplitude, and the sign $\pm$ defines in which direction the wave travels or decays.
For the mixed solutions we have:
\begin{eqnarray}
\mbox{Mix.     } \,  U_{\left\{\eta\right\},0}     (z,t) & = & U_0 \, \exp (\pm k_e z) \cos (\pm k_p z- \omega t) ,\label{mix1} \\
   \mbox{or}     \,  U_{\left\{\eta\right\},\pi/2} (z,t) & = & U_0 \, \exp (\pm k_e z) \sin (\pm k_p z- \omega t), \label{mix2}
\end{eqnarray}
with the complex solutions built as $U_{\left\{\eta\right\},0} \pm  \mathrm{i} \, U_{\left\{\eta\right\},\pi/2}$.
In the above expressions, we shall keep only solutions with $k_e k_p >0$ for which the mixed wave travels and decays in the same direction (exponential growth being unphysical). \\

The functions $\phi_{r \left\{\eta\right\}}(r,\theta)$, $\phi_{\theta \left\{\eta\right\}}(r,\theta)$ and $\phi_{z \left\{\eta\right\}}(r,\theta)$ that satisfy Eq. (\ref{newton}) are described in Appendix \ref{pochhammer}; the impact of in-built stress is discussed therein. They are regrouped in 3 families: torsional (T), longitudinal (L) and flexural (F) which are known as {\it Pochhammer-Chree} waves \cite{poche}. 
Similar results can be obtained for a rectangular beam (with a rather involved formalism describing the $\phi_{i \left\{\eta\right\}}$ functions, $i=x,y,z$), see Ref. \cite{bondarenk}.
For torsional and longitudinal modes, a single index $m\geq0$ describes all the waves of the family; flexural waves require two indexes $n>0,m\geq0$, and are represented in two degenerate sets (defined by a rotation angle $\theta_0=0,+\pi/2$). 
Having chosen a wave within a family from its index(es), and having chosen a type of wave (traveling, evanescent, or mixed), solving the dynamics equation generates a relationship between angular frequency $\omega$ and wavevectors(s) $k_p, k_e$: this is a {\it dispersion relation}.
All dispersion relations can be represented in 3D space with vertical axis $\omega$ and abscissae the complex plane $k_p + \mathrm{i} \, k_e$ (only first quadrant $k_e, k_p>0$ since all properties are propagation-direction independent). This generates rather complex plots that require to be commented \cite{poche,onoe,pao1,pao2}.

		\begin{figure}[t!]
		\centering
	\includegraphics[width=12.1 cm]{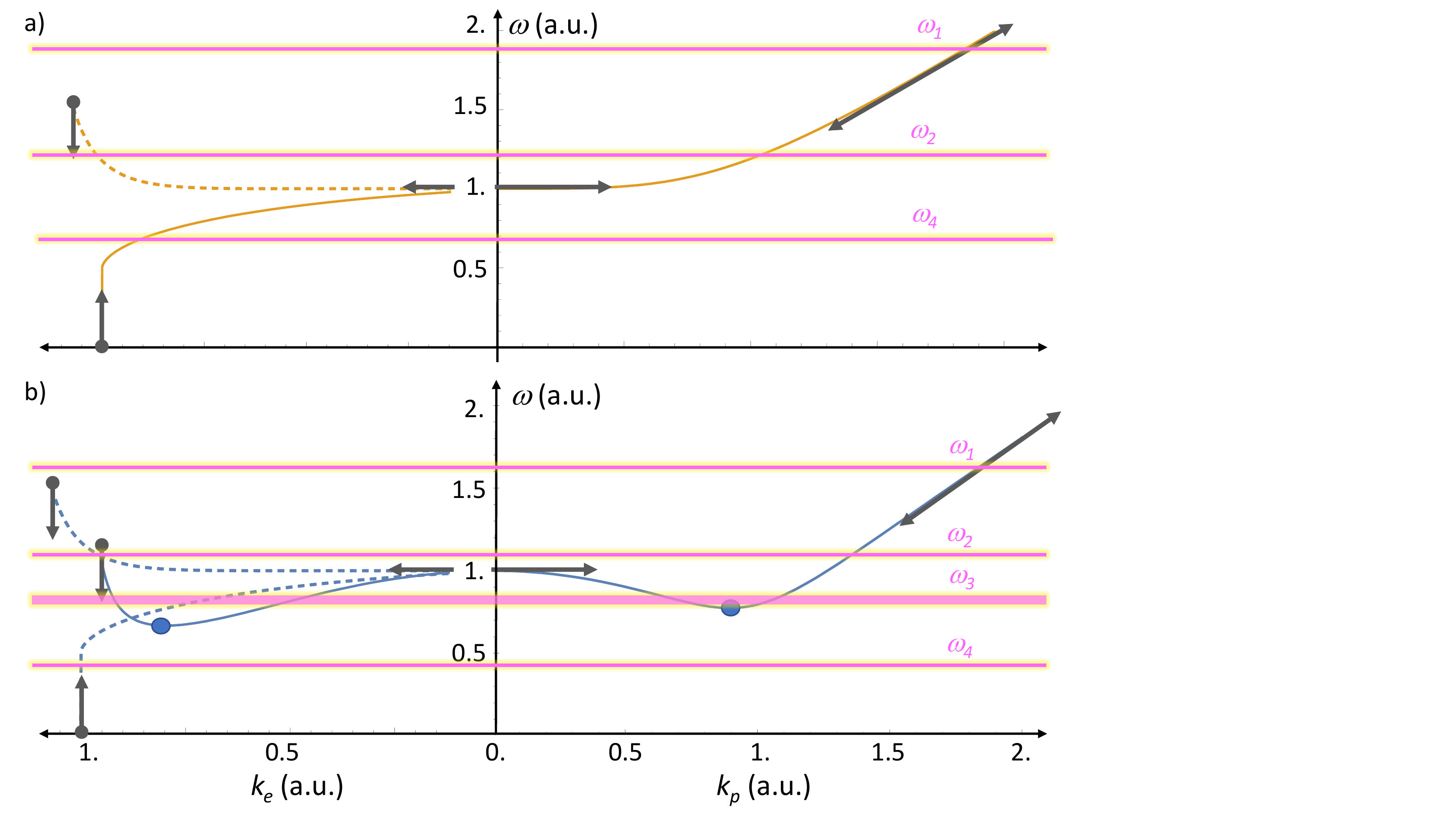}
			\caption{\small{ 
			Typical schematics of wave dispersion relations linking the traveling and evanescent solutions: a {\it branch} (here, strictly higher than the 3 lowest ones; see Appendix \ref{pochhammer} for details). a) The most simple situation in full line; a variant for the evanescent side is shown in dashed. b) Full line: most complex situation that can be imagined, with one minimum on each panel. Dashed lines: simpler variants for the evanescent part. The dot symbols represent the connections to mixed waves (which dispersion relations are out of the planes). Secondary branches not drawn (see text). Asymptotes are shown as arrows; single-sided when the branch terminates (at the dotted end where it connects to another branch or to $\omega=0$, left panel), or double-sided when it does not. Horizontal lines: imposed fixed frequencies $\omega_i$ (see text). }}
			\label{fig_5}
		\end{figure}
		
We write $\omega_{\left\{ m,\mbox{\tiny T} \right\}}(k_p)$, $\omega_{\left\{ m,\mbox{\tiny T}\right\}}(k_e)$ for the traveling and evanescent torsional dispersion relations, and similarly $\omega_{\left\{ m,\mbox{\tiny L}\right\}}(k_p)$, $\omega_{\left\{ m,\mbox{\tiny L}\right\}}(k_e)$ for longitudinal, with index $m$. Flexural dispersion relations are designated by $\omega_{\left\{ n,m,\mbox{\tiny F}\right\}}(k_p)$, $\omega_{\left\{ n,m,\mbox{\tiny F}\right\}}(k_e)$ following the same conventions. Only the lowest torsional and longitudinal waves (i.e. $m=0$) do not support evanescent propagation. 
Traveling wave dispersion relations are drawn in the $(k_p,k_e=0)$ plane, while evanescent ones are drawn in the  $(k_p=0,k_e)$ one. For a given wave, these two join at the vertical axis, when $k_p=k_e=0$. They form what we shall call a {\it branch}, that can be represented by flattening the two planes onto the same page, see Appendix \ref{pochhammer} and schematics in Fig. \ref{fig_5}.
Only the 3 lowest branches ($m=0$ torsional and longitudinal, $n=1,m=0$ flexural) go all the way to $\omega=0$ for $k_p=0$. All other branches have a {\it finite frequency} $\omega \neq 0$ at $k_p=ke=0$ (as is the case in Fig. \ref{fig_5}). \\

The zoology of these modes needs to be discussed, since our modeling requires an exhaustive description of the rod's mechanical solutions.
For longitudinal and flexural waves, branches with different index(es) connect on the evanescent side in a complex manner which depends strongly on the Poisson ratio $\nu$ \cite{onoe,pao2}. They are also interconnected by mixed wave solutions (pieces of branches with both $k_p\neq0, k_e\neq0$ that {\it are not} contained in one of the above mentioned planes), which originate and end up at {\it local minimas} of the branches (dots in Fig. \ref{fig_5}) \cite{poche,onoe}.
Actually, most of the evanescent solutions possess also pieces of branches which are not directly connected to 
the traveling solution, but are linked to the solution network only by other branches or mixed wave solutions, see Refs. \cite{poche,onoe,pao2}. 
We shall call them {\it secondary branches}, as opposed to the {\it main branch} which connects to the  $k_p$ plane.
Mixed waves and secondary branches are not drawn in Appendix \ref{pochhammer}, because they are not 
relevant to our discussion (see Subsection \ref{modes} below): the key argument being that they always link a given main branch to {\it lower ones}, or directly to the $\omega=0$ plane \cite{poche}. 
Standard 
relevant 
configurations are thus shown in Fig. \ref{fig_5} (for branches strictly higher than the 3 lowest ones). At large $k_p \gg 1/R$, the dispersion relations are always linear. On the evanescent side, the branch is always limited in $k_e$ (with a rather weak span of available $\omega$ values). 
At connecting points, 
the group velocity $d \omega/d k_e$ is either zero or infinite. 

A solution on a given branch is thus described by a set of parameters $\left\{\eta\right\}$, as introduced above.
It regroups the index(es) defining the branch plus the wavevector (e.g. $\left\{ m,\mbox{L}, k_p\right\}$ for propagating longitudinal wave $m$ of wavevector $k_p$, or $\left\{ n,m,\mbox{F}, k_e\right\}$ for evanescent flexural wave $n,m$ of wavevector $k_e$). \\
		
\subsection{Normal modes}
\label{modes}

The {\it normal modes} of the rod are standing waves obtained from the previous propagating solutions.
These are due to the boundary conditions chosen at each end of the rod. 
We should thus define these (idealized) boundary conditions for the amplitude parameter that describes each of the wave families (see Appendix \ref{pochhammer} for details on this parametrization). \\

For torsional waves, the amplitude of motion of a mode is defined from the tangential displacement at the periphery $u_\theta(r=R,\theta,z,t)$. By construction, this means that $\phi_{\theta \left\{\eta\right\}}(r=R,\theta)=1$ for any set $\left\{\eta\right\}$.
The boundary conditions then naturally write:
\begin{equation}
u_\theta(r=R,\theta,z=0;L,t) = 0, \label{boundT}
\end{equation}
for any $\theta$ and $t$.

Similarly, the longitudinal waves are characterized by the motion amplitude in the $z$-direction on the axis, $u_z(r=0,\theta,z,t)$, meaning $\phi_{z \left\{\eta\right\}}(r=0,\theta)=1$ for any $\left\{\eta\right\}$. The boundary conditions become:
\begin{equation}
u_z(r=0,\theta,z=0;L,t) = 0. \label{boundL}
\end{equation}
Eqs. (\ref{equr}-\ref{equz}) are formally equivalent under the exchange:
\begin{eqnarray}
U_{\left\{\eta\right\}}(z,t) & \rightarrow &  \int_z U_{\left\{\eta\right\}}(z',t) dz' / R,  \label{replacez1} \\
\frac{\partial U_{\left\{\eta\right\}}}{\partial z} (z,t) & \rightarrow & U_{\left\{\eta\right\}}(z,t) / R, \label{replacez2}
\end{eqnarray}
so that we can still seek solutions for longitudinal motion using the same Eqs. (\ref{travel1}-\ref{mix2});
the boundary condition Eq. (\ref{boundL}) is thus formally equivalent to Eq. (\ref{boundT}).

The flexural waves amplitude is more complex to define: indeed, the motion is characterized through a {\it planar} displacement $\vec{u}_F=u_r(R,\Theta_F,z,t)\vec{e}_r+u_\theta(R,\Theta_F,z,t) \vec{e}_\theta$ of the surface of the rod, defined for a specific direction $\Theta_F$ (see Appendix \ref{pochhammer}), 
which leads to $\phi_{r \left\{\eta\right\}}(R,\Theta_F)^2+\phi_{\theta \left\{\eta\right\}}(R,\Theta_F)^2=1$ for any $\left\{\eta\right\}$.
The boundary conditions should then stipulate that $\vec{u}_F$ vanishes on the two extremities:
\begin{equation}
\left|\vec{u}_F(r=R,\theta=\Theta_F,z=0;L,t)\right| = 0, \label{boundF}
\end{equation}
which involves now both $u_r$ and $u_\theta$.
Eqs. (\ref{boundT},\ref{boundL},\ref{boundF}) are referred to as idealized ``weak clamping'', which limits as little as possible the displacement field $\vec{u}$ at extremities. 
This modeling will be further discussed in Subsection \ref{clamps}. \\

Let us consider now that energy is fed into the rod at a given frequency $\omega$, toward a specific branch. 
The different situations that can be encountered are depicted in Fig. \ref{fig_5}.  
The simplest situation is the one obtained with $\omega_1$ in both a) and b) sub-figures: only one (traveling) wavevector can afford this frequency. 
Since the $\phi_{i \left\{\eta\right\}}$ functions ($i=r,\theta,z$) do not depend on sign $\pm k_p$, using Eqs. (\ref{equr},\ref{equq},\ref{equz}) all the ``weak clamp'' conditions
listed above reduce to $U_{\left\{\eta\right\}}(z=0;L,t)=0$.
It is then matched from Eq. (\ref{travel1}) summing up two waves traveling in opposite directions $\pm k_p$, with opposite amplitudes $U_0$ and $-U_0$, leading to:
\begin{equation}
U_{\left\{\eta\right\},0}     (z,t)  =   2 U_0 \, \sin ( k_p z) \, \sin(\omega t) ,
\end{equation}
for which $2 U_0\, \sin(\omega t)$ factorizes out in Eqs. (\ref{equr}-\ref{equz}).
Similarly form Eq. (\ref{travel2}) one obtains:
\begin{equation}
U_{\left\{\eta\right\},\pi/2}     (z,t)  = 2 U_0 \, \sin ( k_p z) \, \cos(\omega t) ,
\end{equation}
with the complementary phase for the time-oscillation. The wavevector $k_p$ should then verify:
\begin{equation}
k_p   =   \frac{\pi }{L} q \mbox{,  } q \in \mathbb{N}^* . \label{modesol}
\end{equation}

The same situation applies to the case $\omega_2$ in Fig. \ref{fig_5} a) considering the full line, or to the lowest torsional and longitudinal branches ($m=0$) which do not support evanescent solutions (see Appendix \ref{pochhammer}). This is indeed trivially equivalent to what is usually done in textbooks when considering low frequency/long wavelength torsional/longitudinal modes in beam theory, see for instance Ref. \cite{clelandBk}.
Explicitly, the displacement field $\vec{u}_{\left\{\eta\right\}}(r,\theta,z,t)$ reads:

\begin{eqnarray}
u_{r \left\{\eta\right\}}(r,\theta,z,t) & = & \phi_{r \left\{\eta\right\}}(r,\theta) \, \sin(k_p z)\, U_{\omega}(t) , \label{equr1} 
\end{eqnarray}
\newpage
\begin{eqnarray} 
u_{\theta \left\{\eta\right\}}(r,\theta,z,t) & = & \phi_{\theta \left\{\eta\right\}}(r,\theta)\, \sin(k_p z)\, U_{\omega}(t), \label{equq1}\\
u_{z \left\{\eta\right\}}(r,\theta,z,t) & = & \phi_{z \left\{\eta\right\}}(r,\theta) \,R k_p \cos(k_p z)\, U_{\omega}(t), \label{equz1}
\end{eqnarray}
for torsional (T) and flexural (F) solutions, and:
\begin{eqnarray}
u_{r \left\{\eta\right\}}(r,\theta,z,t) & = & -\phi_{r \left\{\eta\right\}}(r,\theta) \, \frac{\cos(k_p z)}{R k_p}\, U_{\omega}(t) , \label{equr2} \\
u_{\theta \left\{\eta\right\}}(r,\theta,z,t) & = & -\phi_{\theta \left\{\eta\right\}}(r,\theta)\, \frac{\cos(k_p z)}{R k_p}\, U_{\omega}(t), \label{equq2}\\
u_{z \left\{\eta\right\}}(r,\theta,z,t) & = & \phi_{z \left\{\eta\right\}}(r,\theta) \, \sin(k_p z)\, U_{\omega}(t), \label{equz2}
\end{eqnarray}
in the longitudinal (L) case. Now on, we
write $U_{\omega}(t) = U_0 \cos(\omega t) + U_{\pi/2} \sin(\omega t)$ the generic time dependence; with 
the $\phi_{i \left\{\eta\right\}}$ functions properly normalized, the amplitude of motion is encoded in the two quadratures of the motion $U_{0},U_{\pi/2}$ (in meters).
A given normal mode motion is {\it completely defined} by these amplitudes, together with the set $\left\{\eta\right\}$ to which we add the standing wave parameter $q$.
These expressions that we call ``string-like'' (in analogy with the high-stress solutions of Euler-Bernoulli theory) are the ones we will use below when defining mode energies. \\

Consider now the case $\omega_2$ in Fig. \ref{fig_5} a) with the dashed line as evanescent contribution, or equivalently $\omega_2$ in Fig. \ref{fig_5} b) with the full line branch (or the dashed one turned upwards).
There are now {\it two} possibilities to support an excitation at $\omega_2$, one at finite $k_p$ and the other one at finite $k_e$.
Following the same philosophy as above, we should now combine Eqs. (\ref{travel1},\ref{travel2},\ref{eva2}) with wavevectors $\pm k_p$, $\pm k_e$, factorizing out the same time-oscillation $\sin (\omega t)$ [and similarly with Eqs. (\ref{travel1},\ref{travel2},\ref{eva1}) for the $\pi/2$ shifted one $\cos (\omega t)$]; however now, the $\phi_{i \left\{\eta\right\}}$ do not factorize out, since they are generally {\it different} for $k_p$ and $k_e$ waves.
Besides, we need an extra boundary condition to fix the amplitudes of the evanescent components.
This condition is chosen to be that the $z$-derivative of the amplitude parameter vanishes at the two extremities:
\begin{eqnarray}
\frac{\partial u_z}{\partial z} (r=0,\theta,z=0;L,t) & = & 0 \,\,\, \mbox{for L-waves} , \\
\left|\frac{\partial \vec{u}_F}{\partial z} \right| (R,\Theta_F,z=0;L,t) & = & 0 \,\,\, \mbox{for F-waves} . \label{bound2F}
\end{eqnarray}
We call this situation idealized ``strong clamping'' (as opposed to the previous one). Such a situation is never encountered by torsional (T) waves (see Appendix \ref{pochhammer}). The approach matches the standard one performed in the framework of the Euler-Bernoulli modeling \cite{clelandBk}. 
Besides, we shall discuss explicitly in Subsection \ref{clamps} how the boundary conditions are related to the {\it energy flow} toward the anchoring points. \\

After some manipulation, the solution writes for longitudinal (L) modes:

\begin{eqnarray}
& & u_{r \left\{\eta\right\}}(r,\theta,z,t)  = U_{\omega}(t) \times \nonumber \\
&& \!\!\!\!\!   \left[ \left( - \phi_{r \left\{\eta_p \right\}}(r,\theta) \frac{\cos[k_p z] }{R k_p} - \phi_{r \left\{\eta_e \right\}}(r,\theta) \frac{k_p \cosh[k_e z]}{R k_e^2}  \right) \right. \nonumber \\
  & & \!\!\!\!\!\!\!\!\!\!    \left. + c_{ep} \left( \phi_{r \left\{\eta_p \right\}}(r,\theta) \frac{\sin[k_p z]}{R k_p} - \phi_{r \left\{\eta_e \right\}}(r,\theta) \frac{\sinh[k_e z]}{R k_e} \right) \right] \! , \\
	& & u_{\theta \left\{\eta\right\}}(r,\theta,z,t)  = U_{\omega}(t) \times \nonumber \\
&& \!\!\!\!\!   \left[ \left( - \phi_{\theta \left\{\eta_p \right\}}(r,\theta) \frac{\cos[k_p z] }{R k_p} - \phi_{\theta \left\{\eta_e \right\}}(r,\theta) \frac{k_p \cosh[k_e z]}{R k_e^2}  \right) \right. \nonumber \\
  & & \!\!\!\!\!\!\!\!\!\!    \left. + c_{ep} \left( \phi_{\theta \left\{\eta_p \right\}}(r,\theta) \frac{\sin[k_p z]}{R k_p} - \phi_{\theta \left\{\eta_e \right\}}(r,\theta) \frac{\sinh[k_e z]}{R k_e} \right) \right] \! , \\
		& & u_{z \left\{\eta\right\}}(r,\theta,z,t)  = U_{\omega}(t) \times \nonumber \\
&& \!\!\!\!\!   \left[ \left(  \phi_{z \left\{\eta_p \right\}}(r,\theta) \sin[k_p z]  - \phi_{z \left\{\eta_e \right\}}(r,\theta) \frac{k_p \sinh[k_e z]}{k_e}  \right) \right. \nonumber \\
  & & \!\!\!\!\!\!\!\!\!\!    \left. + c_{ep} \left( \phi_{z \left\{\eta_p \right\}}(r,\theta) \cos[k_p z] - \phi_{z \left\{\eta_e \right\}}(r,\theta) \cosh[k_e z] \right) \bigg] \right. \! , 
\end{eqnarray}
with the generic time dependence $U_{\omega}(t)$ and the difference between traveling and evanescent functions denoted by $\eta_p,\eta_e$ respectively.

For flexural (F) modes the situation at hand is again more complex. 
Eqs. (\ref{boundF},\ref{bound2F}) are satisfied for $\phi_{r \left\{\eta_p \right\}}(R,\Theta_F)=\phi_{r \left\{\eta_e \right\}}(R,\Theta_F)$ and $\phi_{\theta \left\{\eta_p \right\}}(R,\Theta_F)=\phi_{\theta \left\{\eta_e \right\}}(R,\Theta_F)$ simultaneously.
While it might seem to over-constrain the problem, the choice of the $\Theta_F$ plane reference makes it guaranteed (see Appendix \ref{pochhammer}).
Postponing the discussion of this fact, we obtain:
\begin{eqnarray}
		& & u_{r \left\{\eta\right\}}(r,\theta,z,t)  = U_{\omega}(t) \times \nonumber \\
&& \!\!\!\!\!  \left[ \left(  \phi_{r \left\{\eta_p \right\}}(r,\theta) \sin[k_p z]  - \phi_{r \left\{\eta_e \right\}}(r,\theta) \frac{k_p \sinh[k_e z]}{k_e}  \right) \right. \nonumber \\
  & & \!\!\!\!\!\!\!\!\!\!    \left. + c_{ep} \left( \phi_{r \left\{\eta_p \right\}}(r,\theta) \cos[k_p z] - \phi_{r \left\{\eta_e \right\}}(r,\theta) \cosh[k_e z] \right) \bigg]  \right. \!, \\
			& & u_{\theta \left\{\eta\right\}}(r,\theta,z,t)  = U_{\omega}(t) \times \nonumber \\
&& \!\!\!\!\!   \left[ \left(  \phi_{\theta \left\{\eta_p \right\}}(r,\theta) \sin[k_p z]  - \phi_{\theta \left\{\eta_e \right\}}(r,\theta) \frac{k_p \sinh[k_e z]}{k_e}  \right) \right. \nonumber \\
  & & \!\!\!\!\!\!\!\!\!\!    \left. + c_{ep} \left( \phi_{\theta \left\{\eta_p \right\}}(r,\theta) \cos[k_p z] - \phi_{\theta \left\{\eta_e \right\}}(r,\theta) \cosh[k_e z] \right) \bigg]  \right. \!, \\
	& & u_{z \left\{\eta\right\}}(r,\theta,z,t)  = U_{\omega}(t) \times \nonumber \\
&& \!\!\!\!\!\!\!\!\!\!\!  \left[ \left( \phi_{z \left\{\eta_p \right\}}(r,\theta) R k_p \cos[k_p z] - \phi_{z \left\{\eta_e \right\}}(r,\theta) R k_p \cosh[k_e z]  \right) \right. \nonumber \\
  & & \!\!\!\!\!  \left. - c_{ep} \times \right. \nonumber \\
	&&\!\!\!\!\!\!\!\!\!\!\!\!    \left.  \left( \phi_{z \left\{\eta_p \right\}}(r,\theta) R k_p \sin[k_p z] + \phi_{z \left\{\eta_e \right\}}(r,\theta) R k_e \sinh[k_e z]  \right)   \right] \!. \label{eqbeammode} 
\end{eqnarray}
The constant writes:
\begin{equation}
c_{ep}  =  -\frac{k_e \sin L k_p - k_p \sinh L k_e}{k_e \left( \cos L k_p - \cosh L k_e\right)} ,                                 
\end{equation}
and the wavevectors $k_p, k_e$ should satisfy:
\begin{eqnarray}
1- \cos L k_p \cosh L k_e \mbox{\hspace*{3.5cm}} & & \nonumber \\
+\frac{(k_e - k_p)(k_e+k_p)}{2 \, k_e k_p} \sin L k_p \sinh L k_e & = &  0.         \label{modesol2}                         
\end{eqnarray}
Considering $k_e(k_p)$ as a function implicitly parametrized through $\omega$, Eq. (\ref{modesol2}) is the equivalent of Eq. (\ref{modesol}) leading to a {\it discrete set} of $k_p$ possibilities that we index again with $q > 0$, and conveniently incorporate in $\left\{\eta\right\}$. 
This situation also applies to the first flexural branch $n=1,m=0$ for small enough frequencies, when both traveling and evanescent solutions exist (see Appendix \ref{pochhammer}). It is in direct correspondence with the solving of the Euler-Bernoulli equation (at low in-built stress), which describes the same physics at small wavelengths/low frequencies (with the neat simplification $k_e=k_p$) \cite{clelandBk}. 
For this reason, We shall call ``beam-like'' the above mentioned expressions. \\

The situation encountered for $\omega_4$ in Fig. \ref{fig_5} is rather trivial: only evanescent or mixed waves can sustain this frequency. 
One can show that in this case, it is {\it impossible} to match the boundary conditions; therefore for a given branch, no standing waves can exist as soon as there is no traveling solution available.

The situation depicted in Fig. \ref{fig_5} b) with $\omega_3$ is the most interesting. 
In this case, {\it two} traveling waves match on the right panel.
Consider first an evanescent part of the branch of the type depicted by the dashed curve turned upwards: there is then no evanescent solution.
Strictly speaking, it is impossible to satisfy the boundary conditions at the same time for two $k_p$ having {\it exactly} the same $\omega$: therefore, 
we shall have two distinct waves with very close $\omega$, both verifying Eq. (\ref{modesol}).  
If one considers the dashed evanescent branch turned downwards, there is then {\it also one} evanescent solution. In this case again we shall have two distinct solutions, with very close $\omega$, but verifying this time Eq. (\ref{modesol2}).
The most complex situation arises for the full line evanescent branch where {\it two} evanescent solutions match $\omega_3$. Four possible combinations will then coexist with very close frequencies, each related to a distinct equation of the type of Eq. (\ref{modesol2}).
There is a sort of ``degeneracy lifting'' which we represented in Fig. \ref{fig_5} with a fuzzy and broader line for $\omega_3$ than for the others. \\

Imposing the boundary conditions thus turned all the branches into infinite sets of standing wave modes, indexed through $q>0$: these are the {\it normal modes} $\vec{u}_{\left\{\eta\right\}} (r,\theta,z,t)$ that describe all possible deformations of the rod in the limit of linear elasticity.
Consider now {\it an arbitrary} standing wave $\vec{u}$ within the rod. It is obtained by a linear superposition of all the solutions we constructed, formally:
\begin{equation}
\vec{u} (r,\theta,z,t) = \sum_{\left\{\eta\right\}} \vec{u}_{\left\{\eta\right\}} (r,\theta,z,t). \label{superpose}
\end{equation}
This is our starting point for the mechanical description of the rod energetics.

The vast majority of the normal modes are of the ``string-like'' nature, especially {\it all asymptotic behaviors} at large $k_p$ (which is of importance when dealing with {\it convergence} issues in the following Sections); only some branches support ``beam-like'' modes, at (reasonably) small $k_p$. For this reason (and also pragmatically because all the analytics is much simpler), we will restrict in Sections \ref{ineq} and \ref{gradient}
the mathematical description to the one referring to Eqs. (\ref{equr1}-\ref{equz2}). 
But all the presented concepts also apply to ``beam-like'' solutions, and we leave the derivation of the proper related expressions for a future work. 

\subsection{Stored energy: kinetic and potential}
\label{energy}

We can now define the mechanical energy stored in the rod by an arbitrary motion $\vec{u}$.
We thus start by recalling basics of continuum mechanics that can be found in textbooks, see e.g. Ref. \cite{clelandBk}.
The (volumic) density of kinetic energy writes:
\begin{equation}
{\cal E}_{k} = \frac{1}{2}  \, \rho \left( \dot{u_r}^2 + \dot{u_\theta}^2 +\dot{u_z}^2 \right), \label{kin}
\end{equation}
while the potential energy density ${\cal E}_{p}$ is the sum of a bending term $ {\cal E}_{E_Y}$ and a stretching term $ {\cal E}_\sigma $ defined by Eq. (\ref{eqEsigma}), which is nonzero only when an axial stress $\sigma_0 \neq 0$ is in-built.
The latter corresponds to the elastic energy stored in the rod by the length variations caused by motion, which we model by a body force.
This approach is not exact but preserves the Pochhammer-Chree waves as being solutions. It neglects an addendum $ {\cal E}_0 $ which is essentially irrelevant, see the discussion of Appendix \ref{inbuiltstress} for details.
The bending energy density is defined in continuum mechanics from the strain $[\epsilon_{kl}]$ and stress $[\sigma_{ij}]$ tensors as:
\begin{eqnarray}
 {\cal E}_{E_Y} &=& \frac{1}{2} \left(  \epsilon_{rr} \, \sigma_{rr} + \epsilon_{\theta \theta}  \, \sigma_{\theta \theta} + \epsilon_{zz}  \, \sigma_{zz} \right)  \nonumber \\
& + & \epsilon_{r \theta} \, \sigma_{r \theta} + \epsilon_{r z} \, \sigma_{r z} + \epsilon_{\theta z} \, \sigma_{\theta z} , \label{pot}
\end{eqnarray} 
written in polar coordinates. The strain field is obtained from the displacement vector $\vec{u}$:
\begin{eqnarray}
\epsilon_{rr}  & = & \frac{\partial u_r} {\partial r} , \label{strainr} \\
\epsilon_{\theta \theta}  & = & \frac{u_r} { r} +\frac{1} { r} \frac{\partial u_\theta} {\partial \theta}  , \\
\epsilon_{zz}  & = & \frac{\partial u_z} {\partial z} , \\
\epsilon_{r \theta}  & = & \frac{1}{2} \left( \frac{\partial u_\theta} {\partial r} +\frac{1} { r} \frac{\partial u_r} {\partial \theta} -\frac{u_\theta} { r}\right), \\
\epsilon_{r z}  & = & \frac{1}{2} \left( \frac{\partial u_r} {\partial z} + \frac{\partial u_z} {\partial r} \right), \\
\epsilon_{\theta z}  & = & \frac{1}{2} \left( \frac{\partial u_\theta} {\partial z} + \frac{1} { r} \frac{\partial u_z} {\partial \theta} \right),
\end{eqnarray} 
and the stress field from the elasticity relations, Eq. (\ref{stiff}):
\begin{eqnarray}
\sigma_{rr}  & = & \frac{E_Y} {(1+\nu)(1-2 \nu)}  \left[ (1-\nu) \, \epsilon_{r r} +\nu \, \epsilon_{\theta \theta}+\nu \, \epsilon_{z z} \right] , \label{stress1} \\
\sigma_{\theta \theta}  & = & \frac{E_Y} {(1+\nu)(1-2 \nu)}  \left[ \nu \, \epsilon_{r r} +(1-\nu) \, \epsilon_{\theta \theta}+\nu \, \epsilon_{z z}  \right] , \\
\sigma_{zz}  & = & \frac{E_Y} {(1+\nu)(1-2 \nu)}  \left[ \nu \, \epsilon_{r r} +\nu \, \epsilon_{\theta \theta}+(1-\nu) \, \epsilon_{z z} \right], \\
\sigma_{r \theta}  & = & \frac{E_Y} {(1+\nu)}   \epsilon_{r \theta}  , \\
\sigma_{r z}  & = & \frac{E_Y} {(1+\nu)}    \epsilon_{r z}  , \\
\sigma_{\theta z}  & = & \frac{E_Y} {(1+\nu)}    \epsilon_{\theta z}  , \label{stress6}
\end{eqnarray} 
having introduced the Young's modulus $E_Y$ and Poisson's ratio $\nu$.
Injecting Eq. (\ref{superpose}) in Eqs. (\ref{kin},\ref{pot},\ref{eqEsigma}), the energy contributions can be decomposed onto the normal modes:
\begin{eqnarray}
{\cal E}_{k} & = & \sum_{\left\{\eta\right\}} {\cal E}_{\rho \, \left\{\eta\right\},\left\{\eta \right\} }+  \!\!\!\!\! \sum_{\left\{\eta\right\} \neq \left\{\eta'\right\}}  {\cal E}_{\rho \, \left\{\eta\right\},\left\{\eta'\right\} } \, , \label{erho} \\
{\cal E}_{p} & = & \sum_{\left\{\eta\right\}} \left( {\cal E}_{\sigma \, \left\{\eta\right\},\left\{\eta\right\} } +{\cal E}_{E_Y \, \left\{\eta\right\},\left\{\eta\right\} } \right) \nonumber \\
 &+&   \!\!\! \sum_{\left\{\eta\right\} \neq \left\{\eta'\right\}} \left( {\cal E}_{\sigma \, \left\{\eta\right\},\left\{\eta'\right\} } +{\cal E}_{E_Y \, \left\{\eta\right\},\left\{\eta'\right\} }\right), \label{epe}
\end{eqnarray}
with the bilinear forms $ {\cal E}_{A\, \left\{\eta\right\},\left\{\eta'\right\} }$ ($A=\rho,\sigma,E_Y$) constructed from the expansions of the above expressions with respect to the $\vec{u}_{\left\{\eta\right\}}$ of the sums. 
One can show that these expressions are symmetric under the exchange $\left\{\eta\right\} \longleftrightarrow \left\{\eta'\right\}$ (Appendix \ref{coefII}). 
The total energy density is simply:
\begin{equation}
{\cal E}_{tot} = {\cal E}_{k} + {\cal E}_{p}. \label{energytot}
\end{equation}
The bilinear forms introduced here are explicitly given in the following Sections \ref{ineq} and \ref{gradient} for the simple case of ``string-like'' modes.

\subsection{Dissipation: internal and clamp friction}
\label{clamps}

Energy can be fed into a mode (or can decay away) via two paths: through {\it internal degrees of freedom} or towards the {\it two clamping ends}.
The internal 
contribution is chosen to reproduce a conventional viscous damping; in mechanics terms, this is usually realized by introducing a complex Young's modulus with an imaginary part representing the damping coefficient \cite{kotthaus}. 
This simply means that locally, friction is modeled by a stress field $[\chi_{ij}]$ proportional to the rate of change of the strain \cite{ilya_nanolett}:
\begin{eqnarray}
\chi_{rr} & = &  \frac{\Lambda_Y} {(1+\nu_\lambda)(1-2 \nu_\lambda)} \times \nonumber \\
& & \left[ (1-\nu_\lambda) \, \frac{\partial \epsilon_{r r}}{\partial t} +\nu_\lambda \, \frac{\partial \epsilon_{\theta \theta}}{\partial t}+\nu_\lambda \, \frac{\partial \epsilon_{z z}}{\partial t} \right]   , \\
\chi_{\theta \theta}  & = &  \frac{\Lambda_Y} {(1+\nu_\lambda)(1-2 \nu_\lambda)} \times \nonumber \\
&&  \left[ \nu_\lambda \, \frac{\partial \epsilon_{r r}}{\partial t} +(1-\nu_\lambda) \, \frac{\partial \epsilon_{\theta \theta}}{\partial t}+\nu_\lambda \, \frac{\partial \epsilon_{z z}}{\partial t}  \right]  , \\
\chi_{zz}  &= &  \frac{\Lambda_Y} {(1+\nu_\lambda)(1-2 \nu_\lambda)} \times \nonumber \\
&& \left[ \nu_\lambda \, \frac{\partial \epsilon_{r r}}{\partial t} +\nu_\lambda \, \frac{\partial \epsilon_{\theta \theta}}{\partial t}+(1-\nu_\lambda) \, \frac{\partial \epsilon_{z z}}{\partial t} \right]  , \\
\chi_{r \theta}  & = &  \frac{\Lambda_Y} {(1+\nu_\lambda)}   \frac{\partial \epsilon_{r \theta}}{\partial t}   , \\
\chi_{r z}  & = &  \frac{\Lambda_Y} {(1+\nu_\lambda)}    \frac{\partial \epsilon_{r z}}{\partial t}  , \\
\chi_{\theta z}  & = &  \frac{\Lambda_Y} {(1+\nu_\lambda)}    \frac{\partial \epsilon_{\theta z}}{\partial t}   ,
\end{eqnarray} 
for which we chose the same parametrization as for Eqs. (\ref{stress1}-\ref{stress6}) for simplicity. 
Two constants are thus introduced, the damping parameter 
$\Lambda_Y$ in Pa.s and $\nu_\lambda$ that quantifies the directionality of the friction field with respect to the strain field
 (similarly to $\nu$).
The power lost through friction (per unit volume) is then obtained as:
\begin{eqnarray}
 \dot{\cal D}_{int} &=&  \frac{\partial \epsilon_{rr}}{\partial t} \, \chi_{rr} +  \frac{\partial  \epsilon_{\theta \theta}}{\partial t}  \, \chi_{\theta \theta} +  \frac{\partial  \epsilon_{zz}}{\partial t}  \, \chi_{zz}   \nonumber \\
& + & 2 \left(  \frac{\partial  \epsilon_{r \theta}}{\partial t} \, \chi_{r \theta} +  \frac{\partial  \epsilon_{r z}}{\partial t} \, \chi_{r z} +  \frac{\partial  \epsilon_{\theta z}}{\partial t} \, \chi_{\theta z} \right) , \label{dissipbulk}
\end{eqnarray} 
similarly to the stored bending energy Eq. (\ref{pot}). 
Note however the factor of 2 difference in definitions.
The internal mechanisms leading to the definition of $\Lambda_Y$ are outside of the scope of this article. An introduction to high-temperature mechanisms can be found in Ref. \cite{clelandBk}; 
conversely, internal two-level systems (defects for crystals, or structural for amorphous matter) are usually invoked as the dominant mechanism of low-temperature friction in NEMS and MEMS. For an illustration of this, see e.g. Ref. 
\cite{olivePRB}. \\

The modeling of the anchoring points (see Fig. \ref{fig_2}) requires specific attention. 
First, one should point out that for $- \varepsilon < z < 0$ and $L< z < L+\varepsilon$, the writing of $\vec{u}$ based on Eqs. (\ref{equr}-\ref{equz}) {\it cannot be valid anymore}, since precisely in these regions of space the motion should gradually switch from rod-like waves to bulk-like waves.  
As such, we recast formally these equations into:
\begin{eqnarray}
u_r (r, \theta, z,t) & = & \varphi_{r \left\{\eta\right\}}(r,\theta,z) \, U_{\left\{\eta\right\}}(z,t) , \label{varphi1}\\
u_\theta (r, \theta, z,t) & = & \varphi_{\theta \left\{\eta\right\}}(r,\theta,z) \, U_{\left\{\eta\right\}}(z,t) , \\
u_z (r, \theta, z,t) & = & \varphi_{z \left\{\eta\right\}}(r,\theta,z) \, R \frac{\partial U_{\left\{\eta\right\}}}{\partial z}(z,t) , \label{varphi3}
\end{eqnarray}
where the ``modal shape'' now depends also on $z$. We assume that all the functions introduced are regular, so that we can write:
\begin{equation} 
\varphi_{i \left\{\eta\right\}}(r,\theta,z) = \sum_{n=0}^{+\infty} \varphi_{i \left\{\eta\right\}}^{(n)}(r,\theta) \frac{(z/R)^n}{n\mathrm{!} } , \label{phiexpand}
\end{equation}
for $z$ around 0 (and a similar expansion for $z \approx L$), and $i=r, \theta,z$. Obviously $\varphi_{i \left\{\eta\right\}}(r,\theta,0)=\varphi_{i \left\{\eta\right\}}^{(0)}(r,\theta)=\phi_{i \left\{\eta\right\}}(r,\theta)$. 
Higher order functions $n \neq 0$ depend on the actual geometry of the clamp, and their description is outside of the scope of this manuscript.
From these expressions, we can define the time-dependent stress field $[\sigma_{c\,ij}]$ within the clamp, injecting the above in Eqs. (\ref{strainr}-\ref{stress6}).
The power density flowing through the two regions of length $\varepsilon$ is thus expressed similarly to Eq. (\ref{dissipbulk}):
\begin{eqnarray}
 \dot{\cal D}_{clamp} &=& \frac{\partial \epsilon_{c\, rr}}{\partial t} \, \sigma_{c\,rr} +  \frac{\partial  \epsilon_{c\, \theta \theta}}{\partial t}  \, \sigma_{c\, \theta \theta} +  \frac{\partial  \epsilon_{c\, zz}}{\partial t}  \, \sigma_{c\, zz}   \nonumber \\
& \!\!\!\!\!\!\!\!\!\!\!\!\!\!\!\!\!\!\!\!\!\!\!\!\! + & \!\!\!\!\!\!\!\!\!\!\!\!\!\!\!   2 \left(  \frac{\partial  \epsilon_{c\, r \theta}}{\partial t} \, \sigma_{c \, r \theta} +  \frac{\partial  \epsilon_{c\, r z}}{\partial t} \, \sigma_{c\, r z} +  \frac{\partial  \epsilon_{c\, \theta z}}{\partial t} \, \sigma_{c\, \theta z} \right), \label{dissipclamps}
\end{eqnarray} 
valid only for $-\varepsilon < z <0$ and $L < z < L+\varepsilon$ (and zero elsewhere by definition).
The rate of change of the strain $[\partial \epsilon_{c\, kl} / \partial t]$ within the clamp is again defined from a linear response perspective. 
We therefore introduce two friction constants $\Lambda_c, \nu_c$ that fully characterize the ability of energy to radiate into the bulk supports.  
Formally, this is given by:
\begin{eqnarray}
\frac{\partial \epsilon_{c\, rr}}{\partial t} & = & \Lambda_c^{-1} \left[ \sigma_{c\,r r} -\nu_c \left( \sigma_{c\,\theta \theta} + \sigma_{c\,z z}\right) \right], \label{epsc1} \\
\frac{\partial  \epsilon_{c\, \theta \theta}}{\partial t} & = & \Lambda_c^{-1} \left[ \sigma_{c\,\theta \theta} -\nu_c \left( \sigma_{c\,r r} + \sigma_{c\,z z}\right) \right], \\
\frac{\partial  \epsilon_{c\, z z}}{\partial t} & = & \Lambda_c^{-1} \left[ \sigma_{c\,z z} -\nu_c \left( \sigma_{c\,r r} + \sigma_{c\,\theta \theta}\right) \right], \\
\frac{\partial  \epsilon_{c\, r \theta}}{\partial t} & = & \Lambda_c^{-1} \left(1+\nu_c \right) \sigma_{c\,r \theta}, \\
\frac{\partial  \epsilon_{c\, r z}}{\partial t} & = & \Lambda_c^{-1}  \left(1+\nu_c \right) \sigma_{c\,r z}, \\
\frac{\partial  \epsilon_{c\, \theta z}}{\partial t} & = & \Lambda_c^{-1}  \left(1+\nu_c \right) \sigma_{c\,\theta z}, \label{epsc6}
\end{eqnarray}
keeping again the same type of parametrization as in the above.
This approach generalizes the introduction of an {\it admittance matrix} when modeling only the lowest mechanical branches \cite{judge1,judge2,cross,Ignacio}; however, the calculation of this coefficient is outside of the scope of our manuscript. 
It suffice to point out that leakage of energy towards the anchor can be mininmised through a proper design of the clamp; this leads to the creation of "phononic shields" \cite{regal,kipp}.
Note that Eq. (\ref{dissipclamps}) contains {\it all} contributions that arise from the distortion, including 
the shear generated by the angle appearing between the anchoring point and the
 rod experiencing flexure \cite{suhel,biswas}.
Again, through a clever design distortions at the anchor can be minimised, which leads to a decrease in friction named "soft clamping" \cite{schliesser,ilya_nanolett}.

The final ingredient of the clamp modeling consists in the application of the {boundary conditions} already introduced in 
Subsection \ref{modes}.
Let us consider first the case of T and F solutions, for which we only impose $U_{\left\{\eta\right\}}(z=0;L,t)=0$ yet.
This leads to the expansions:
\begin{eqnarray}
U_{\left\{\eta\right\}}(z,t) &\approx& z \frac{\partial U_{\left\{\eta\right\}}}{\partial z}(0,t)+ \frac{z^2}{2} \frac{\partial^2 U_{\left\{\eta\right\}}}{\partial z^2}(0,t), \label{bound1} \\
\frac{\partial U_{\left\{\eta\right\}}}{\partial z}(z,t) &\approx& \frac{\partial U_{\left\{\eta\right\}}}{\partial z}(0,t)+ z \frac{\partial^2 U_{\left\{\eta\right\}}}{\partial z^2}(0,t), \\
\frac{\partial^2 U_{\left\{\eta\right\}}}{\partial z^2}(z,t) &\approx& \frac{\partial^2 U_{\left\{\eta\right\}}}{\partial z^2}(0,t), 
\end{eqnarray}
in the vicinity of $z \approx 0$ (and similar expressions for $z \approx L$).
Using the replacement Eqs. (\ref{replacez1}-\ref{replacez2}), the equivalent expansions for L modes write:
\begin{eqnarray}
R \, U_{\left\{\eta\right\}} & \rightarrow & \int_z U_{\left\{\eta\right\}}(z',t) dz'  \approx  \nonumber \\
&& \,\, \tilde{U}_{\left\{\eta\right\}}(0,t) + \frac{z^2}{2} \frac{\partial U_{\left\{\eta\right\}}}{\partial z}(0,t), \label{zeroFirst}  \\
R \frac{\partial U_{\left\{\eta\right\}}}{\partial z} & \rightarrow &  U_{\left\{\eta\right\}}(z,t)  \approx  z \frac{\partial U_{\left\{\eta\right\}}}{\partial z}(0,t), \\
R \frac{\partial^2 U_{\left\{\eta\right\}}}{\partial z^2} & \rightarrow & \frac{\partial U_{\left\{\eta\right\}}}{\partial z}(z,t)  \approx  \frac{\partial U_{\left\{\eta\right\}}}{\partial z}(0,t), \label{bound6}
\end{eqnarray}
the $\tilde{U}$ in Eq. (\ref{zeroFirst}) referring to the primitive function with respect to the $z$ variable (and evaluated at $z=0$ here).
In each set of equations, two functions which are {\it constants-in-$z$} (but functions of  $t$) appear, which are defined from the mode shape function $U_{\left\{\eta\right\}}$. 
One can argue that they represent two possible ``loss channels'' (for mode $\left\{\eta\right\}$), linked to the distortion pattern of the wave.
Depending on the clamp condition (so-called ``weak'' or ``strong''), which defines the type of mode confined in the rod (``string-like'' or ``beam-like'' respectively), different ``loss channels'' are selected. This is intimately linked to the nature of the forces and torques applied onto the anchoring point by the rod's distortion. The concept of ``channels'' for energy flow is discussed further in the following Sections.

Similarly to Eqs. (\ref{erho},\ref{epe}), the friction terms can be expanded onto the normal modes as:
\begin{eqnarray}
\!\!\! \dot{\cal D}_{int} & \!\!\! = & \!\!\! \sum_{\left\{\eta\right\}} \dot{\cal D}_{int \, \left\{\eta\right\},\left\{\eta \right\} }+   \!\!\!\!\! \sum_{\left\{\eta\right\} \neq \left\{\eta'\right\}} \!\!\!\! \dot{\cal D}_{int \, \left\{\eta\right\},\left\{\eta'\right\} } \, , \label{bulkD} \\
\!\!\! \dot{\cal D}_{clamp} & \!\!\! = & \!\!\! \sum_{\left\{\eta\right\}}   \dot{\cal D}_{clamp \, \left\{\eta\right\},\left\{\eta \right\} } 
+     \!\!\!\!\! \sum_{\left\{\eta\right\} \neq \left\{\eta'\right\}}  \dot{\cal D}_{clamp \, \left\{\eta\right\},\left\{\eta'\right\} }  , \label{bulkC}
\end{eqnarray}
with coefficients resembling those found for the bending expansion ${\cal E}_{E_Y}$ within Eq. (\ref{epe}).
The total power density lost through friction is then:
\begin{equation}
\dot{\cal D}_{tot} = \dot{\cal D}_{int} + \dot{\cal D}_{clamp}, \label{totdissipe}
\end{equation}
reminding the specificity that the power density $\dot{\cal D}_{clamp}$ is nonzero only for $-\varepsilon < z <0$ and $L < z < L+\varepsilon$ (while $\dot{\cal D}_{int}$ is defined over $0<z<L$ strictly).
Note that for the modes to be well-defined, the total damping rate $\Gamma_{\left\{\eta\right\}}$ derived from $\dot{\cal D}_{tot}$ should remain much smaller than the mode resonance frequency $\omega_{\left\{\eta\right\}}$ (the {\it high-Q limit}, see following Subsection \ref{ineq}).
Also, the friction coefficients $\Lambda_Y$ and $\Lambda_c$ which model specific damping mechanisms may appear to be frequency-dependent in the most generic case (as well as their respective $\nu_\lambda$ and $\nu_c$):
we write them $\Lambda_Y(\omega)$, $\Lambda_c(\omega)$. 
For simplicity, we shall drop the frequency dependence of the $\nu_\lambda, \nu_c$ parameters (see Appendix \ref{coefs}).
Practically, within a high-Q approximation they can be treated as constant for each specific mode, 
and we simply define $\Lambda_Y \approx \Lambda_{Y\, \left\{\eta\right\}}$ (and similarly $\Lambda_c \approx \Lambda_{c\, \left\{\eta\right\}}$) when $\omega \approx \omega_{\left\{\eta\right\}}$.
This is to be contrasted with the hypothesis made on the elastic constants $E_Y, \nu$ which are assumed to be frequency-independent.
As a consequence (and contrarily to the energetics coefficients discussed in Subsection \ref{energy}), the difference between damping coefficients $\Lambda_{Y\, \left\{\eta\right\}}, \Lambda_{Y\, \left\{\eta'\right\}}$ and $\Lambda_{c\, \left\{\eta\right\}}, \Lambda_{c\, \left\{\eta'\right\}}$ leads in principle to an asymmetry in the bilinear forms introduced here under the exchange $ \left\{\eta\right\} \leftrightarrow \left\{\eta'\right\}$; these refined aspects are discussed in Appendix \ref{coefII}.
As for the energy expansions, the friction coefficients are detailed in Sections \ref{ineq} and \ref{gradient} below for the simple case of ``string-like'' modes.

\section{in-equilibrium situation}
\label{ineq}

Let us first describe the simple case of thermal equilibrium. Each mode $\left\{\eta\right\}$ 
experiences {\it Brownian motion}, such that the time-dependent motion $U_{\omega}(t)$ in Eqs. (\ref{equr1}-\ref{eqbeammode}) 
is not a simple sine-wave (at frequency $\omega$) imposed by an external drive of some sort. It is from now on a stochastic time-dependent variable $U_{\omega_{\left\{\eta\right\}}}(t)$ which spectral weight lies mainly around $\omega_{\left\{\eta\right\}}$ (Section \ref{FDT}).
Consider the energy density expansions Eqs. (\ref{erho},\ref{epe}). As will be shown in the following
Section \ref{gradient}, the sums $\left\{\eta\right\} \neq \left\{\eta'\right\}$ 
are the consequence of the presence of a thermal gradient; we shall thus take them to be zero for the discussion on which we focus here.
For ``string-like'' solutions, injecting Eqs. (\ref{equr1}-\ref{equz2}) into Eqs. (\ref{kin},\ref{eqEsigma},\ref{pot}), and integrating over the cross-section we obtain respectively:
\begin{eqnarray}
\int_0^{2 \pi} \!\!\!  \int_0^R {\cal E}_{\rho \, \left\{\eta\right\},\left\{\eta \right\} } \, r dr d\theta & = & \frac{1}{2} {\cal M}_{\left\{\eta\right\}} \sin^2 (k_p z) \, \dot{U}_{\omega_{\left\{\eta\right\}}}(t)^2 \nonumber \\
&\!\!\!\!\!\!\!\!\!\!\!\!\!\!\!\! 
\!\!\!\!\!\!\!\!\!\!\!\!\!\!\!\!\!\!\!\!\!\!\!\!\!\!\!\!\!\!\!\!\!\!\!\!\!\!\!\!\!\!\!\!\!\!\!\!\!\!\!\!\!\!\!\!\! \!\!\!\!\!\!\!\!\!\!\!\!\!\!\!\!\!\!\!\!\!\!\!\!\!\!\!\!\!\!\! +& \!\!\!\!\!\!\!\!\!\!\!\!\!\!\!\!\!\!\!\!\!\!\! 
\!\!\!\!\!\!\!\!\!\!\!\!\!\!\!\!\! \!\!\!\!\!\!\!\!\!\!\!\!\!\!\! \frac{1}{2} \Delta {\cal M}_{\left\{\eta\right\}} \left[\cos^2 (k_p z)- \sin^2 (k_p z) \right] \dot{U}_{\omega_{\left\{\eta\right\}}}(t)^2 , \label{modekin} \\
\int_0^{2 \pi} \!\!\!  \int_0^R {\cal E}_{\sigma \, \left\{\eta\right\},\left\{\eta \right\} } \, r dr d\theta & = & \frac{1}{2} {\cal S}_{\left\{\eta\right\}} \cos^2 (k_p z) \,  U_{\omega_{\left\{\eta\right\}}}(t)^2 \nonumber \\
&\!\!\!\!\!\!\!\!\!\!\!\!\!\!\!\! 
\!\!\!\!\!\!\!\!\!\!\!\!\!\!\!\!\!\!\!\!\!\!\!\!\!\!\!\!\!\!\!\!\!\!\!\!\!\!\!\!\!\!\!\!\!\!\!\!\!\!\!\!\!\!\!\!\! \!\!\!\!\!\!\!\!\!\!\!\!\!\!\!\!\!\!\!\!\!\!\!\!\!\!\!\!\!\!\! +& \!\!\!\!\!\!\!\!\!\!\!\!\!\!\!\!\!\!\!\!\!\!\! 
\!\!\!\!\!\!\!\!\!\!\!\!\!\!\!\!\! \!\!\!\!\!\!\!\!\!\!\!\!\!\!\! \frac{1}{2} \Delta {\cal S}_{\left\{\eta\right\}} \left[\cos^2 (k_p z)- \sin^2 (k_p z) \right] U_{\omega_{\left\{\eta\right\}}}(t)^2 , \label{modestress} \\
\int_0^{2 \pi} \!\!\!  \int_0^R {\cal E}_{E_Y \, \left\{\eta\right\},\left\{\eta \right\} } \, r dr d\theta& = & \frac{1}{2} {\cal B}_{\left\{\eta\right\}}  
\cos^2 (k_p z) \, U_{\omega_{\left\{\eta\right\}}}(t)^2 \nonumber \\
&\!\!\!\!\!\!\!\!\!\!\!\!\!\!\!\! 
\!\!\!\!\!\!\!\!\!\!\!\!\!\!\!\!\!\!\!\!\!\!\!\!\!\!\!\!\!\!\!\!\!\!\!\!\!\!\!\!\!\!\!\!\!\!\!\!\!\!\!\!\!\!\!\!\! \!\!\!\!\!\!\!\!\!\!\!\!\!\!\!\!\!\!\!\!\!\!\!\!\!\!\!\!\!\!\! +& \!\!\!\!\!\!\!\!\!\!\!\!\!\!\!\!\!\!\!\!\!\!\! 
\!\!\!\!\!\!\!\!\!\!\!\!\!\!\!\!\! \!\!\!\!\!\!\!\!\!\!\!\!\!\!\! \frac{1}{2} \Delta {\cal B}_{\left\{\eta\right\}} \left[\cos^2 (k_p z)- \sin^2 (k_p z) \right] U_{\omega_{\left\{\eta\right\}}}(t)^2 , \label{modebend}
\end{eqnarray}
for torsional (T) and flexural (F) modes. For longitudinal (L) modes, one should swap $\cos \leftrightarrow \sin$. The writing of the introduced coefficients (which are $k_p$-dependent) is given explicitly in Appendix \ref{coefs}. We shall not discuss the much more involed situation encountered for ``beam-like'' solutions.
We define:
\begin{equation}
{\cal K}_{\left\{\eta\right\}} = {\cal S}_{\left\{\eta\right\}}+ {\cal B}_{\left\{\eta\right\}}, \label{spring}
\end{equation}
the spring constant per unit length of mode $\left\{\eta\right\}$. The parameter ${\cal M}_{\left\{\eta\right\}} $ is its mass per unit length.
Integrating Eq. (\ref{modekin}) over the length $L$, one derives an {\it effective inertia}  $ ({\cal M}_{\left\{\eta\right\}} L/2) \, \ddot{U}_{\omega_{\left\{\eta\right\}}}$ and similarly with 
Eqs. (\ref{modebend},\ref{modestress}) an {\it effective restoring force} $-({\cal K}_{\left\{\eta\right\}}L/2) \, U_{\omega_{\left\{\eta\right\}}}$.
These define the mode effective mass ${\cal M}_{\left\{\eta\right\}} L/2$ and spring constant ${\cal K}_{\left\{\eta\right\}}L/2$, where the factor $1/2$ appears because of the sinusoidal mode shape. These will enable to build the mode's {\it effective dynamics equation} in the following Section \ref{FDT}.
Per construction these quantities verify:
\begin{equation}
{\cal M}_{\left\{\eta\right\}} \, \omega_{\left\{\eta\right\}}^2 = {\cal K}_{\left\{\eta\right\}} . \label{property}
\end{equation}
We write $\left\langle \cdots \right\rangle$ the ensemble average on the stochastic variables.
Summing up Eqs. (\ref{modekin}-\ref{modebend}), and using the property $\left\langle \dot{U}_{\omega_{\left\{\eta\right\}}}^2 \right\rangle = \omega_{\left\{\eta\right\}}^2   \left\langle U_{\omega_{\left\{\eta\right\}}}^2 \right\rangle$  justified by the spectral concentration of $U_{\omega_{\left\{\eta\right\}}}$ around $\omega_{\left\{\eta\right\}}$ (see Section \ref{FDT}), we obtain the total average energy per unit length of the mode:
\begin{eqnarray}
\frac{d E_{tot\, \left\{\eta\right\}} (z)}{d z} & = & \left( \frac{1}{2} + \frac{1}{2} \right) \frac{ {\cal K}_{\left\{\eta\right\}} }{2}  \left\langle U_{\omega_{\left\{\eta\right\}}}^2 \right\rangle  \nonumber \\
& \!\!\!\!\!\!\!\!\!\!\!\! 
\!\!\!\!\!\!\!\!\!\!\!\!\!\!\!\!\!\!\!\!\!\!\!\!\!\!\!\!\!\!\!\!\!\!\!\!\!\!\!\!\!\!\!\!\!\!\!\!\!\!\!\!\!\!\!\!\! \!\!\!\!\!\!\!\!\!\!\!\!\!\!\!\!\!\!\!\!\!\!\!\!\!\!\!\!\!\!\! +&  \!\!\!\!\!\!\!\!\!\!\!\!\!\!\!\!\!\!\! 
\!\!\!\!\!\!\!\!\!\!\!\!\!\!\!\!\! \!\!\!\!\!\!\!\!\!\!\!\!\!\!\! \frac{1}{2} \left(\frac{\Delta {\cal K}_{\left\{\eta\right\}}}{{\cal K}_{\left\{\eta\right\}}}+ \frac{\Delta {\cal M}_{\left\{\eta\right\}}}{{\cal M}_{\left\{\eta\right\}}} \right) \cos (2 k_p z)  \, {\cal K}_{\left\{\eta\right\}}  \left\langle U_{\omega_{\left\{\eta\right\}}}^2 \right\rangle, \label{finalenergy}
\end{eqnarray}
where we made use of Eqs. (\ref{spring},\ref{property}), and $\Delta {\cal K}_{\left\{\eta\right\}} =\Delta {\cal S}_{\left\{\eta\right\}}+ \Delta {\cal B}_{\left\{\eta\right\}}$. We kept the two $1/2$ in this writing to underline that one term refers to kinetic energy, and the other potential.

Postponing the discussion of the mode dynamics (Section \ref{FDT}), the {\it equipartition theorem} applied to mode $\left\{\eta\right\}$ states that:
\begin{equation}
 \frac{{\cal K}_{\left\{\eta\right\}} L}{2} \, \left\langle U_{\omega_{\left\{\eta\right\}}}^2 \right\rangle = k_B T_{\left\{\eta\right\}} , \label{equiparti}
\end{equation}
with $T_{\left\{\eta\right\}}$ the temperature associated to its Brownian motion (and $L$ the length of the rod).
On a given branch, the ``string-like'' solutions verify $k_p = q \, \pi/L$ with $q>0$ (and $\left\{\eta\right\}$ containing both this index and the branch label, $\left\{ m,\mbox{T}, q \right\}$, $\left\{ m,\mbox{L}, q \right\}$ or $\left\{ m,n \neq 0,\mbox{F}, q \right\}$ by definition; remember also that they are two degenerate flexural families $\theta_0=0$ and $+\pi/2$, see Appendix \ref{pochhammer}).
Eq. (\ref{finalenergy}) finally writes:
\begin{eqnarray}
\frac{d E_{tot\, \left\{\eta\right\}} (z)}{d z} & = & \frac{k_B T_{\left\{\eta\right\}}}{L} \nonumber \\
& \!\!\!\!\!\!\!\!\!\!\!\! 
\!\!\!\!\!\!\!\!\!\!\!\!\!\!\!\!\!\!\!\!\!\!\!\!\!\!\!\!\!\!\!\!\!\!\!\!\!\!\!\!\!\!\!\!\!\!\!\!\!\!\!\!\!\!\!\!\! \!\!\!\!\!\!\!\!\!\!\!\!\!\!\!\!\!\!\!\!\!\!\!\!\!\!\!\!\!\!\! +&  \!\!\!\!\!\!\!\!\!\!\!\!\!\!\!\!\!\!\! 
\!\!\!\!\!\!\!\!\!\!\!\!\!\!\!\!\! \!\!\!\!\!\!\!\!\!\!\!\!\!\!\! \frac{1}{2} \left(\frac{\Delta {\cal K}_{\left\{\eta\right\}}}{{\cal K}_{\left\{\eta\right\}}}+ \frac{\Delta {\cal M}_{\left\{\eta\right\}}}{{\cal M}_{\left\{\eta\right\}}} \right) \cos \left(2 \pi \, q \frac{z}{L}\right)  \, \frac{k_B T_{\left\{\eta\right\}}}{L}  . \label{thesoluce}
\end{eqnarray}
Noticing that $\left| \Delta {\cal M}_{\left\{\eta\right\}}/{\cal M}_{\left\{\eta\right\}} \right| <1$ and $\left| \Delta {\cal K}_{\left\{\eta\right\}}/{\cal K}_{\left\{\eta\right\}} \right| <1$, the second line of this expression verifies:
\begin{eqnarray}
-\cos \left(2 \pi \, q \frac{z}{L}\right) &<& \nonumber \\
& &\!\!\!\!\!\!\!\!\!\!\!\!\!\!\!\!\!\!\!\!\!\!\!\!\!\!\!\!\!\!\!\!\!\!\!\!\!\!\!\!\!\!\!\!\!\!\!\!   \frac{1}{2} \left(\frac{\Delta {\cal K}_{\left\{\eta\right\}}}{{\cal K}_{\left\{\eta\right\}}}+ \frac{\Delta {\cal M}_{\left\{\eta\right\}}}{{\cal M}_{\left\{\eta\right\}}} \right) \cos \left(2 \pi \, q \frac{z}{L}\right) \nonumber \\
& < &+ \cos \left(2 \pi \, q \frac{z}{L}\right) . \label{inequality}
\end{eqnarray}

Summing up to an upper cutoff $N_{cut\,i} \gg 1$, we produce the total energy per unit length corresponding to this specific branch $i$ 
(with $i$ representing the branch label; now on we do not use the writing in $\left\{ \cdots \right\}$ for it, in order to keep it clearly distinct from the mode labels):
\begin{equation}
\frac{d E_{tot\, branch \, i} (z)}{d z}  = N_{cut \,i} \frac{k_B T_{branch\, i}}{L} \left( 1 + 0 \right), \label{equilE}
\end{equation}
where $T_{branch\, i}=1/N_{cut\, i} \sum_{q=1}^{N_{cut\, i}} T_{\left\{\eta\right\}}$ is defined as the average temperature of the branch.
For $T_{branch\, i}$ to be finite, all temperatures $T_{\left\{\eta\right\}}$ have to be bound by an upper limit.
Therefore, from Eq. (\ref{inequality}) we realize that 
the sum $1/N_{cut\, i} \sum_{q=1}^{N_{cut\, i}}$ performed over the second line of 
Eq. (\ref{thesoluce}) tends to zero for all $ 0 < z < L$ since $1/N_{cut \, i} \sum_{q=1}^{N_{cut \, i}} \cos \left(2 \pi \, q \frac{z}{L}\right) \rightarrow 0$ for $N_{cut\, i} \rightarrow + \infty$, excluding the two extremes $z=0;L$. This is what is reminded by the 0 in the above expression.

We can now add together the energy per unit length of all the branches, leading to the total energy integrated over the cross-section as defined by Eq. (\ref{energytot}). For simplicity, we will again neglect ``beam-like'' solutions and assume that all branches are described by ``string-like'' modes; this shall be enough to demonstrate our methodology.
The calculated energy {\it should match} the macroscopic description's result $3N \, k_B T_{avg}/L$ of Eq. (\ref{energydens}), here within the assumption $T_c=T_h=T_{avg}$.
It imposes that:
\begin{eqnarray}
\sum_{\mbox{all branches $i$}} \!\!\!\!\!\!\!\! N_{cut\, i} &= & 3 N ,  \label{2Nequa} \\
\frac{1}{3 N} \!\!\!\! \sum_{\substack{\mbox{all branches $i$} \\ \mbox{and wavevectors $q$}\\ \mbox{i.e. all   modes  $\left\{\eta\right\}$}}}  \!\!\!\!\!\!\!\! T_{\left\{\eta\right\}} & = & T_{avg} , \label{Tequil}
\end{eqnarray}
with $N$ the number of atoms in the solid rod.
The first equation simply states that the total number of degrees of freedom is preserved, which is also what is performed in Debye's approach (see Section \ref{phonons}).
Physically, the existence of a finite number of branches and of (very large but) finite cutoffs $N_{cut\, i}$ is precisely due to the fact that a mode's shape cannot present distortions on a scale smaller than the inter-atomic distance (see Appendices \ref{pochhammer} and \ref{coefs}). 
The second equation means that {\it the mode's mean temperature equals the macroscopic temperature} of the two baths $T_{avg}$.  \\

We now compute the friction experienced by each mode, restricting again the discussion to ``string-like'' solutions.
As previously, the $\left\{\eta\right\} \neq \left\{\eta'\right\}$ terms in our expressions shall be set to zero.
After integration over the cross-section, the internal contribution Eq. (\ref{dissipbulk}) leads to:
\begin{eqnarray}
\int_0^{2 \pi} \!\!\!  \int_0^R {\cal \dot{D}}_{int \, \left\{\eta\right\},\left\{\eta \right\} } \, r dr d\theta & = & {\cal L}_{\left\{\eta\right\}} \left[ 1 \pm \cos (2 k_p z) \right] \dot{U}_{\omega_{\left\{\eta\right\}}}(t)^2 \nonumber \\
&\!\!\!\!\!\!\!\!\!\!\!\!\!\!\!\! 
\!\!\!\!\!\!\!\!\!\!\!\!\!\!\!\!\!\!\!\!\!\!\!\!\!\!\!\!\!\!\!\!\!\!\!\!\!\!\!\!\!\!\!\!\!\!\!\!\!\!\!\!\!\!\!\!\! \!\!\!\!\!\!\!\!\!\!\!\!\!\!\!\!\!\!\!\!\!\!\!\!\!\!\!\!\!  +& \!\!\!\!\!\!\!\!\!\!\!\!\!\!\!\!\!\!\!\!\!  
\!\!\!\!\!\!\!\!\!\!\!\!\!\!\!\!\! \!\!\!\!\!\!\!\!\!\!\!\!\!\!\!  2 \Delta {\cal L}_{\left\{\eta\right\}} \left[\pm \cos (2 k_p z) \right] \dot{U}_{\omega_{\left\{\eta\right\}}}(t)^2 , \label{dissintg} 
\end{eqnarray}
which is the power dissipated per unit length due to internal degrees of freedom.
The $\pm$ sign stands for T and F modes ($+$), and L modes ($-$) respectively.
As for the energetic coefficients, ${\cal L}_{\left\{\eta\right\}} $ and $\Delta {\cal L}_{\left\{\eta\right\}} $ are given explicitly in Appendix \ref{coefs}.
After integration over the length $L$ of the rod, the $\cos (2 k_p z)$ expressions of Eq. (\ref{dissintg}) disappear. We can then identify an {\it effective viscous force} acting on mode $\left\{\eta\right\}$:
\begin{equation}
F_{int\,\left\{\eta\right\}} (t) = - {\cal L}_{\left\{\eta\right\}}  L \, \dot{U}_{\omega_{\left\{\eta\right\}}}(t), \label{force1}
\end{equation}
such that $F_{int\,\left\{\eta\right\}} \dot{U}_{\omega_{\left\{\eta\right\}}}$ equals the dissipated power. ${\cal L}_{\left\{\eta\right\}}$ is our internal friction coefficient per unit length. 

The clamp dissipated power density Eq. (\ref{dissipclamps}) is defined only in the vicinity of the anchoring points. 
From the expressions of the displacement field profile Eqs. (\ref{varphi1}-\ref{varphi3}) and the boundary conditions Eqs. (\ref{bound1}-\ref{bound6}), we see that 
at lowest order this term is {\it z-independent}. Integrating over the clamp volumes $0<z<\varepsilon$ and $L<z<L+\varepsilon$ thus leads to:
\begin{eqnarray}
\int_{\substack{\mbox{clamps} \\ 0<z<\varepsilon \\ L<z<L+\varepsilon}}  \int_0^{2 \pi} \!\!\!  \int_0^R {\cal \dot{D}}_{clamp \, \left\{\eta\right\},\left\{\eta \right\} } \, r dr d\theta dz & = & \nonumber \\
&&\!\!\!\!\!\!\!\!\!\!\!\!\!\!\!\!\!\!\!\!\!  \!\!\!\!\!\!\!\!\!\!\!\!\!\!\!\!\!\!\!\!\! \!\!\!\!\!\!\!\!\!\!\!\!\!\!\!\!\!\!\!\!\!  {\cal C}_{\left\{\eta\right\}} \varepsilon \left[ \omega_{\left\{\eta\right\}} \, U_{\omega_{\left\{\eta\right\}}}(t) \right]^2 , \label{dissclamp} 
\end{eqnarray}
for $\varepsilon \ll R$; higher order terms are outside of the scope of this article, and do not impact our methodology. The coefficient ${\cal C}_{\left\{\eta\right\}}$ (clamp friction per unit length) is given in Appendix \ref{coefs} for T, L and F modes.
Depending on the type of solution considered (T, L or F, ``string-like'' or beam-like'' with the corresponding boundary conditions), the expression of 
${\cal C}_{\left\{\eta\right\}}$ is {\it different}, selecting thus for each case a specific ``channel'' for the energy leakage towards the outside. 
This can be interpreted in terms of forces and torques applied onto the anchoring point, which depend strongly on the nature of the standing wave localized in the rod; an explicit discussion is presented in Appendix \ref{coefs}.
Since the dynamics of $U_{\omega_{\left\{\eta\right\}}}(t)$ is spectrally localized around $\omega_{\left\{\eta\right\}}$ (Section \ref{FDT}), 
we can substitute $\omega_{\left\{\eta\right\}} \, U_{\omega_{\left\{\eta\right\}}}(t) \approx \dot{U}_{\omega_{\left\{\eta\right\}}}(t)$ within a term that does not carry energy on average.
Eq. (\ref{dissclamp}) therefore leads to the identification of 
an {\it effective clamping viscous force} acting on mode $\left\{\eta\right\}$:
\begin{equation}
F_{clamp\,\left\{\eta\right\}} (t) = - {\cal C}_{\left\{\eta\right\}}  \varepsilon \, \dot{U}_{\omega_{\left\{\eta\right\}}}(t) , \label{force2}
\end{equation}
which physically decomposes symmetrically in two halves applied on each end of the rod;
note that for a perfect clamping $\varepsilon \rightarrow 0$, this term vanishes (see Appendix \ref{coefs} and Section \ref{FDT}). \\

The two damping forces Eqs. (\ref{force1},\ref{force2}) add up and fully characterize the total friction, Eq. (\ref{totdissipe}), which will appear in the effective dynamics equation of mode $\left\{\eta\right\}$ (Section \ref{FDT}).
The {\it fluctuation-dissipation theorem} states that they are related to two stochastic {\it Langevin forces},
characteristic of each of the baths they refer to. Their correlation functions verify ($\delta_0$ being Dirac's function):
\begin{eqnarray}
\!\!\!\!\!\!\!\!\!\!\!\!\!\!\!\!\!\!\!\! \left\langle \delta f_{int\,\left\{\eta\right\}}(t)  \delta  f_{int\,\left\{\eta\right\}}(t')  \right\rangle & \!\!\!=& \!\!\! 2 {\cal L}_{\left\{\eta\right\}} L \, k_B T_{int} \delta_0 (t-t'), \\
\!\!\!\!\!\!\!\!\!\!\!\!\!\!\!\!\!\!\!\! \left\langle \delta f_{clamp\,\left\{\eta\right\}}(t) \delta  f_{clamp\,\left\{\eta\right\}}(t') \right\rangle & \!\!\!=& \!\!\! 2 {\cal C}_{\left\{\eta\right\}} \varepsilon \, k_B T_{avg} \delta_0 (t-t'), \\
\!\!\!\!\!\!\!\!\!\!\!\!\!\!\!\!\!\!\!\! \left\langle \delta f_{clamp\,\left\{\eta\right\}}(t)  \delta f_{int\,\left\{\eta\right\}}(t')  \right\rangle & \!\!\!=& \!\!\! 0,
\end{eqnarray}
which simply means that the intrinsic response times of these baths are infinitely faster than any of the modes, and that they are uncorrelated. These forces are responsible for the Brownian motion of mode $\left\{\eta\right\}$; by definition the clamp is thermalised at a temperature $T_{avg}$, while we define $T_{int}$ for the internal degrees of freedom. 
These two contributions can be conveniently recast in a single one, with:
\begin{eqnarray}
\!\!\!\!\!\! F_{\left\{\eta\right\}} & = & - \left( {\cal L}_{\left\{\eta\right\}} L + {\cal C}_{\left\{\eta\right\}}  \varepsilon \right)\, \dot{U}_{\omega_{\left\{\eta\right\}}}  , \\
\!\!\!\!\!\! \delta f_{\left\{\eta\right\}} & = & \delta f_{int\,\left\{\eta\right\}} + \delta f_{clamp\,\left\{\eta\right\}} , \\
\!\!\!\!\!\! \left\langle \delta f_{ \left\{\eta\right\}}(t) \delta  f_{ \left\{\eta\right\}}(t') \right\rangle  & = & 2\left( {\cal L}_{\left\{\eta\right\}} L + {\cal C}_{\left\{\eta\right\}}  \varepsilon \right) \times \nonumber \\
&&\!\!\!\!\!\!  k_B T_{\left\{\eta\right\}} \delta_0 (t-t'), \label{stdFDT}
\end{eqnarray}
in which appears the total friction coefficient ${\cal L}_{\left\{\eta\right\}} L + {\cal C}_{\left\{\eta\right\}}  \varepsilon$ and the mode temperature written as:
\begin{equation}
T_{\left\{\eta\right\}} = \frac{ {\cal L}_{\left\{\eta\right\}} L \, T_{int} + {\cal C}_{\left\{\eta\right\}}  \varepsilon \, T_{avg} }{ {\cal L}_{\left\{\eta\right\}} L + {\cal C}_{\left\{\eta\right\}}  \varepsilon  } .
\end{equation}
Therefore, when the internal degrees of freedom of the rod are thermalized to the same temperature as its two ends (which is the case if there is no extra source of heat or cooling acting upon them), we recover the simple case of $T_{int} = T_{avg}$.
This leads to the {\it in-equilibrium} result:
\begin{equation}
T_{\left\{\eta\right\}} =  T_{avg} ,
\end{equation}
which obviously implies $T_{branch\,i}=T_{avg}$ for all branches $i$, and verifies our former condition Eq. (\ref{Tequil}).
The quality factor $Q_{\left\{\eta\right\}}$ associated to mode $\left\{\eta\right\}$ is defined as:
\begin{eqnarray}
Q_{\left\{\eta\right\}} & = & \frac{ \omega_{\left\{\eta\right\}} }{ \Gamma_{\left\{\eta\right\}} } , \label{QuFact} \\
\Gamma_{\left\{\eta\right\}} & = &  \frac{ {\cal L}_{\left\{\eta\right\}} + {\cal C}_{\left\{\eta\right\}} \, \varepsilon/L }{ {\cal M}_{\left\{\eta\right\}}/2 } , \label{gamma}
\end{eqnarray}
with $\Gamma_{\left\{\eta\right\}}$ the relaxation rate, and should verify $Q_{\left\{\eta\right\}} \gg 1$ in order to guarantee a well-defined resonance; which poses a condition to be fulfilled for the friction coefficients.

\section{Thermal gradient situation}
\label{gradient}

Consider now that a weak thermal gradient has been established between the two ends of the rod. In order for the energy density to reproduce Eq. (\ref{energydens}), the cross-terms in Eqs. (\ref{erho},\ref{epe}) {\it must be nonzero}. Integrating over $r-\theta$, we obtain:
\begin{eqnarray}
\int_0^{2 \pi} \!\!\!  \int_0^R {\cal E}_{\rho \, \left\{\eta\right\},\left\{\eta' \right\} } \, r dr d\theta & = & \frac{1}{2} {\cal M}_{\left\{\eta\right\},\left\{\eta' \right\} } \left[\frac{1}{2}\cos [(k_p-k_p') z] \right.  \nonumber \\
&\!\!\!\!\!\!\!\!\!\!\!\!\!\!\!\! 
\!\!\!\!\!\!\!\!\!\!\!\!\!\!\!\!\!\!\!\!\!\!\!\!\!\!\!\!\!\!\!\!\!\!\!\!\!\!\!\!\!\!\!\!\!\!\!\!\!\!\!\!\!\!\!\!\! \!\!\!\!\!\!\!\!\!\!\!\!\!\!\!\!\!\!\!\!\!\!\!\!\!\!\!\!\!\!\!\!\!\!\!\!\!\!\!\!\!\!\!\!\!\!\!\!\!\!\!\!\!\!\!\!\!\!\!\!\! \pm & \!\!\!\!\!\!\!\!\!\!\!\!\!\!\!\!\!\!\!\!\!\!\!\!\!\!\!\!\!\!\!\!\!\!\! \!\!\!\!\! 
\!\!\!\!\!\!\!\!\!\!\!\!\!\!\!\!\! \!\!\!\!\!\!\!\!\!\!\!\!\!\!\! \left. \left(  \frac{\Delta {\cal M}_{\left\{\eta\right\},\left\{\eta'\right\}} }{{\cal M}_{\left\{\eta\right\},\left\{\eta' \right\} }} -\frac{1}{2} \right)   \cos ([k_p+k_p'] z ) \right]  \dot{U}_{\omega_{\left\{\eta\right\}}} \dot{U}_{\omega_{\left\{\eta'\right\}}}    , \label{modekin2}  
\end{eqnarray}
for the kinetic part, and:
\begin{eqnarray}
\int_0^{2 \pi} \!\!\!  \int_0^R \left( {\cal E}_{\sigma \, \left\{\eta\right\},\left\{\eta' \right\} }+{\cal E}_{E_Y \, \left\{\eta\right\},\left\{\eta' \right\} } \right) \, r dr d\theta & = & \frac{1}{2} {\cal K}_{\left\{\eta\right\}, \left\{\eta' \right\}} \nonumber \\
&\!\!\!\!\!\!\!\!\!\!\!\!\!\!\!\! 
\!\!\!\!\!\!\!\!\!\!\!\!\!\!\!\!\!\!\!\!\!\!\!\!\!\!\!\!\!\!\!\!\!\!\!\!\!\!\!\!\!\!\!\!\!\!\!\!\!\!\!\!\!\!\!\!\! \!\!\!\!\!\!\!\!\!\!\!\!\!\!\!\!\!\!\!\!\!\!\!\!\!\!\!\!\!\!\!\!\!\!\!\!\!\!\!\!\!\!\!\!\!\!\!\!\!\!\!\!\!\!\!\!\!\!\!\!\!\!\!\!\!\!\!\!\!\!\!\!\!  & \!\!\!\!\!\!\!\!\!\!\!\!\!\!\!\!\!\!\!\!\!\!\!\!\!\!\!\!\!\!\!\!\!\!\!\!\!\!\!\!\!\!\!\!\!\! \!\!\!\!\!\!\!\!\!\!\!\!\!\!\!\!\!\!\! \!\!\!\!\!\!\!\!\!\!\!\!\!\!\!\!\!\!\!
\!\!\!\!\!\!\!\!\!\!\!\!\!\!\!\!\! \!\!\!\!\!\!\!\!\!\!\!\!\!\!\! \times \left[\frac{1}{2}\cos [(k_p-k_p') z] \pm \left(  \frac{\Delta {\cal K}_{\left\{\eta\right\},\left\{\eta'\right\}} }{{\cal K}_{\left\{\eta\right\},\left\{\eta' \right\} }} +\frac{1}{2} \right)   \cos ([k_p+k_p'] z ) \right] \nonumber \\
& & \!\!\!\!\!\!\!\!\!\!\!\!\!\!\!\!\!\!\!\!\!\!\!\!\!\!\!\!\!\!\!\!\!\!\!\!\!\!\!\!\!\!\!\!\!\!\!\!\!\!\!\!\!\!\!\!\!\!\!\!\!\!\!\!\!\!\!\!\!\!\!\!\!\!\!\! \!\!\!\!\!\!\!\!\!\! \!\!\!\!\!\!\!\!\!\!\!\!\!\!\! \!\!\!\!\!\!\!\!\!\!\!\!\!\!\! \times   U_{\omega_{\left\{\eta\right\}}} U_{\omega_{\left\{\eta'\right\}}}    , \label{modebend2}
\end{eqnarray}
for the potential part; we define $k_p=q \, \pi/L$ and $k_p'=q' \, \pi/L$ with both integers $q, q'>0$ for each mode. The $\pm$ sign stands for T and F modes (+) and L modes (-). The modal coefficients introduced in this Section are listed in Appendix \ref{coefII}.

Reminding that energy is localized around specific frequencies within the motion spectrum (see Section \ref{FDT}), we extend the in-equilibrium property mentioned in the previous Section: 
\begin{equation}
\left\langle  \dot{U}_{\omega_{\left\{\eta\right\}}} \dot{U}_{\omega_{\left\{\eta'\right\}}} \right\rangle  \approx  \omega_{\left\{\eta\right\}} \omega_{\left\{\eta'\right\}} \, \left\langle  U_{\omega_{\left\{\eta\right\}}} U_{\omega_{\left\{\eta'\right\}}} \right\rangle ,
\end{equation}
which we insert in Eq. (\ref{modekin2}) while taking the ensemble average of both contributions of the energy density.
Besides, the scalar product $< \cdots \bullet \cdots >$ defined in Appendix \ref{pochhammer} (not to be confused with the ensemble average just mentioned)     
tells us that modes from different branches are {\it orthogonal}, which means that they verify ${\cal M}_{\left\{\eta\right\},\left\{\eta' \right\} } =0$ (and $\Delta {\cal M}_{\left\{\eta\right\},\left\{\eta' \right\} }$ as well).
By construction, this implies that Eq. (\ref{modekin2}) is zero: such cross-terms {\it cannot carry energy} to create a thermal gradient, and shall disappear from the calculation.
Only modes belonging to the same branch should be kept in our expansion, leading to the expressions in Eqs. (\ref{grad1},\ref{grad2}) written below.
The first term in Eq. (\ref{grad2}) corresponds to our previous contribution Eq. (\ref{equilE}), which is due to the mean temperatures $T_{\left\{\eta\right\}}$; the $\pm$ sign has the same meaning as above.
We define:
\begin{equation}
\frac{ {\cal K}_{\left\{\eta\right\},\left\{\eta' \right\} } L }{2} \left\langle  U_{\omega_{\left\{\eta\right\}}} U_{\omega_{\left\{\eta'\right\}}} \right\rangle  = k_B T_{\left\{\eta\right\},\left\{\eta' \right\} } , \label{equiparticross}
\end{equation}
which conveniently extends the equipartition expression Eq. (\ref{equiparti}).
$T_{\left\{\eta\right\},\left\{\eta' \right\} }$ corresponds to {\it correlations } existing between the modes, which are discussed in detail in the following Section \ref{FDT}.
Note that $T_{\left\{\eta\right\},\left\{\eta' \right\} }=T_{\left\{\eta'\right\},\left\{\eta\right\} }$ per construction.
As for the in-equilibrium situation, all $T_{\left\{\eta\right\},\left\{\eta' \right\} }$ should be finite which means that the second sum in Eq. (\ref{grad2})
vanishes for $N_{cut \,i} \rightarrow + \infty$.
Defining $q' = q + \Delta q$ (with $\Delta q \neq 0$), we re-write $T_{\left\{\eta\right\},\left\{\eta' \right\} }=T_{\left\{\eta\right\},\Delta q } $ with $\left\{\eta\right\}$ containing (as already explained) both the branch label and $q$.
\begin{widetext}
\begin{eqnarray} 
\frac{d E_{tot} (z)}{d z}  & = & \sum_{\mbox{all branches $i$}} \!\!\!\!\!\!\!\! \frac{d E_{tot\, branch \, i} (z)}{d z}, \label{grad1} \\
\frac{d E_{tot\, branch \, i} (z)}{d z}  & = & N_{cut \,i} \frac{k_B T_{branch\, i}}{L} \nonumber \\
& & \!\!\!\!\!\! + \, N_{cut \,i} \Bigg\{ \frac{1}{N_{cut \,i}} \! \sum_{q=1}^{N_{cut \,i}} \sum_{\substack{q'=1 \\q' \neq q }}^{N_{cut \,i}} 
\left[ 1+ \frac{\omega_{\left\{\eta \right\}} \omega_{\left\{\eta'\right\}} \, {\cal M}_{\left\{\eta\right\},\left\{\eta' \right\} } }{ {\cal K}_{\left\{\eta\right\},\left\{\eta' \right\} } } \right] \left[ \frac{1}{2}\cos [\frac{\pi}{L} (q-q') z]  \right] \frac{1}{2} {\cal K}_{\left\{\eta\right\},\left\{\eta' \right\} } \left\langle  U_{\omega_{\left\{\eta\right\}}} U_{\omega_{\left\{\eta'\right\}}} \right\rangle   \nonumber \\
\!\!\!\!\!\!\!\!\!\!\!\!\!\!\!\!\!\!\!\!\!\!\!\!\!\!\!\!\!\!\!\!\!\!\!\!\!\!\!\!\!\!\!\!\!\!\!\!\!\!\!\!\!\!\!\!\!\!\!\!\!\!\!\!\!\!\!\!\!\!\!\!&  & \!\!\!\!\!\!\!\!     \pm \,  \frac{1}{N_{cut \,i}} \! \sum_{q=1}^{N_{cut \,i}} \sum_{\substack{q'=1 \\q' \neq q }}^{N_{cut \,i}}  \left[ \left(  \frac{\Delta {\cal K}_{\left\{\eta\right\},\left\{\eta'\right\}} }{{\cal K}_{\left\{\eta\right\},\left\{\eta' \right\} }} +\frac{1}{2} \right) + \frac{\omega_{\left\{\eta \right\}} \omega_{\left\{\eta'\right\}} \, {\cal M}_{\left\{\eta\right\},\left\{\eta' \right\} } }{ {\cal K}_{\left\{\eta\right\},\left\{\eta' \right\} } }\left(  \frac{\Delta {\cal M}_{\left\{\eta\right\},\left\{\eta'\right\}} }{{\cal M}_{\left\{\eta\right\},\left\{\eta' \right\} }} -\frac{1}{2} \right) \right]  \nonumber \\
& &\!\!\!\!\!\!\!\!   \times \left[ \cos (\frac{\pi}{L} [q+q'] z)  \right] \frac{1}{2} {\cal K}_{\left\{\eta\right\},\left\{\eta' \right\} } \left\langle  U_{\omega_{\left\{\eta\right\}}} U_{\omega_{\left\{\eta'\right\}}} \right\rangle \Bigg\} . \label{grad2}
\end{eqnarray}
\end{widetext}
%

An educated guess already tells us that these correlations should quickly decay as $|\Delta q |$ increases.
The neighboring modes $\left\{\eta\right\},\left\{\eta'\right\}$ in a pair with non-vanishing contribution to Eq. (\ref{grad2}) should therefore be very similar;
this actually means that 
the result Eq. (\ref{property}) can be extended to the present case, 
 with $\omega_{\left\{\eta\right\}}   \omega_{\left\{\eta' \right\}} {\cal M}_{\left\{\eta\right\},\left\{\eta' \right\} }/{\cal K}_{\left\{\eta\right\},\left\{\eta' \right\} } \approx 1$ (see Appendix \ref{coefII} for details).
Furthermore, the sum on $\Delta q = q'-q$ runs on negative and positive integers, and for most modes the actual boundary is irrelevant since correlations die away rapidly: only the very first modes on a branch (small $q \approx 1$) and the very last ($q \approx N_{cut\,i}$) will have a truncated, asymmetric sum on $\Delta q$. This has no weight in the total expression Eq. (\ref{grad2}) and can be safely neglected, which allows us to write:
\begin{eqnarray}
\frac{d E_{tot\, branch \, i} (z)}{d z}  & = & N_{cut \,i} \frac{k_B T_{branch\, i}}{L} \nonumber \\
 &  &\!\!\!\!\!\!\!\!\!\!\!\!\!\!\!\!\!\!\!\!\!\!\!\!\!\!\!\!\!\!\!\!\!  + N_{cut \,i} \frac{k_B}{L} \!\!\! \sum_{\substack{\Delta q=-\infty \\ \Delta q \neq 0  }}^{+ \infty} \!\!\!\!  \cos [\frac{\pi}{L} \Delta q \, z] \, T_{branch\, i, \Delta q}  ,
\end{eqnarray}
which is the generalization of Eq. (\ref{equilE}), with $T_{branch\, i, \Delta q} = 1/N_{cut\, i} \sum_{q=1}^{N_{cut\, i}} T_{\left\{\eta\right\}, \Delta q}$ (the average correlation function for branch $i$). 

Summing-up all branches, and having imposed that the total number of degrees of freedom is preserved [as stated by Eq. (\ref{2Nequa})], we finally reproduce the thermal gradient shape Eq. (\ref{profile}) appearing in the total energy density expression Eq. (\ref{energydens}) from the equality:
\begin{equation}
 \Delta T \left( \frac{z}{L} - \frac{1}{2} \right) = \sum_{\substack{\Delta q=-\infty \\ \Delta q \neq 0  }}^{+ \infty} \!\!\!\!  \cos [\frac{\pi}{L} \Delta q \, z] \, T_{avg , \Delta q}  ,
\end{equation} 
where $T_{avg , \Delta q} = 1/3N \sum_{\mbox{all modes $\left\{\eta\right\}$}} T_{\left\{\eta\right\}, \Delta q}$ extends the previous in-equilibrium expression Eq. (\ref{Tequil}).
The links between the different correlation parameters $T_{\left\{\eta\right\}, \Delta q}$, $T_{branch\, i, \Delta q}$, and $T_{avg , \Delta q}$ shall be discussed in the following Section \ref{FDT}. 
Since the $ \cos [\frac{\pi}{L} \Delta q \, z]$ functions form a free family, this is easily solved producing:
\begin{equation}
T_{avg , \Delta q} =-\frac{  \left[1 - (-1)^{\Delta q} \right]}{\pi^2 \left(\Delta q\right)^2} \Delta T, \label{main}
\end{equation}
with $\Delta T$ the macroscopic temperature gradient between the two ends.
All even terms are zero, and only odd $\Delta q$ allow correlations, which fall as $\propto 1/ \Delta q^2 $ (such that only nearest neighboring modes are really relevant). 
Note that
when integrating Eqs. (\ref{modekin2},\ref{modebend2}) over the length of the rod, these energy contributions vanish: there is therefore {\it no overall 
cross-terms in the inertia or restoring forces} 
 that appear in the mode's effective dynamics equation
 [Eq. (\ref{Dyn1}), see Section \ref{FDT} below]. \\

Similarly to the energy densities, the power friction densities Eqs. (\ref{bulkD},\ref{bulkC}) also have $\left\{\eta\right\} \neq \left\{\eta'\right\}$ cross-terms. The internal friction contribution leads to:
\begin{eqnarray}
&& \int_0^{2 \pi} \!\!\!  \int_0^R {\cal \dot{D}}_{int \, \left\{\eta\right\},\left\{\eta' \right\} } \, r dr d\theta  =  \nonumber \\
&& \Bigg\{ {\cal L}_{\left\{\eta\right\}, \left\{\eta'\right\}}  \left( \cos [( k_p-k_p') z] \pm \cos [( k_p+k_p') z] \right)  \nonumber \\
&& \pm  2 \Delta {\cal L}_{\left\{\eta\right\}, \left\{\eta'\right\}}    \cos [( k_p+k_p') z]   \Bigg\} \dot{U}_{\omega_{\left\{\eta\right\}}} \dot{U}_{\omega_{\left\{\eta'\right\}}} , \label{bulkcross}
\end{eqnarray}
when integrated over the cross-section. The $\pm$ corresponds to T and F modes with +, and L modes with - (as above).
The ${\cal L}_{\left\{\eta\right\}, \left\{\eta'\right\}} $, $\Delta {\cal L}_{\left\{\eta\right\}, \left\{\eta'\right\}} $ coefficients are given in Appendix \ref{coefII} explicitly.
Inspecting Eq. (\ref{bulkcross}), we see that the internal friction density has a $z$-dependence similar to the potential energy Eq. (\ref{modebend2}).
Again, integrating over the length of the rod these terms vanish from the effective dynamics equation (Section \ref{FDT}).

The situation encountered with the clamping terms $\left\{\eta\right\} \neq \left\{\eta'\right\}$ is
in this respect slightly different. One obtains:
\begin{eqnarray}
\int_{\substack{\mbox{clamps} \\ 0<z<\varepsilon \\ L<z<L+\varepsilon}}  \int_0^{2 \pi} \!\!\!  \int_0^R {\cal \dot{D}}_{clamp \, \left\{\eta\right\},\left\{\eta' \right\} } \, r dr d\theta dz & = & \nonumber \\
&&\!\!\!\!\!\!\!\!\!\!\!\!\!\!\!\!\!\!\!\!\!  \!\!\!\!\!\!\!\!\!\!\! \!\!\!\!\!\!\!\!\!\!\!\!\!\!\!\!\!\!\!\!\! \!\!\!\!\!\!\!\!\!\!\!\!\!\!\!\!\!\!\!\!\! \!\!\!\!\!\!\!\!\!\!\!\!\!\!\!\!\!\!\!\!\!  \!\!\!\!\!\!\!\!\!\!\! \!\!\!\!\!\!\!\!\!\!\!\!\!\!\!\!\!\!\!\!\!  \left( \frac{1 + (-1)^{\Delta q}}{2} \right) \! {\cal C}_{\left\{\eta\right\},\left\{\eta'\right\}} \varepsilon \left[ \omega_{\left\{\eta\right\}}  \omega_{\left\{\eta'\right\}} \, U_{\omega_{\left\{\eta\right\}}} U_{\omega_{\left\{\eta'\right\}}} \right]  , \label{dissclampcross} 
\end{eqnarray}
with the ${\cal C}_{\left\{\eta\right\},\left\{\eta'\right\}}$ coefficients discussed in Appendix \ref{coefII}.
Eq. (\ref{dissclampcross}) means that for odd $\Delta q$ (the ones that carry correlations), these {\it clamping cross-terms vanish} too: again, the mode dynamics equation is free from such contributions (see Section \ref{FDT}).
However, even though there is no net related viscous clamping force acting on the mode, there is a finite force acting on {\it each end} of the rod
(with two opposite signs).
This has to be contrasted with the in-equilibrium treatment which results in a finite clamping friction force (the two anchoring contributions have the same sign). 
Replacing Eq. (\ref{equiparticross}) in Eq. (\ref{dissclampcross}), the {\it total energy flow} $\dot{Q}$ can thus be written as:
\begin{equation}
\dot{Q} = +\!\! \sum_{\left\{\eta\right\} \neq \left\{\eta'\right\}} \frac{{\cal C}_{\left\{\eta\right\},\left\{\eta'\right\}} \varepsilon \, \omega_{\left\{\eta\right\}}  \omega_{\left\{\eta'\right\}} }{{\cal K}_{\left\{\eta\right\},\left\{\eta'\right\}} L} \, k_B T_{\left\{\eta\right\},\left\{\eta'\right\}}\, , \label{flow}
\end{equation}
considering only one end of the rod (in this writing, the sign refers to flow in the direction of $\vec{z}$), and summing up all contributions. 
%
The findings of this Section shall finally be analyzed in the following one, in the framework of an extension of the {\it fluctuation-dissipation theorem}.

\section{Extended Fluctuation-Dissipation Theorem}
\label{FDT}

The aim of this final Section is to interpret the previous findings in terms of
{\it effective stochastic forces} $\delta f_{\left\{\eta\right\}}(t)$ acting upon each of the modes $\left\{\eta\right\}$.
Putting together the ingredients derived in the two previous Sections and applying Newton's law, we derive the mode's effective 
dynamics equation:
\begin{eqnarray}
& & \!\!\!\!\!\!\!\!\!\! \frac{{\cal M}_{\left\{\eta\right\} } L}{2} \,\, \ddot{U}_{\omega_{\left\{\eta\right\}}} + \left( {\cal L}_{\left\{\eta\right\} } L + {\cal C}_{\left\{\eta\right\} } \varepsilon \right) \dot{U}_{\omega_{\left\{\eta\right\}}} + \frac{{\cal K}_{ \left\{\eta\right\} } L}{2} \,\, U_{\omega_{\left\{\eta\right\}}} \nonumber \\
 & = & \delta f_{\left\{\eta\right\}} . \label{Dyn1} 
\end{eqnarray}
The standard procedure when considering a single mode is then to perform a Rotating Wave Approximation (RWA); we shall not do that here, and instead look for the exact solution. The RWA is discussed explicitly in Appendix \ref{beatings}.

Applying the same approach as the one of the standard Fluctuation-Dissipation Theorem, let us assume that the $\delta f_{\left\{\eta\right\}}$ forces are all centered ($\left\langle \delta f_{\left\{\eta\right\}}(t)\right\rangle=0$) and stationary. We define:
\begin{eqnarray}
C_{\delta f_{\left\{\eta\right\}}, \delta f_{\left\{\eta'\right\}}}(\tau) & = & \left\langle \delta f_{\left\{\eta\right\}}(t) \delta f_{\left\{\eta'\right\}}(t-\tau) \right\rangle , \\
S_{\delta f_{\left\{\eta\right\}}, \delta f_{\left\{\eta'\right\}}}(\omega) & = & {\cal F T } \left[ C_{\delta f_{\left\{\eta\right\}},\delta f_{\left\{\eta'\right\}}} \right] \! \left( \omega\right),
\end{eqnarray}
the correlation function and corresponding spectrum of the force fluctuations.  ${\cal F T }[\cdots]$ is the Fourier Transform (defined as $\int_{-\infty}^{+\infty} \cdots \, \exp[-\mathrm{i} \, \omega \tau] d\tau$).
Similarly, we write:
\begin{eqnarray}
C_{U_{\omega_{\left\{\eta\right\}}}\!,\,U_{\omega_{\left\{\eta'\right\}}}}(\tau) & = & \left\langle U_{\omega_{\left\{\eta\right\}}}(t)\, U_{\omega_{\left\{\eta'\right\}}}(t-\tau) \right\rangle , \\
S_{U_{\omega_{\left\{\eta\right\}}}\!,\,U_{\omega_{\left\{\eta'\right\}}}}(\omega) & = & {\cal F T } \left[C_{U_{\omega_{\left\{\eta\right\}}}\!,\,U_{\omega_{\left\{\eta'\right\}}}} \right] \! \left( \omega\right),
\end{eqnarray}
for the motion amplitudes of two arbitrary modes $\left\{\eta\right\}, \left\{\eta'\right\}$ (which might be equal or not). Eq. (\ref{Dyn1}) leads to:
\begin{eqnarray}
& & S_{U_{\omega_{\left\{\eta\right\}}}\!,\,U_{\omega_{\left\{\eta'\right\}}}}(\omega)  =  \nonumber \\
& & \chi_{\left\{\eta\right\}}(\omega) \, \chi_{\left\{\eta'\right\}}(\omega)^{*} \, \frac{S_{\delta f_{\left\{\eta\right\}},\delta f_{\left\{\eta'\right\}}}(\omega)}{({\cal K}_{ \left\{\eta\right\} } L)/2 \,({\cal K}_{ \left\{\eta'\right\} } L)/2 } , \\
& & \chi_{\left\{\eta\right\}}(\omega)  =  
 \frac{\omega_{\left\{\eta\right\}}^2}{-\omega^2 +\omega_{\left\{\eta\right\}}^2 + \mathrm{i} \, \omega \Gamma_{\left\{\eta\right\}}} , \label{xi}
\end{eqnarray}
with $*$ the complex conjugate.
The frequency $\omega_{\left\{\eta\right\}}$ is defined from the mode mass and spring constant with Eq. (\ref{property}), 
and the relaxation rate $\Gamma_{\left\{\eta\right\}}$ with Eq. (\ref{gamma}).
The $\chi_{\left\{\eta\right\}}$ (complex) functions are normalized mode susceptibilities. 
The correlation function of velocities is easily obtained from $C_{\dot{U}_{\omega_{\left\{\eta\right\}}}\!,\,\dot{U}_{\omega_{\left\{\eta'\right\}}}}(\tau)=- d^2 C_{U_{\omega_{\left\{\eta\right\}}}\!,\,U_{\omega_{\left\{\eta'\right\}}}}(\tau)/d\tau^2$, which leads to the simple relationship for the spectra:
\begin{equation}
S_{\dot{U}_{\omega_{\left\{\eta\right\}}}\!,\,\dot{U}_{\omega_{\left\{\eta'\right\}}}}(\omega) = \omega^2 S_{U_{\omega_{\left\{\eta\right\}}}\!,\,U_{\omega_{\left\{\eta'\right\}}}}(\omega).
\end{equation}
We then have:  
\begin{eqnarray}
\left\langle U_{\omega_{\left\{\eta\right\}}}(t)\, U_{\omega_{\left\{\eta'\right\}}}(t) \right\rangle &= &C_{U_{\omega_{\left\{\eta\right\}}}\!,\,U_{\omega_{\left\{\eta'\right\}}}}(0) \nonumber \\
&& \!\!\!\!\!\!\!\!\!\! \!\!\!\!\!\!\!\!\!\! \!\!\!\!\!\!\!\!\!\! =  \frac{1}{2 \pi} \int_{-\infty}^{+\infty} S_{U_{\omega_{\left\{\eta\right\}}}\!,\,U_{\omega_{\left\{\eta'\right\}}}}(\omega) d \omega , \label{intX} \\
\left\langle \dot{U}_{\omega_{\left\{\eta\right\}}}(t)\, \dot{U}_{\omega_{\left\{\eta'\right\}}}(t) \right\rangle &= &C_{\dot{U}_{\omega_{\left\{\eta\right\}}}\!,\,\dot{U}_{\omega_{\left\{\eta'\right\}}}}(0) \nonumber \\
&& \!\!\!\!\!\!\!\!\!\! \!\!\!\!\!\!\!\!\!\! \!\!\!\!\!\!\!\!\!\! =  \frac{1}{2 \pi} \int_{-\infty}^{+\infty} \omega^2 \, S_{U_{\omega_{\left\{\eta\right\}}}\!,\,U_{\omega_{\left\{\eta'\right\}}}}(\omega) d \omega , \label{intV}
\end{eqnarray}
for which we shall impose Eq. (\ref{equiparti}) when $\left\{\eta\right\} = \left\{\eta'\right\}$ and Eq. (\ref{equiparticross}) when $\left\{\eta\right\} \neq \left\{\eta'\right\}$.
Following again the standard procedure, we shall assume that the force noise spectrum $S_{\delta f_{\left\{\eta\right\}}, \delta f_{\left\{\eta'\right\}}}$ is flat on the scale of the two mechanical resonances at $\omega_{\left\{\eta\right\}}$ and $\omega_{\left\{\eta'\right\}}$.
This is certainly adapted to high-Q modes, and simply states that {\it there is no specific correlation time} characterizing the forces $\delta f_{\left\{\eta\right\}}(t)$. 
Since $\delta f_{\left\{\eta\right\}}$ are real-valued, we introduce two real-valued variables per $\left\{\eta\right\}, \left\{\eta'\right\}$ couple:
\begin{eqnarray}
S_{\delta f_{\left\{\eta\right\}}, \delta f_{\left\{\eta'\right\}}} (\omega) & = & {\cal X}_{\left\{\eta\right\}, \left\{\eta'\right\}} + \mathrm{i} \, \mbox{sign}(\omega) \, {\cal Y}_{\left\{\eta\right\}, \left\{\eta'\right\}} , \,\,\,\,  \\
{\cal X}_{\left\{\eta\right\}, \left\{\eta'\right\}} & = & {\cal X}_{\left\{\eta' \right\}, \left\{\eta\right\}} , \\
{\cal Y}_{\left\{\eta\right\}, \left\{\eta'\right\}} & = & - {\cal Y}_{\left\{\eta' \right\}, \left\{\eta\right\}} .
\end{eqnarray}
Finding ${\cal X}_{\left\{\eta\right\}, \left\{\eta'\right\}}, {\cal Y}_{\left\{\eta\right\}, \left\{\eta'\right\}}$ finally solves the problem at hand.
Note that ${\cal Y}_{\left\{\eta\right\}, \left\{\eta'\right\}} \neq 0$ breaks time-reversal invariance [$C_{\delta f_{\left\{\eta\right\}}, \delta f_{\left\{\eta'\right\}}}(\tau) \neq C_{\delta f_{\left\{\eta\right\}}, \delta f_{\left\{\eta'\right\}}}(-\tau)$], which we did not impose to the modeling (see discussion below). \\

The properties of the integrals Eqs. (\ref{intX},\ref{intV}) are described in Appendix \ref{integrals}.
One should distinguish two situations:
\begin{itemize}
\item When the frequencies $\omega_{\left\{\eta\right\}} \approx \omega_{\left\{\eta'\right\}}$, meaning that the two resonance peaks described by the susceptibilities Eq. (\ref{xi}) overlap: $\left|\omega_{\left\{\eta\right\}}- \omega_{\left\{\eta'\right\}}\right|$ is much smaller than their respective damping rates $\Gamma_{\left\{\eta\right\}},\Gamma_{\left\{\eta'\right\}}$ (which obviously applies to the case $\left\{\eta\right\}=\left\{\eta'\right\}$). 
For $\left\{\eta\right\} \neq \left\{\eta'\right\}$, this limit is reached on any branch for large wavevectors $k_p R, k_p' R \gg 1$, when the frequency difference between the resonances becomes infinitely small. On high order branches, this is also true for small wavevectors where the frequency is essentially constant (see Appendix \ref{pochhammer}).
Then:
\begin{eqnarray}
&& \left\langle \dot{U}_{\omega_{\left\{\eta\right\}}}\, \dot{U}_{\omega_{\left\{\eta'\right\}}} \right\rangle  =  \omega_{\left\{\eta\right\}}^2 \, \left\langle U_{\omega_{\left\{\eta\right\}}}\, U_{\omega_{\left\{\eta'\right\}}} \right\rangle , \label{VtoX} \\
&& \left\langle U_{\omega_{\left\{\eta\right\}}}\, U_{\omega_{\left\{\eta'\right\}}} \right\rangle  =  \frac{\omega_{\left\{\eta\right\}}}{2} \frac{ \sqrt{Q_{\left\{\eta\right\}} Q_{\left\{\eta'\right\}}}}{{\frac{{\cal K}_{ \left\{\eta\right\} } L}{2} \,\frac{{\cal K}_{ \left\{\eta'\right\} } L}{2} }} \nonumber \\
&&  \,\,\,\,\,  \times \left[\mbox{I}_{\mbox{Re}} \, {\cal X}_{\left\{\eta\right\}, \left\{\eta'\right\}} + \mbox{I}_{\mbox{Im}} \frac{{\cal Y}_{\left\{\eta\right\}, \left\{\eta'\right\}}}{\sqrt{Q_{\left\{\eta\right\}} Q_{\left\{\eta'\right\}}}}\right], \label{SpectreI}
\end{eqnarray}
in the high-Q limit, with $Q_{\left\{\eta\right\}}$ defined in Eq. (\ref{QuFact}). $\mbox{I}_{\mbox{Re}}$ and $\mbox{I}_{\mbox{Im}}$ are two functions of $Q_{\left\{\eta\right\}}/Q_{\left\{\eta'\right\}}$ described in Appendix \ref{integrals}.
\item When the frequencies $\omega_{\left\{\eta\right\}} \neq \omega_{\left\{\eta'\right\}}$, the two resonance peaks are clearly separated: $\left|\omega_{\left\{\eta\right\}}- \omega_{\left\{\eta'\right\}}\right|$ is much larger than their respective damping rates $\Gamma_{\left\{\eta\right\}},\Gamma_{\left\{\eta'\right\}}$. This situation is encountered {\it only on the lowest branches, at small wavevectors $k_p R, k_p' R \ll 1$}.
In this case one obtains:
\begin{eqnarray}
&& \left\langle \dot{U}_{\omega_{\left\{\eta\right\}}}\, \dot{U}_{\omega_{\left\{\eta'\right\}}} \right\rangle  =  \omega_{\left\{\eta\right\}} \omega_{\left\{\eta'\right\}} \,  \frac{1}{2} \frac{ \sqrt{\omega_{\left\{\eta\right\}} \omega_{\left\{\eta'\right\}}}}{{\frac{{\cal K}_{ \left\{\eta\right\} } L}{2} \,\frac{{\cal K}_{ \left\{\eta'\right\} } L}{2} }} \nonumber \\
&&  \times \left[\mbox{J}_{\mbox{Re}} \frac{{\cal X}_{\left\{\eta\right\}, \left\{\eta'\right\}}}{\sqrt{Q_{\left\{\eta\right\}} Q_{\left\{\eta'\right\}}}} + \mbox{J}_{\mbox{Im}} \, {\cal Y}_{\left\{\eta\right\}, \left\{\eta'\right\}}\right], \\
&& \left\langle U_{\omega_{\left\{\eta\right\}}}\, U_{\omega_{\left\{\eta'\right\}}} \right\rangle  =   \frac{1}{2} \frac{\sqrt{\omega_{\left\{\eta\right\}} \omega_{\left\{\eta'\right\}}}}{{\frac{{\cal K}_{ \left\{\eta\right\} } L}{2} \,\frac{{\cal K}_{ \left\{\eta'\right\} } L}{2} }} \nonumber \\
&& \,\,\,\,\,  \times \left[\mbox{I}_{\mbox{Re}} \frac{{\cal X}_{\left\{\eta\right\}, \left\{\eta'\right\}}}{\sqrt{Q_{\left\{\eta\right\}} Q_{\left\{\eta'\right\}}}} + \mbox{I}_{\mbox{Im}} \, {\cal Y}_{\left\{\eta\right\}, \left\{\eta'\right\}}\right], \label{SpectreII}
\end{eqnarray}
again in the high-Q limit. The functions  $\mbox{I}_{\mbox{Re}}, \mbox{I}_{\mbox{Im}}, \mbox{J}_{\mbox{Re}}$ and $\mbox{J}_{\mbox{Im}}$ are presented in Appendix \ref{integrals}. They depend in principle on two variables, $Q_{\left\{\eta\right\}}/Q_{\left\{\eta'\right\}}$ and $\omega_{\left\{\eta\right\}}/\omega_{\left\{\eta'\right\}}$.
\end{itemize}

Considering $Q_{\left\{\eta\right\}},Q_{\left\{\eta'\right\}} \rightarrow +\infty$, the ${\cal Y}_{\left\{\eta\right\}, \left\{\eta'\right\}}$ term vanishes from Eq. (\ref{SpectreI}) [and can thus be neglected], while it is the ${\cal X}_{\left\{\eta\right\}, \left\{\eta'\right\}}$ one that vanishes from Eq. (\ref{SpectreII}) [and shall be neglected].
Besides, it turns out that when  $\omega_{\left\{\eta\right\}} \neq \omega_{\left\{\eta'\right\}}$, the two functions $\mbox{I}_{\mbox{Im}} \approx \mbox{J}_{\mbox{Im}}$: this leads to $\left\langle \dot{U}_{\omega_{\left\{\eta\right\}}}\, \dot{U}_{\omega_{\left\{\eta'\right\}}} \right\rangle \approx \omega_{\left\{\eta\right\}} \omega_{\left\{\eta'\right\}} \left\langle U_{\omega_{\left\{\eta\right\}}}\, U_{\omega_{\left\{\eta'\right\}}} \right\rangle $, a property which extends Eq. (\ref{VtoX}) and was applied in Section \ref{gradient}.
Injecting Eqs. (\ref{equiparti},\ref{equiparticross}) in Eqs. (\ref{SpectreI},\ref{SpectreII}), we finally obtain:
\begin{eqnarray}
&& {\cal X}_{\left\{\eta\right\}, \left\{\eta'\right\}} \approx {\cal X}_{\left\{\eta\right\}, \left\{\eta'\right\}} \frac{{\cal K}_{\left\{\eta'\right\} , \left\{\eta'\right\} }}{\sqrt{ {\cal K}_{ \left\{\eta\right\} } {\cal K}_{ \left\{\eta'\right\} } }}  =  \nonumber \\
&& \frac{2  \sqrt{\frac{ {\cal M}_{ \left\{\eta\right\} } L}{2} \frac{{\cal M}_{ \left\{\eta'\right\} } L}{2} \, \Gamma_{\left\{\eta\right\}} \Gamma_{\left\{\eta'\right\}}} \,\,k_B T_{\left\{\eta\right\}, \left\{\eta'\right\}}}{\mbox{I}_{\mbox{Re}}\left(\frac{Q_{\left\{\eta\right\}}}{Q_{\left\{\eta'\right\}}} \right)}  , \label{geneFDT} \\
&& {\cal Y}_{\left\{\eta\right\}, \left\{\eta'\right\}}  \approx 0 ,
\end{eqnarray}
for modes $\omega_{\left\{\eta\right\}} \approx \omega_{\left\{\eta'\right\}}$. We used the property ${\cal K}_{\left\{\eta'\right\} , \left\{\eta'\right\} } \approx {\cal K}_{ \left\{\eta\right\} } \approx {\cal K}_{ \left\{\eta'\right\} }$ valid for nearby modes.
This is the situation encountered in the vast majority of cases; especially for $\left\{\eta\right\}=\left\{\eta'\right\}$, one recovers the standard fluctuation-dissipation result Eq. (\ref{stdFDT}) since $\mbox{I}_{\mbox{Re}}\left(1 \right)=1$ (see Appendix \ref{integrals}).
Also, the stochastic force properties {\it are then invariant by time-reversal} with $C_{\delta f_{\left\{\eta\right\}}, \delta f_{\left\{\eta'\right\}}}(\tau) = C_{\delta f_{\left\{\eta\right\}}, \delta f_{\left\{\eta'\right\}}}(-\tau)$, as one would expect. This situation should be possible to demonstrate experimentally, using high-order modes, with not too large quality factors.

The marginal situation $\omega_{\left\{\eta\right\}} \neq \omega_{\left\{\eta'\right\}}$ should however be commented, for practical reasons: this is the situation encountered with the modes that are easily at reach for the experimentalist \cite{bellon1,bellon2,bellon3,TLSus}.  
In this case, the model leads to:
\begin{eqnarray}
&& {\cal Y}_{\left\{\eta\right\}, \left\{\eta'\right\}} \approx {\cal Y}_{\left\{\eta\right\}, \left\{\eta'\right\}} \frac{{\cal K}_{\left\{\eta'\right\} , \left\{\eta'\right\} }}{\sqrt{ {\cal K}_{ \left\{\eta\right\} } {\cal K}_{ \left\{\eta'\right\} } }}  =  \nonumber \\
&& \frac{2 \sqrt{\frac{ {\cal M}_{ \left\{\eta\right\} } L}{2} \frac{{\cal M}_{ \left\{\eta'\right\} } L}{2} \, \omega_{\left\{\eta\right\}} \omega_{\left\{\eta'\right\}}}\, \,k_B T_{\left\{\eta\right\}, \left\{\eta'\right\}}}{\mbox{I}_{\mbox{Im}}\left(\frac{\omega_{\left\{\eta\right\}}}{\omega_{\left\{\eta'\right\}}} \right)}  , \label{geneFDT2} \\
&& {\cal X}_{\left\{\eta\right\}, \left\{\eta'\right\}}  \approx 0 ,
\end{eqnarray}
with the normalization $\mbox{I}_{\mbox{Im}} (2.61) \approx -1$ chosen arbitrarily (Appendix \ref{integrals}). This means that the stochastic force correlators verify $C_{\delta f_{\left\{\eta\right\}}, \delta f_{\left\{\eta'\right\}}}(\tau) = - C_{\delta f_{\left\{\eta\right\}}, \delta f_{\left\{\eta'\right\}}}(-\tau)$: they are {\it anti-symmetric} by time-reversal. 
Such kinds of correlations have been introduced in the context of exotic superconductivity or magnetic orders \cite{antisym_corr}, with a specific broken symmetry.
If this is a plausible physical result for mechanical states that could lead to measurable correlations remains to be demonstrated. \\

To conclude the modeling, let us consider the energy flow $\dot{Q}$ as defined from Eq. (\ref{flow}).
Inserting in this expression the mode parameters given in Appendix \ref{coefII}, it can be recast into:
\begin{equation}
\dot{Q} = +\!\! \sum_{\left\{\eta\right\} \neq \left\{\eta'\right\}} k_B \frac{E_Y}{\Lambda_{c\,\left\{\eta\right\} }} \frac{\varepsilon}{L} \, g_{\left\{\eta\right\}, \left\{\eta'\right\}  } \, T_{\left\{\eta\right\},\left\{\eta'\right\}} \, , \label{kappadef}
\end{equation}
with $g_{\left\{\eta\right\}, \left\{\eta'\right\}  }$ a new set of adimentional parameters. Using again the argument that modes are very close, we can approximate $g_{\left\{\eta\right\}, \left\{\eta'\right\}  } \approx g_{\left\{\eta\right\}} $  and rewrite the sum over $q$ and $\Delta q$ following the same approach as in the previous Section \ref{gradient}:
\begin{equation}
\dot{Q} \approx +\!\!\!\!\!\!\!\! \sum_{\substack{\mbox{all} \\ \mbox{ branches $i$}}} \sum_{q=1}^{N_{cut \,i}} \sum_{\substack{\Delta q=-\infty \\ \Delta q \neq 0  }}^{+ \infty} \!\!\!\! k_B \frac{E_Y}{\Lambda_{c\,\left\{\eta\right\} }} \frac{\varepsilon}{L} \, g_{\left\{\eta\right\}} \, T_{\left\{\eta\right\},\Delta q } .
\end{equation}
Each term of the sum essentially acts as a {\it conduction channel} for the energy flow: the modes themselves are not truly transporting energy, but their correlations do. 
These correlations are nothing but the amplitudes of the mode's 2-wave mixing, 
appearing because of the quadratic form of the energetic functionals (see Appendix \ref{beatings}). 
Besides, surprisingly
Eq. (\ref{geneFDT}) {\it does not} depend specifically on the couplings $ {\cal C}_{ \left\{\eta\right\},  \left\{\eta'\right\}}$ to the anchoring ends (which are our two baths), but only on the overall relaxation rates.
There is therefore no reason to assume any specific difference between modes in their ability to carry the heat, and we will simply assume $T_{\left\{\eta\right\},\Delta q } = T_{branch\, i, \Delta q} = T_{avg , \Delta q}$.
The calculation simplifies then into:
\begin{equation}
\dot{Q} \approx +3N \, k_B \frac{E_Y}{\bar{\Lambda}_{c}} \frac{\varepsilon}{L} \sum_{\substack{\Delta q=-\infty \\ \Delta q \neq 0  }}^{+ \infty} \!\!\!\! T_{avg,\Delta q } , \label{beflast}
\end{equation}
having introduced:
\begin{equation}
\frac{1}{\bar{\Lambda}_{c} } = \frac{1}{3N} \!\!\!\!\!\!\!\! \sum_{\substack{\mbox{all branches $i$} \\ \mbox{and wavevectors $q$}\\ \mbox{i.e. all   modes  $\left\{\eta\right\}$}}} \frac{ g_{\left\{\eta\right\}} }{\Lambda_{c\,\left\{\eta\right\} }}, \label{deflambdabar}
\end{equation}
which defines the average clamping loss parameter $\bar{\Lambda}_{c}$ (in Pa.s), discussed in more details in Appendix \ref{coefII}.
The only requirement here is that it should be finite; injecting Eq. (\ref{main}) into Eq. (\ref{beflast}) we finally write:
\begin{equation}
\dot{Q} \approx - \frac{ \pi R^2}{L} \left[ \frac{3}{2} \frac{ n k_B \, E_Y}{\bar{\Lambda}_{c}} \left(\varepsilon L \right) \right] \Delta T , \label{finalQ}
\end{equation}
with $n$ the atom volumic density.
When modeling the clamps, we assumed that $\varepsilon \ll L$; actually for the bracket to be finite, we realize that $\varepsilon $ should even be second-order 
 $o(1/L^2)$. We therefore introduce a new lengthscale $ \ell$:
\begin{equation}
\varepsilon = \frac{\ell^2}{L} ,
\end{equation}
which characterizes the anchoring zone. Both $\bar{\Lambda}_{c}$ and $\ell$ are nonzero for a physical realistic system, and the ideal clamp limit $\varepsilon \rightarrow 0$ appears as a trivial consequence of considering the rod infinitely long $L \rightarrow + \infty$.
Comparing expression Eq. (\ref{finalQ}) to Eq. (\ref{heat}), we identify the thermal conductivity $\kappa$:
\begin{equation}
\kappa = \frac{3}{2} \frac{ n k_B \, E_Y  \ell^2}{\bar{\Lambda}_{c}} = \frac{1}{3} C_v \, \bar{c}\left[ \frac{3}{2} \frac{ E_Y  \ell^2}{\bar{c} \, \bar{\Lambda}_{c}}\right] , \label{kappaTheor}
\end{equation}
having introduced the Dulong-Petit specific heat $C_v = 3 n k_B$, and the mean sound velocity $\bar{c}$.
The material-dependent intensive parameter in brackets corresponds to the microscopic parameter $\bar{\lambda}_{mfp}$, the {\it phonon mean-free-path} reminded in Section \ref{phonons}.
Eq. (\ref{kappaTheor}) therefore eventually links a microscopic lengthscale, the mean-free-path, to a macroscopic one (but nonetheless very small) which is the size that defines the clamping region $\ell$. 
This is somehow a consequence of the diffusive limit of phonon transport: 
within this region, the nonlinear processes responsible for $\bar{\lambda}_{mfp}$ generate the cross-correlations between modes; while the rod itself is treated within a completely linear theory.

\section{Conclusion}		
\label{conclu}

We present a theoretical treatment that links mechanics to thermal properties for a dielectric (thin-and-long) rod connected to two heat reservoirs.
It is based initially on the basic observation that thermodynamics applied to mechanical modes {\it must reproduce}
the outcomes of the microscopic phononic theory at the macroscopic level, in thermal equilibrium but also when a small heat flow is present.
The key hypotheses on which we built are: for mechanics, linear response theory up to the atomic inter-spacing, and a high enough temperature to ensure classicality to all the mechanical modes. For phononics, local equilibrium should be guaranteed within the system.
From the Pochhammer-Chree waves that exhaustively describe the rod's mechanics \cite{poche}, we reconstruct both the specific heat (mean temperature) {\it and the thermal conductivity} (from the $T$-gradient profile).
The mathematics presented is limited 
 to ``string-like'' modes, namely the ones constructed from pure traveling waves (and no evanescent contributions).
As a result, the consequences of the model are captured within a {\it modified version of the Fluctuation-Dissipation Theorem}.
It predicts cross-correlations between nearby mechanical modes, which should be measurable.
These cross-correlations arise because what transports energy is not the modes themselves, but actually their mixing. 
This counter-intuitive outcome redefines what one calls a ``propagation channel'' for heat.
Besides, the theory produces a fundamental link between a microscopic lengthscale, the phononic mean-free-path, and a macroscopic one which defines physically the clamping region (which is assumed to be small). 
The nonzero mixing amplitudes are generated there, through phonon-phonon interactions which are outside of the scope of the manuscript. \\

The theory should apply equally well to mechanical standing waves constructed from both traveling and evanescent ones (``beam-like''  modes). 
Calculating the coefficients appearing in the model within this situation is of practical interest (but is particularly more difficult), since such modes are the ones which are the easiest to measure, like the first flexures. If such modes, which do not overlap in frequency-space, can really present correlations is an open question that should be answered experimentally.
Finally, the theory can also be extended to cantilevers and 2D systems (plates and membranes), which are extensively studied experimentally \cite{bellon3,TLSus}.
No matter what are the actual geometries and materials considered, the prediction of cross-correlations has to be robust; only the quantitative magnitude should depend on the details of each specific system considered.
Eventually, applying these concepts to the quantum case would be extremely enlightening \cite{meAQS,Jukka}, making the bridge with the conventional mesoscopic physics treatment of conduction channels.

\section*{Acknowledgements}

\small{We acknowledge the support of the ERC CoG grant ULT-NEMS No. 647917. The research leading to these results has received funding from the European Union's Horizon 2020 Research and Innovation Programme, under grant agreement No. 824109, the European Microkelvin Platform (EMP).}

\appendix
\section{Pochhammer-Chree waves}
\label{pochhammer}
 
We look for propagative solutions to Eq. (\ref{newton}), and we will restrict our discussion to traveling and evanescent waves.
The $\phi_{i \left\{\eta\right\}}(r,\theta)$ functions ($i=r,\theta,z$) introduced in Eqs. (\ref{equr}-\ref{equz})
can be written \cite{poche}:
\begin{eqnarray}
\phi_{r \left\{\eta\right\}}(r,\theta) & = & \sin(n \theta +\theta_0) \times  \nonumber \\
& & \!\!\!\!\!\!\!\!\!\!\!\!\!\!\!\!\!\!\!\!\!\!\!\!\!\!\! \left( A_{\left\{\eta\right\}} \, R \alpha_{\left\{\eta\right\}} J_n'\left[\alpha_{\left\{\eta\right\}} r \right] + B_{\left\{\eta\right\}} \, R k_{e,p} \, J_n'\left[\beta_{\left\{\eta\right\}} r \right] \right.  \nonumber \\
& & \left. +C_{\left\{\eta\right\}} \, \frac{n R}{r} J_n \left[\beta_{\left\{\eta\right\}} r \right] \right) , \\
\phi_{\theta \left\{\eta\right\}}(r,\theta) & = & \cos(n \theta +\theta_0)\times \nonumber \\
& & \!\!\!\!\!\!\!\!\!\!\!\!\!\!\!\!\!\!\!\!\!\!\!\!\!\!\!  \left( A_{\left\{\eta\right\}} \,  \frac{n R}{r} J_n  \left[\alpha_{\left\{\eta\right\}} r\right] + B_{\left\{\eta\right\}} \, \frac{n R}{r} \frac{  k_{e,p}}{\beta_{\left\{\eta\right\}}} J_n \left[\beta_{\left\{\eta\right\}} r\right] \right. \nonumber \\
& & \left. + C_{\left\{\eta\right\}} \, R \beta_{\left\{\eta\right\}} J_n'\left[\beta_{\left\{\eta\right\}} r\right] \right) , \\
\phi_{z \left\{\eta\right\}}(r,\theta) & = &  \sin(n \theta +\theta_0) \times \nonumber \\
&& \!\!\!\!\!\!\!\!\!\!\!\!\!\!\!\!\!\!\!\!\!\!\!\!\!\!\!   \left( A_{\left\{\eta\right\}} \, J_n  \left[\alpha_{\left\{\eta\right\}} r\right]  - B_{\left\{\eta\right\}} \, \frac{\beta_{\left\{\eta\right\}} }{\pm k_{e,p}}  J_n  \left[\beta_{\left\{\eta\right\}} r\right]\right), 
\end{eqnarray}
with $\theta_0 = 0$ or $+\pi/2$, $J_n$ Bessel's function of the first kind (of order $n \in \mathbb{N}$) and $J_n'$ its derivative.
 $k_{e,p}$ stands for either $k_e$ or $k_p$, and the $\pm$ sign reads $+$ for traveling and $-$ for evanescent.
We shall already point out that the wavevectors $k_{e,p}$ extend up to about $\propto 1/a$ where $a$ is the inter-atomic distance; beyond this limit Eq. (\ref{newton}) is not relevant anymore.
The parameters $\alpha_{\left\{\eta\right\}}$ and $\beta_{\left\{\eta\right\}}$ verify:
\begin{eqnarray}
\alpha_{\left\{\eta\right\}}^2 \pm k_{e,p}^2 & = & \frac{\omega_{ \left\{\eta\right\}}^2}{c_l^2}, \label{alphaEq} \\
\beta_{\left\{\eta\right\}}^2 \pm k_{e,p}^2  & = & \frac{\omega_{ \left\{\eta\right\}}^2}{c_t^2},  \label{betaEq}
\end{eqnarray}
where $c_l= c_0 \sqrt{(1-\nu)/[(1+\nu)(1-2\nu)]}$ and $c_t=c_0/ \sqrt{2(1+\nu)}$ are the longitudinal and transverse velocities  respectively, with $c_0=\sqrt{E_Y/\rho}$. $A_{\left\{\eta\right\}}$, $B_{\left\{\eta\right\}}$ and $C_{\left\{\eta\right\}}$ are (dimensionless) constants which will be defined below from boundary conditions and normalization.
Here we discuss only the situation where no body forces are present in the rod; the impact of in-built stress is explicitly addressed in Subsection \ref{inbuiltstress} below.

		\begin{figure*}[t!]
		\center
	\includegraphics[width=17.cm]{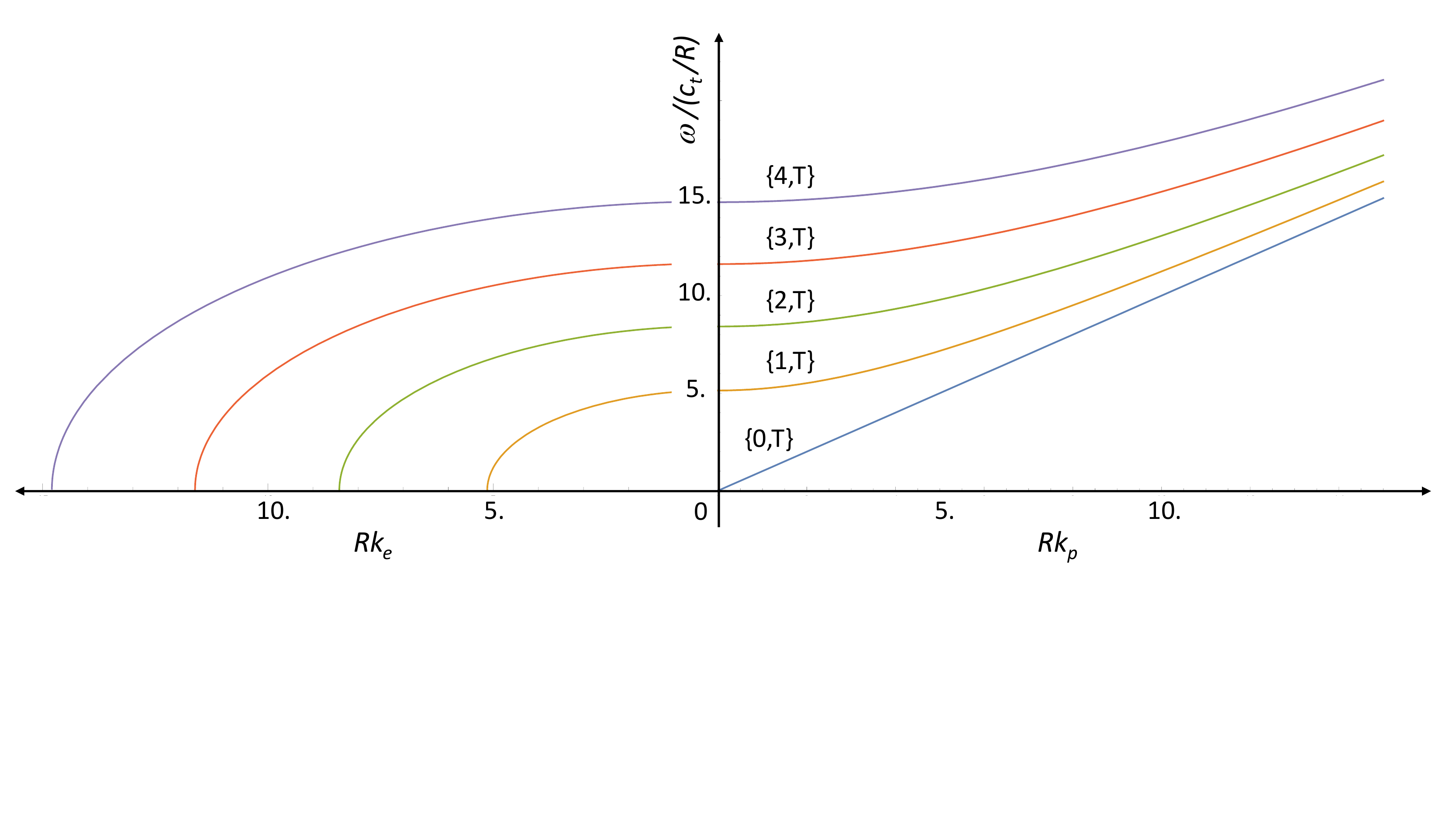}     
	\vspace*{-3cm}
			\caption{\small{Branches obtained for torsional (T) waves, in normalized units. Left: evanescent solutions. Right: traveling solutions. The numbering $\left\{m,\mbox{ T}\right\}$ is indicated in the graph. The breaks in the lines on the left of the vertical axis are there for readability of the labels.}}
			\label{fig_6}
		\end{figure*}

For a given $n$ and $\theta_0$, imposing the free-stress boundary condition on the periphery of the rod leads to {\it families} of solutions described in the following Subsections: T, L and F branches \cite{poche,onoe,pao1,pao2}. From Eqs. (\ref{alphaEq},\ref{betaEq}), 
one finds the dispersion relations $\omega_{ \left\{\eta\right\}}(k_e)$ and $\omega_{ \left\{\eta\right\}}(k_p)$ which are depicted in this Appendix, which in turn define the $\alpha_{\left\{\eta\right\}}, \beta_{\left\{\eta\right\}}$ constants. Each can be indexed by $m \in \mathbb{N}$.

On the vectorial space that contains the displacement field solutions (namely 3D-space vector $\vec{u}(r,\theta,z,t)$ functions with $r \in [0,R]$, $\theta \in [0,2\pi]$ and $\left\{ z,t \right\} \in \mathbb{R}^2$) we define the scalar product (not to be confused with the ensemble average):
\begin{eqnarray}
& & <  \vec{u}_{\left\{\eta\right\}} \bullet \vec{u}_{\left\{\eta'\right\}}>  =  \int \!\!\! \int  \vec{u}_{\left\{\eta\right\}} . \vec{u}_{\left\{\eta'\right\}} \, r dr d\theta   \nonumber \\
  & = & U_{\left\{\eta\right\}} (z,t) U_{\left\{\eta'\right\}} (z,t) \int_0^{2 \pi} \!\!\! \int_0^R \left[ \phi_{r \left\{\eta\right\}}(r,\theta) \phi_{r \left\{\eta'\right\}}(r,\theta) \right. \nonumber \\
  & + & \left. \phi_{\theta \left\{\eta\right\}}(r,\theta) \phi_{\theta \left\{\eta'\right\}}(r,\theta)  \right] r dr d\theta \nonumber \\
	& + & R^2 \frac{\partial U_{\left\{\eta\right\}}(z,t)}{\partial z} \frac{\partial U_{\left\{\eta'\right\}}(z,t)}{\partial z} \times \nonumber \\
	&   &   \int_0^{2 \pi} \!\!\! \int_0^R \left[ \phi_{z \left\{\eta\right\}}(r,\theta) \phi_{z \left\{\eta'\right\}}(r,\theta)  \right] r dr d\theta, \label{scalar}
\end{eqnarray}
written for solutions of the type described above.
The explicit dependence to the $z-$derivative of $U_{\left\{\eta\right\}}$ functions comes from our writing using only real-valued solutions. When combining those in complex form (i.e. $\partial /\partial z \rightarrow \pm k_e$ or $\pm \mathrm{i} k_p$ for evanescent and traveling respectively), $U_{\left\{\eta\right\}}   U_{\left\{\eta'\right\}}$ can be factorized out. 
The key fact that should be pointed out is that {\it all the branches are orthogonal} with respect to this scalar product, namely:
\begin{equation}
<  \vec{u}_{\left\{\eta\right\}} \bullet \vec{u}_{\left\{\eta'\right\}}>  = 0,  
\end{equation}
if $\eta$ and $\eta'$ correspond to two distinct sets of $\left\{\theta_0,n,m\right\}$, even within the same family. Only within a branch do we have $<  \vec{u}_{\left\{\eta\right\}} \bullet \vec{u}_{\left\{\eta'\right\}}>  \neq 0$, obviously when $\eta=\eta'$ but also when considering two distinct wavevectors $k_{e,p}$ and $k_{e,p}'$ of the same branch (T, L or F). 
This actually justifies why standing wave modes where created in Subsection \ref{modes} within a single branch, and not mixing up the branches (and even the families); as a direct consequence, normal modes of different branches are also orthogonal.
Implications of orthogonality are further discussed in Subsection \ref{gradient}. \\

These solutions were first studied by Pochhammer and Chree \cite{ini1,ini2}.
An obvious question that comes to mind is: are there any other possibilities?
In a more mathematical language, the solutions described being orthogonal they form a free family in the $\left\{ \vec{u} \right\}$-Hilbert space. The question is then: is this a complete set that forms a {\it basis} of this Hilbert space? This is indeed mandatory in order to write {\it any} solution $\vec{u}$ in the form of Eq. (\ref{superpose}).
When we constructed the solutions, imposing the $z$-derivative in Eq. (\ref{equz}) was a necessity to obtain propagating wave solutions, with both Eq. (\ref{zderive}) and Eq. (\ref{tderive}). The question then shifts to the construction of the $\phi_{i \left\{\eta\right\}}$ functions.
Counting these functions, we see that we have $2 \times \left\{n,m\right\}$ solutions listed, with $\left\{n,m \right\}\in \mathbb{N}^2$.
This means that the free family is (infinitely) countable, through a bijection with $\mathbb{N}$; the {\it cardinality} of our set is $\aleph_0$.
Comparing with other problems involving second order differential equations with boundary conditions (e.g. Laplace's equation), which have the same cardinality (their solution space has dimension $\aleph_0$), 
we can safely consider that our set is indeed complete and forms a basis; this is a simple (yet not fully rigorous) implication of the 
 {\it dimension theorem}.

We now turn to explicit Pochhammer-Chree solutions. These functions are used to compute the energetics coefficients introduced in the core of the paper (and listed in their respective Appendices). They are produced using Mathematica$^{\textregistered}$; the presented numerical results are obtained with $\nu=0.20$. 

\subsection {Torsional (T) waves}

The simplest solutions are torsional waves (T) obtained for $\theta_0=0$ and $n=0$.
Then, only $\phi_{\theta \left\{\eta\right\}}$ is nonzero, and reads:
\begin{equation}
\phi_{\theta \left\{m,\mbox{\tiny T} \right\}} (r,\theta) = \frac{-1}{J_1 \left[\beta_{\left\{m,\mbox{\tiny T} \right\}} R \right]} J_0'\left[\beta_{\left\{m,\mbox{\tiny T} \right\}} r \right] ,
\end{equation}
with the proper normalization $\phi_{\theta \left\{\eta\right\}} (r=R,\theta)=1$ for any $m$: the parametrization of the motion is done through the tangential displacement at the periphery of the rod. Note that wavevectors (and $\nu$) do not appear yet explicitly; the angular dependence to $\theta$ also vanished.
The boundary conditions lead in this case to the simple equation:
\begin{equation}
J_2 \left[\beta_{\left\{m,\mbox{\tiny T}\right\}} R \right]=0,
\end{equation}
which defines $\beta_{\left\{m,\mbox{\tiny T}\right\}}$ for $m \in \mathbb{N}$, and thus:
\begin{equation}
\omega_{\left\{m,\mbox{\tiny T}\right\}} (k_{e,p})^2 = c_t^2 \left[  \beta_{\left\{m,\mbox{\tiny T}\right\}}^2 \pm k_{e,p}^2  \right], \label{dispT}
\end{equation}
with $+$ for traveling and $-$ for evanescent solutions.
The first solutions are $\beta_{\left\{0,\mbox{\tiny T}\right\}}=0$ [then $\phi_{\theta \left\{0,\mbox{\tiny T} \right\}} (r,\theta) = r/R$], $\beta_{\left\{1,\mbox{\tiny T}\right\}}=5.1356 \cdots / R$, $\beta_{\left\{2,\mbox{\tiny T}\right\}}=8.4172 \cdots / R$.
The lowest dispersion relations Eq. (\ref{dispT}) are plotted in Fig. \ref{fig_6}.
The first branch is strictly linear (velocity $c_t$) and verifies $\omega_{\left\{0,\mbox{\tiny T}\right\}} (k_{p}=0)=0$; it does not support evanescent solutions.
This is the one discussed in beam theory, see e.g. Ref. \cite{clelandBk}.
The others have finite frequency at $k_{p}=0$; all evanescent solutions link their branches to $\omega = 0$.
The asymptotic behavior of all branches is $d \omega /d k_p = c_t$ at large $k_p$: we then recover a linear dispersion law.
Finally, note that none of the branches reconnect to one another, and that no local minima exist (apart at $k_{e,p}=0$); the branches never cross each other. 

		\begin{figure*}[t!]
		\center
	\includegraphics[width=17.cm]{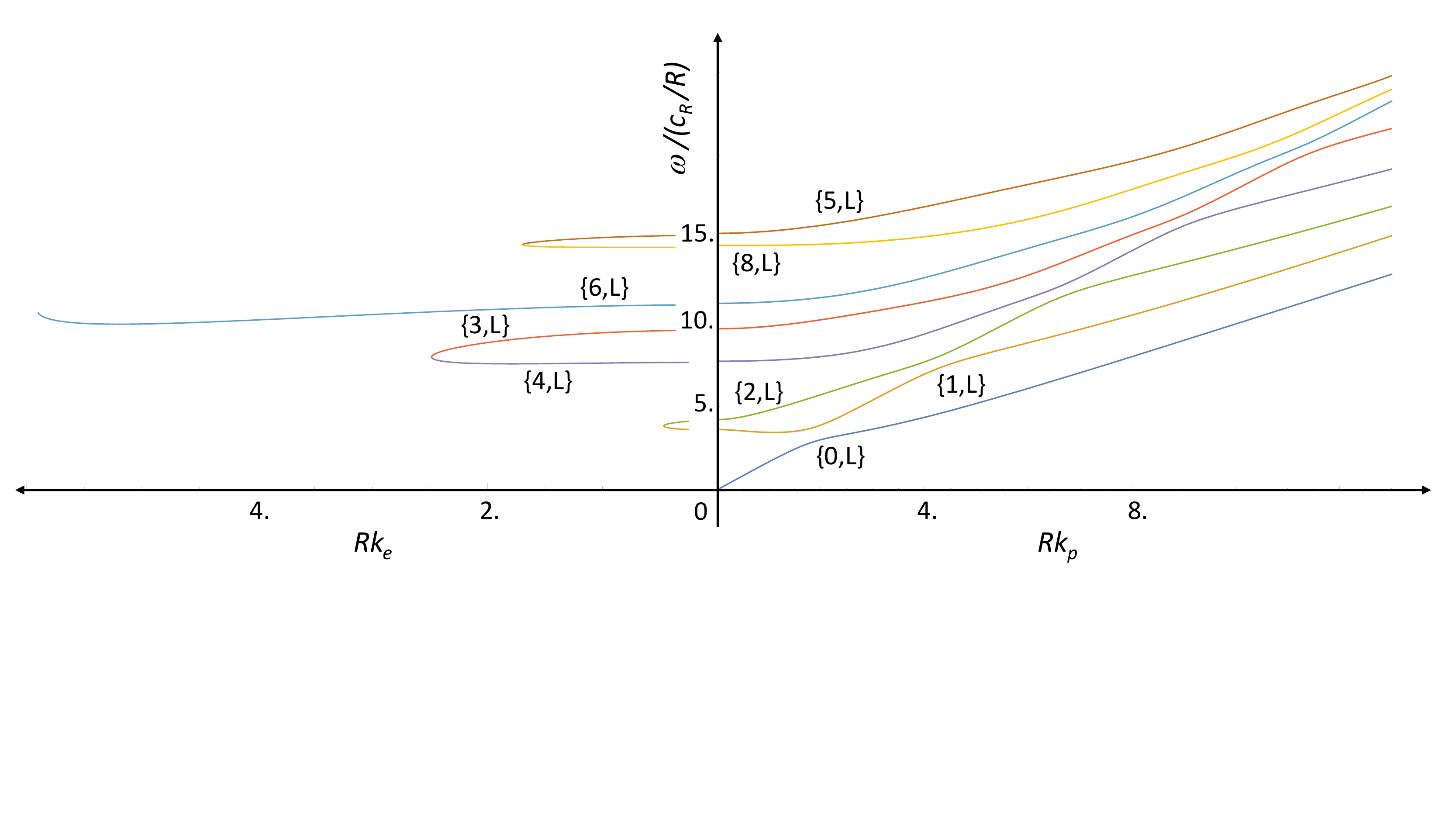}     
	\vspace*{-3cm}
			\caption{\small{Branches obtained for longitudinal (L) waves, in normalized units ($c_R$ stands for Rayleigh velocity, see text). Left: evanescent solutions. Right: traveling solutions. The numbering $\left\{m,\mbox{L}\right\}$ is indicated in the graph (see text for details). Calculation performed for $\nu=0.20$. Note the reconnections between branches on the left panel. The breaks in the lines on the left of the vertical axis are there for readability of the labels. }}
			\label{fig_7}
		\end{figure*}

\subsection {Longitudinal (L) waves}

The next family is found by imposing $\theta_0=+\pi/2$ and $n=0$: these are longitudinal (L) waves.
Then, only $\phi_{\theta \left\{\eta\right\}}$ is zero and motion exists {\it both} on the axis and radially. 
As for T waves, there is no $\theta$ dependence, but the solution is more complex.

The boundary conditions involve now two constants, $A_{\left\{\eta\right\}}$ and $B_{\left\{\eta\right\}}$. They can be written in matrix form, and a solution exists only if the determinant is zero:

\begin{widetext}

\begin{equation}
\begin{vmatrix}
 \pm 2 R^2 k_{e,p} \alpha_{\left\{\eta\right\}} J_1 \left[\alpha_{\left\{\eta\right\}}R \right] &  R^2\left( \pm k_{e,p}^2 -\beta_{\left\{\eta\right\}}^2\right)   J_1 \left[\beta_{\left\{\eta\right\}}R \right] \\
J_0 \left[\alpha_{\left\{\eta\right\}}R \right] R^2 \left(\pm k_{e,p}^2 \nu +\left[ 1-\nu\right] \alpha_{\left\{\eta\right\}}^2 \right) - R \alpha_{\left\{\eta\right\}} \left(1-2 \nu \right) J_1 \left[\alpha_{\left\{\eta\right\}}R \right] \,\,\,\,\,\,  & R k_{e,p} \left( 1- 2 \nu \right) \left(R\beta_{\left\{\eta\right\}}   J_0 \left[\beta_{\left\{\eta\right\}}R \right]-J_1 \left[\beta_{\left\{\eta\right\}}R \right]\right) \\
\end{vmatrix}
=0 . \label{detL}
\end{equation}
\vspace*{0.5cm}

\end{widetext}

Inserting in the above Eqs. (\ref{alphaEq},\ref{betaEq}), this defines the possible solutions $\omega_{\left\{m,\mbox{\tiny L}\right\}} (k_{e,p})$ that we index again with $m$. 
The wave equation now depends explicitly on wavevector $k_{e,p}$ (and $\nu$), with $+$ for traveling and $-$ for evanescent. 
From Eqs. (\ref{alphaEq},\ref{betaEq}) we obtain $\alpha_{\left\{m,\mbox{\tiny L}, k_{e,p} \right\}},\beta_{\left\{m,\mbox{\tiny L}, k_{e,p} \right\}}$, and solving Eq. (\ref{detL}) enables to express the $B_{\left\{\eta\right\}}$ constant as a function of $A_{\left\{\eta\right\}}$.
The first branches are shown in Fig. \ref{fig_7}.  

\begin{figure}[t!]		 
			 \includegraphics[width=11cm]{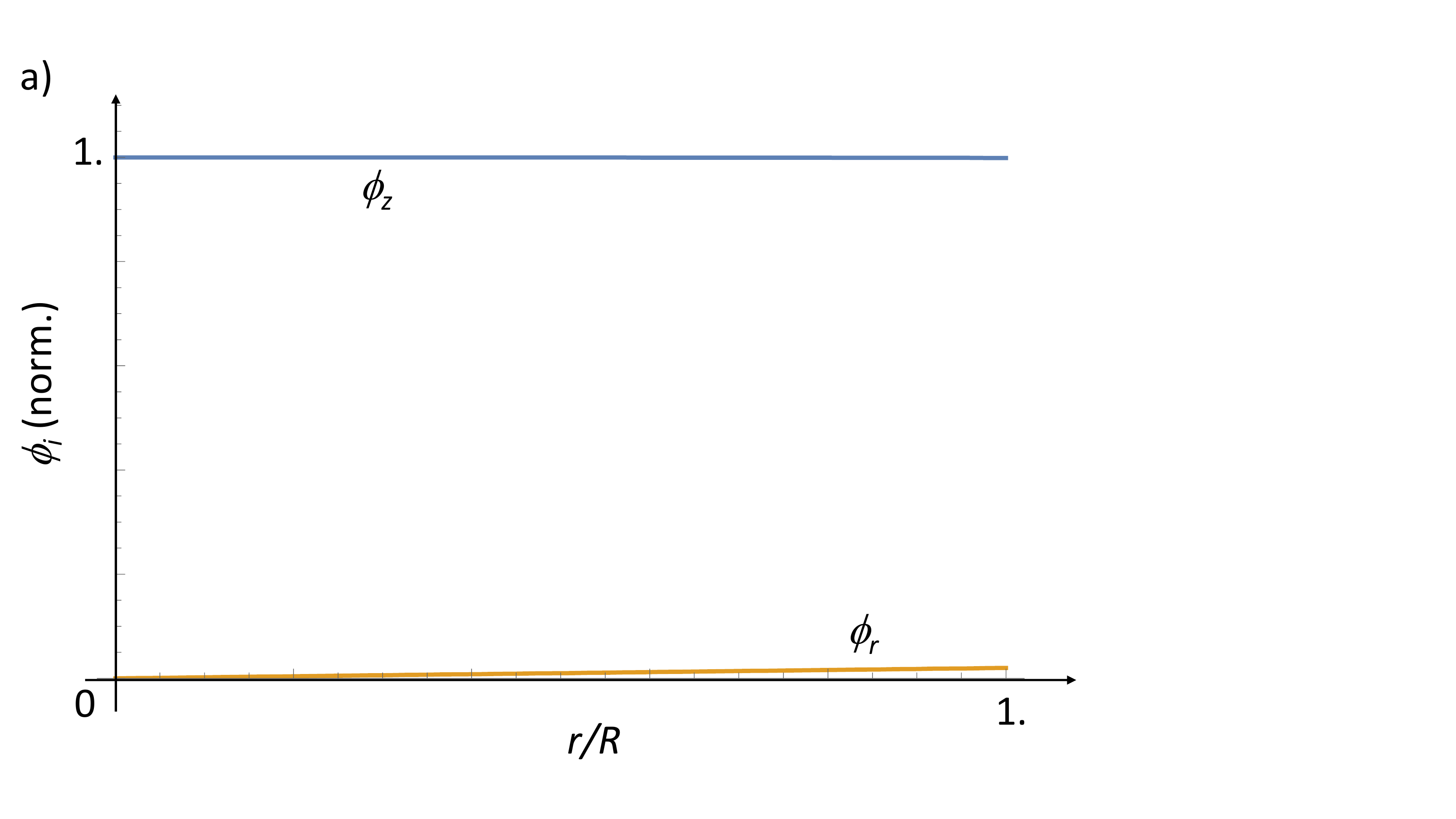}			 			 \includegraphics[width=11cm]{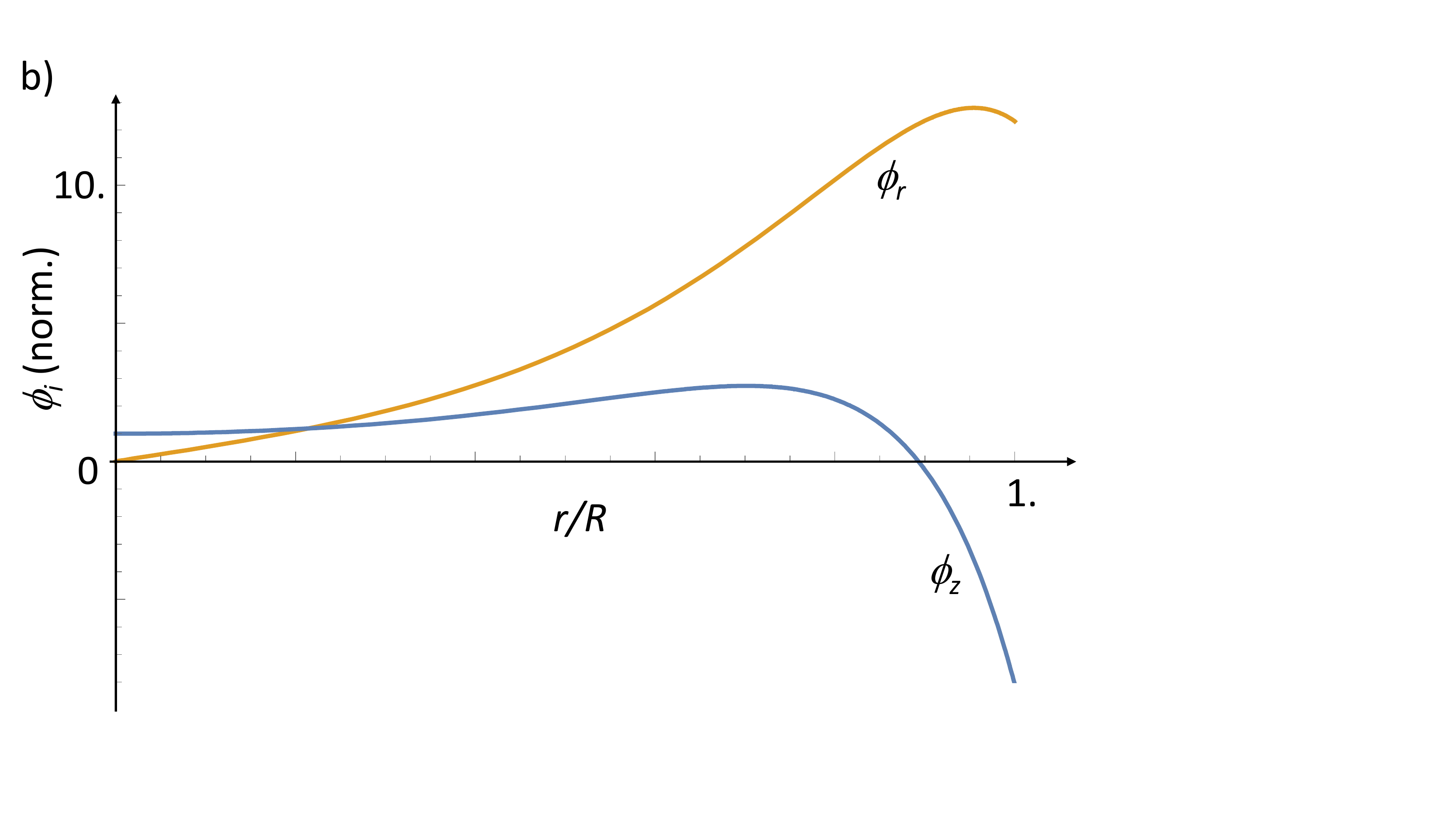}
			\caption{\small{
			Normalized displacement components for longitudinal (L) mode $m=0$ (computed for $\nu=0.20$). a) Solution with $k_p R = 0.10$. b) Solution with $k_p R = 10$. For small wavevectors, the motion is mostly a bulk translation along $\vec{z}$, matching the name {\it longitudinal}. For large wavevectors, the motion gets concentrated on the periphery with {\it both} $u_z$ and $u_r$ components. Higher modes display even more complex patterns. }}
			\label{fig_8}
		\end{figure}

The last constant $A_{\left\{\eta\right\}}$ is defined through a proper normalization of the wave shape. Since the displacement along $z$ on the axis essentially never vanishes (see final comment of this Subsection), and in order to match the common definition used for the first mode at small wavevectors, we chose $\phi_{z \left\{m,\mbox{\tiny L}, k_{e,p}\right\}}(r=0,\theta)=1$ for {\it all} modes $m$ (and {\it all} $k_{e,p}$). This means that applying the replacement procedure Eqs. (\ref{replacez1},\ref{replacez2}), the amplitude of motion is actually encoded in $U_{\left\{m,\mbox{\tiny L}, k_{e,p}\right\}}(z,t) \rightarrow R d U_{\left\{m,\mbox{\tiny L}, k_{e,p}\right\}}(z,t)/dz$. 
Example of displacement fields $u_z, u_r$ are shown in Fig. \ref{fig_8}.

The first branch $m=0$ does not support evanescent solutions. It is linear at small wavevectors with velocity $c_l$ (and goes to $\omega=0$ at $k_p=0$), while for large wavevectors it is linear again with velocity $c_R$, the Rayleigh velocity (the one of surface waves) \cite{poche}. 
A fairly good approximation (within typically $10^{-4}$ for $-0.45 < \nu < +0.45$) can be produced using:
\begin{equation}
c_R \approx c_t \left( 0.87398 + 0.19489 \, \nu - 0.0379 \, \nu^2 - 0.05679 \, \nu^3 \right).
\end{equation}
The $k_p R \ll 1$ situation is the one presented in the framework of beam theory in textbooks \cite{clelandBk}.
All higher branches have finite frequency at $k_p,k_e=0$, with a linear asymptotic dependence at large $k_p$ reaching the transverse velocity $d\omega/d k_p=c_t$.
Fig. \ref{fig_7} shows a complex pattern of reconnections on the evanescent side, which strongly depends on $\nu$ \cite{onoe}.
On both $k_p$ and $k_e$ panels, one can see local minima $d\omega/d k_p=0$ at finite wavevectors in some of the wave dispersions (e.g. $\left\{1,\mbox{L} \right\}$ right panel). These points connect to mixed wave solutions, which are outside of the scope of this paper \cite{poche}. Note that none of the branches cross each other. \\

A final comment on the numbering of the branches is in order.
Actually, the value of $\omega_{\left\{m,\mbox{\tiny L} \right\}}(k_p=0)$ for $m>0$ does depend on the Poisson ratio $\nu$, so there is 
no specific ordering {\it per se}: depending on $\nu$, branches can exchange positions with respect to $\omega$.
On the other hand, there is a difference {\it in nature} between nearby branches which we will rely on for our labeling. We thus 
count branches while increasing frequency, ordering them with the extra condition:
\begin{itemize}
\item  even $m$ ($0,2,4, \cdots$) are strongly longitudinal,
\item odd $m$ ($1,3,5, \cdots$) are strongly radial.
\end{itemize}
The resulting labeling shown in Fig. \ref{fig_7} is {\it not} the one usually found in the literature, e.g. in Refs. \cite{poche,onoe}.
Also as a consequence, our fixed parametrization of motion amplitude based on $\phi_{z \left\{m,\mbox{\tiny L} , k_{e,p} \right\}}(r=0,\theta)=1$ is perfectly adapted to even $m$ branches, but not really to odd ones. 
It can also happen that for very specific $k_{e,p}$ values (on specific branches), the $z$ component vanishes. But this happens only in a single point, and in all the vicinity around this point all calculations are well defined.
Since in the end, the modeling presented in the paper is concerned only with {\it intensive} properties, the actual parametrization does not matter (and final  parameters quoted in the core of the manuscript are always finite by construction), and we kept a fixed parametrization of motion amplitude (independent of $m$) for simplicity.

\subsection {Flexural (F) waves}

The situation encountered with flexural (F) waves is the most complex, with two indexes $n>0,m \geq 0$ and {\it all three displacement components nonzero}.
The solutions produced by the $\pi/2$ rotation (i.e. $\theta_0 = 0$ or $+\pi/2$) are {\it degenerate}: they correspond to the same displacement field patterns (but rotated from each other), having the same frequencies.

\begin{figure*}[t!]
		\center
	\includegraphics[width=17cm]{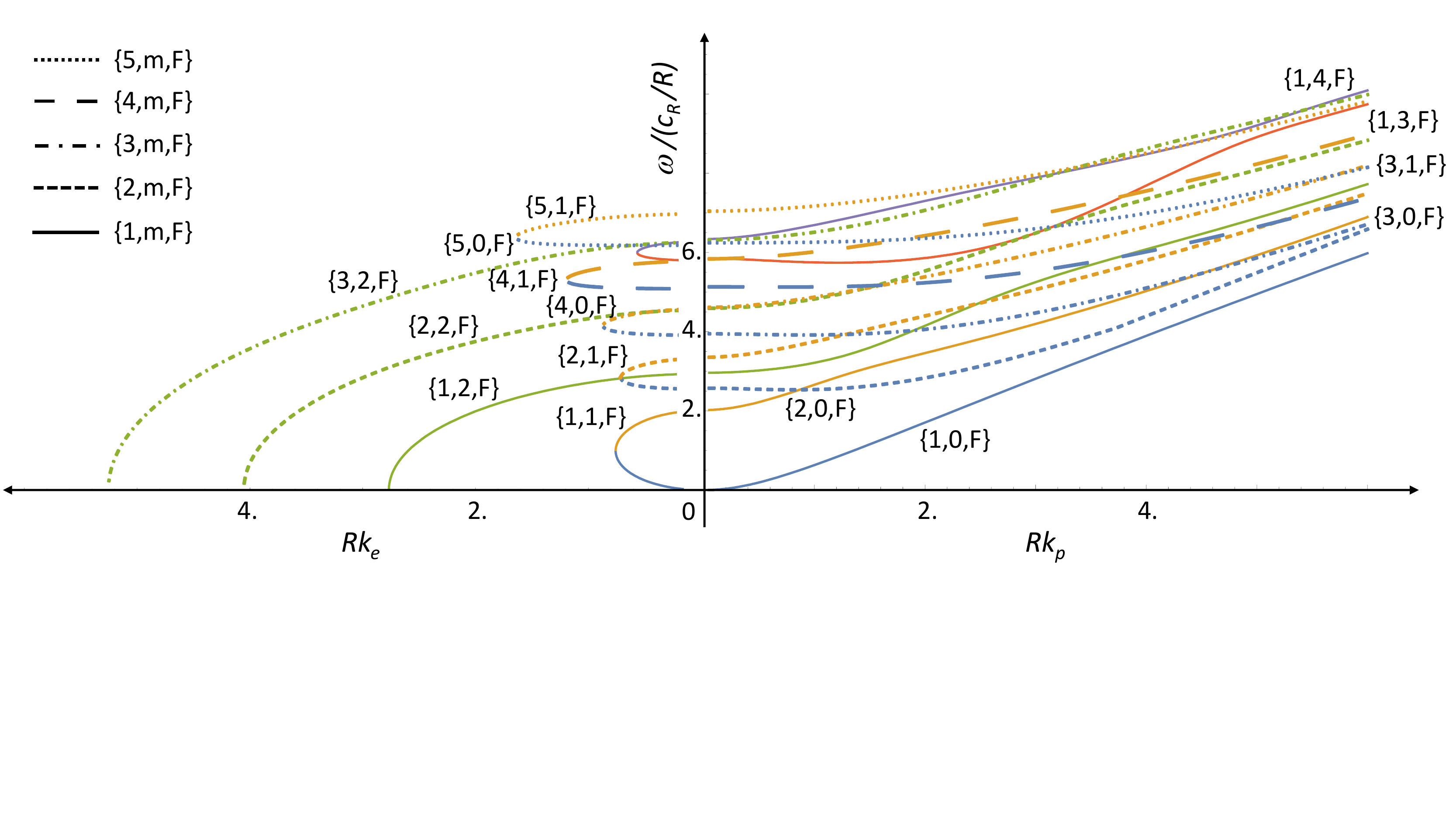}     
	\vspace*{-3cm}
			\caption{\small{Branches obtained for flexural (F) waves, in normalized units ($c_R$ stands for Rayleigh velocity, see text). Left: evanescent solutions. Right: traveling solutions. The numbering $\left\{n,m,\mbox{F}\right\}$ is indicated in the graph (see text for details). Calculation performed for $\nu=0.20$. Note the reconnections between branches on the left panel. The breaks in the lines on the left of the vertical axis are there for readability of the labels. 
}}
			\label{fig_9}
		\end{figure*}

The boundary conditions affect now 3 constants, which can be put in matrix form in a similar fashion to Eq. (\ref{detL}). We write the resulting determinant as Eq. (\ref{detF}) below, similarly to Ref. \cite{poche}. 
The same reasoning defines the dispersion laws $\omega_{\left\{n,m,\mbox{\tiny F} \right\}}(k_e)$ and $\omega_{\left\{n,m,\mbox{\tiny F} \right\}}(k_p)$, which then fix the $\alpha_{\left\{n,m,\mbox{\tiny F}, k_{e,p} \right\}}$ and $\beta_{\left\{n,m,\mbox{\tiny F} , k_{e,p}\right\}}$ constants.
From the matrix equation, we can define the $A_{\left\{\eta\right\}}$ and $B_{\left\{\eta\right\}}$ constants as a function of the last one $C_{\left\{\eta\right\}}$.
The lowest F branches are shown in Fig. \ref{fig_9}

\begin{figure}[t!]		 
			 \includegraphics[width=11cm]{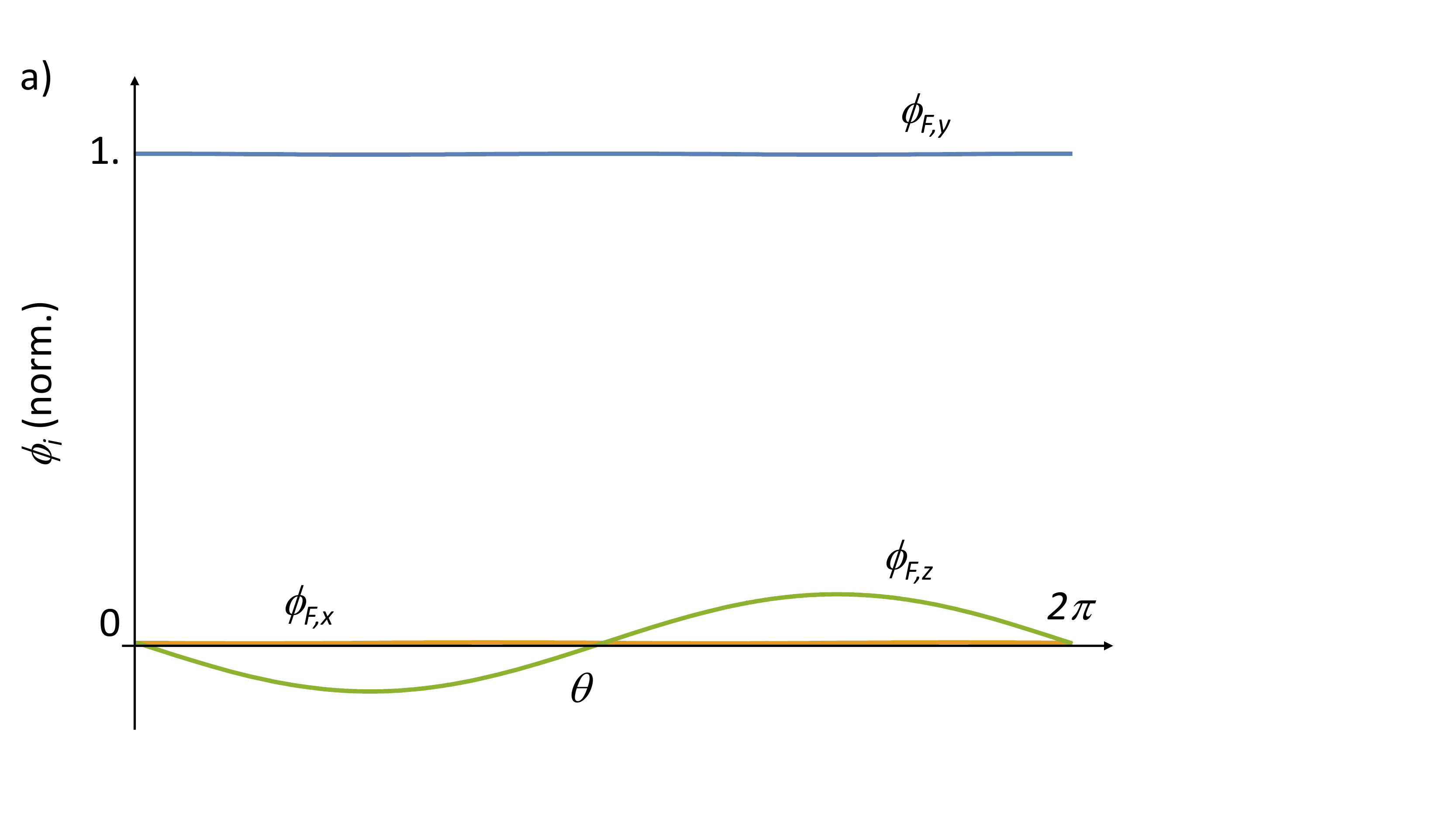}			 			 \includegraphics[width=11cm]{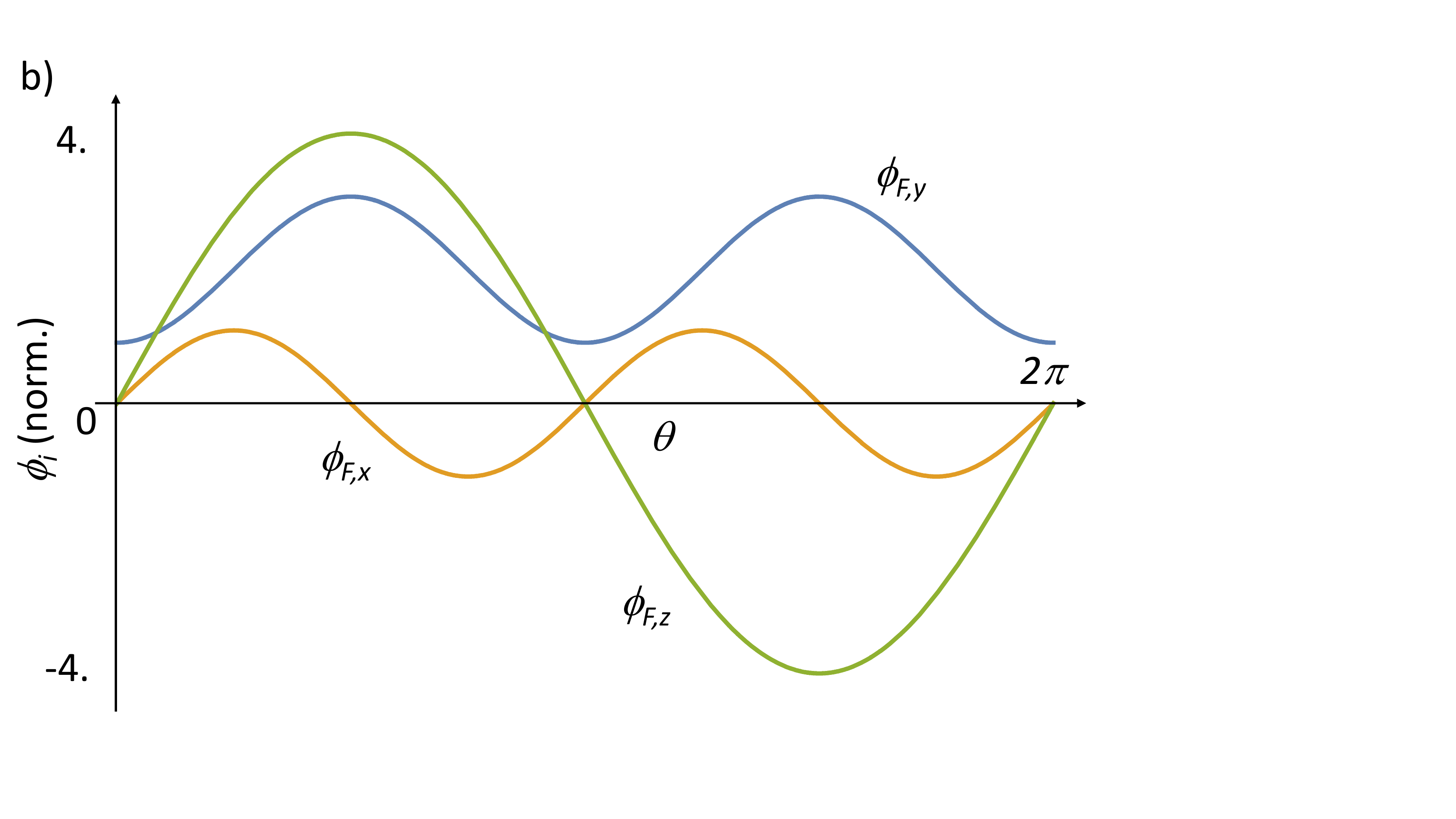}
			\caption{\small{
			Normalized displacement components for flexural (F) wave $n=1$,$m=0$ taken at $r=R$ (computed for $\nu=0.20$). 
			a) Traveling solution with $k_p R = 0.10$. b) Traveling solution with $k_p R = 10$. 
			For small wavevectors, the motion is mostly a bulk translation along $\vec{y}$, matching the name {\it flexural}. For large wavevectors, the motion gets mixed with all $u_x, u_y$ and $u_z$ components, showing as well a $\theta$ dependence. Higher waves display even more complex patterns. 
			}}
			\label{fig_10}
		\end{figure}

Again, the last constant (here, $C_{\left\{\eta\right\}}$) is defined from the normalization procedure.
In order to match the intuitive result that suits the lowest F branch, the amplitude of motion has to be defined from the (planar) displacement $\vec{u}_F=u_r(r=R,\theta,z,t)\vec{e}_r+u_\theta(r=R,\theta,z,t) \vec{e}_\theta$ occurring on the periphery of the rod. This can be projected on the two $\vec{x}, \vec{y}$ axes defining $\phi_{F,x}$ and $\phi_{F,y}$ (from $\phi_r$ and $\phi_\theta$) respectively; we show as an example these components for the $\left\{1,0,\mbox{F}\right\}$ wave in Fig. \ref{fig_10}, calculated for the $\theta_0 = 0$ family.
And indeed for small wavevectors, we see that the planar component of the motion is almost uniform in $\theta$ with $\phi_{F,x} \approx 0$ and $\phi_{F,y}$ constant: this is just the well-known flexure (along $\vec{y}$) of the rod (with nonetheless a small $\phi_{F,z}$ component, see Fig. \ref{fig_10}). Similarly with $\theta_0 = + \pi/2$ one obtains the same type of motion, but along $\vec{x}$. This is what is studied using the simpler Euler-Bernoulli formalism \cite{clelandBk}.

\begin{widetext}

\vspace*{-0.5cm}
\begin{eqnarray}
& & \begin{vmatrix}
a_{11}  & a_{12}  & a_{13}   \\
a_{21}  & a_{22}  & a_{23}   \\
a_{31}  & a_{32}  & a_{33}   \\
\end{vmatrix} = 0 , \label{detF} \\
a_{11} & = & 2 n \, \beta_{\left\{\eta\right\}} \! R \, \left( \alpha_{\left\{\eta\right\}}\! R \, J_{n-1} \left[\alpha_{\left\{\eta\right\}}\!R \right]- \left(n+1\right) J_{n} \left[\alpha_{\left\{\eta\right\}}\!R \right] \right) , \nonumber \\
a_{12} & = &  2 n \, R k_{e,p} \, \left( \beta_{\left\{\eta\right\}}\! R \, J_{n-1} \left[\beta_{\left\{\eta\right\}}\! R \right] - \left( n+1 \right) J_{n} \left[\beta_{\left\{\eta\right\}}\! R \right] \right) , \nonumber \\
a_{13} & = & \beta_{\left\{\eta\right\}} \! R \, J_{n} \left[\beta_{\left\{\eta\right\}}\! R \right] \left(2n [n+1] - R^2 \! \beta_{\left\{\eta\right\}}^2\right) - 2 R^2 \! \beta_{\left\{\eta\right\}}^2 \, J_{n-1} \left[\beta_{\left\{\eta\right\}}\! R \right] , \nonumber \\
a_{21} & = &\beta_{\left\{\eta\right\}} \! R \, \left( J_{n-1} \left[\alpha_{\left\{\eta\right\}}\!R \right] \left(2 n [1-n^2] - 2 \nu (1-n)[2n (1+n) \pm R^2 k_{e,p}^2] + \left(2n [1-\nu] -1 \right) R^2 \! \alpha_{\left\{\eta\right\}}^2 \right) \right. \nonumber \\
& & \left. +  \alpha_{\left\{\eta\right\}} \! R \,\left(n [n+1] -\nu \left[2n (n+1) \pm k_{e,p}^2\right] -(1-\nu) R^2  \alpha_{\left\{\eta\right\}}^2 \right) J_{n-2} \left[\alpha_{\left\{\eta\right\}}\!R \right]\right) , \nonumber \\
a_{22} & = & R^2 k_{e,p} \alpha_{\left\{\eta\right\}} \, \left( 1-2 \nu \right) \left( \beta_{\left\{\eta\right\}} \! R \, J_{n-1} \left[\beta_{\left\{\eta\right\}}\! R \right] - \left( n[n+1] - R^2 \!\beta_{\left\{\eta\right\}}^2 \right) J_{n} \left[\beta_{\left\{\eta\right\}}\! R \right]\right) , \nonumber \\
a_{23} & = & -n \,\alpha_{\left\{\eta\right\}}\! R\,\left( 1-2 \nu \right)\left( \beta_{\left\{\eta\right\}} \! R \, (n+1) \, J_{n-2} \left[\beta_{\left\{\eta\right\}}\! R \right]+\left(2[1-n^2]+ R^2 \! \beta_{\left\{\eta\right\}}^2 \right) J_{n-1} \left[\beta_{\left\{\eta\right\}}\! R \right]\right), \nonumber \\
a_{31} & = & \pm 2 \, R^2 k_{e,p} \alpha_{\left\{\eta\right\}} \, \left( J_{n-1} \left[\alpha_{\left\{\eta\right\}}\!R \right] - J_{n+1} \left[\alpha_{\left\{\eta\right\}}\!R \right]\right) , \nonumber \\
a_{32} & = & R^2\left( \pm k_{e,p}^2 - \beta_{\left\{\eta\right\}}^2 \right) \left( J_{n-1} \left[\beta_{\left\{\eta\right\}}\!R \right] - J_{n+1} \left[\beta_{\left\{\eta\right\}}\!R \right] \right), \nonumber \\
a_{33} & = &   \pm 2 n R k_{e,p} \, J_{n} \left[\beta_{\left\{\eta\right\}}\!R \right] . \nonumber
\end{eqnarray}

\end{widetext}

However for higher branches (or larger $k_p$), the motion is much more complicated; in particular, there is a clear $\theta$-dependent pattern (Fig. \ref{fig_10}).
Therefore, the normalization procedure {\it must} chose a given $\theta=\Theta_F$ angle.
 Again to keep it simple, we take $\Theta_F=0$ for {\it all} $n,m$ (and both evanescent or propagating solutions).
The amplitude definition of $U_{\left\{\eta\right\}}$ is then fixed by constraining $\phi_{r \left\{\eta\right\}}(R,\Theta_F)^2+\phi_{\theta \left\{\eta\right\}}(R,\Theta_F)^2=1$ (or equivalently using the $x,y$ notations).
This is in particular what has been done for Fig. \ref{fig_10}.
Note that similarly to the L waves normalization, $\vec{u}_F$ essentially never vanishes, which makes the procedure well defined for any branch (and almost any $k_{e,p}$, see discussion of previous Subsection).
As a result, $\phi_{\theta \left\{\eta \right\}}(r=R,\theta )= +1 \, \cos \left(n \, \theta \right)$ and $\phi_{r \left\{\eta \right\}}(r=R,\theta ) \propto \sin \left(n \, \theta \right)$ for any $\left\{\eta \right\}$, which actually guarantees the boundary conditions introduced for the construction of standing waves, Subsection \ref{modes}.

For each $n$ subset, the F branches plotted in Fig. \ref{fig_9} display properties similar to the ones of the L branches shown in Fig. \ref{fig_8}. 
We also clearly see on the left panel branches that directly reconnect to $\omega=0$, as in Fig. \ref{fig_7}.
However, noticeably the lowest branch $\left\{1,0,\mbox{F}\right\}$ does support an evanescent part. For small $k_p,k_e \ll 1/R$, these dispersion relations are quadratic, which is what one expects from Euler-Bernoulli (with $\omega_{\left\{1,0,\mbox{\tiny F}\right\}}[k_{e,p}=0]=0$) \cite{clelandBk}.
All $\left\{n,0,\mbox{F}\right\}$ ($n>0$) tend asymptotically at large $k_p$ toward a linear dispersion law with $d \omega/d k_p = c_R$ (the Rayleigh velocity). The other ones tend to the same dependence with $c_t$ velocity \cite{poche}. For a given $n$, the branches never cross.

Finally, we should mention that as for L waves the numbering chosen for the branches 
is again related to the specific nature of the motions, namely:
\begin{itemize}
\item  even $m$ ($0,2,4, \cdots$) are strongly planar,
\item odd $m$ ($1,3,5, \cdots$) are strongly longitudinal.
\end{itemize}
Such labeling issues are usually not discussed in the literature \cite{poche,pao1,pao2}.
As in the previous Subsection, this means that the chosen normalization is not specifically adapted to odd $m$ branches; but we use it for simplicity. 

\subsection {Effect of in-built stress}
\label{inbuiltstress}

Consider that the rod is subject to an in-built axial stress $\sigma_0$ (along $\vec{z}$). 
We assume that the length $L$ of the rod corresponds to the static rest shape under this load.
An arbitrary displacement $\vec{u}(r,\theta,z,t)$ causes then a stretching $\Delta L= \int \! \delta L$, which stores an elastic energy (per unit volume) $ \sigma_0 \, \delta L/\delta z$. The stretched length $L+\Delta L = \int \! d L$ is defined from:
\begin{equation}
d L^2 = \left(dz +  \frac{\partial u_z}{\partial z} dz\right)^2 + \left(\frac{\partial u_r}{\partial z} dz\right)^2 + \left( \frac{\partial u_\theta}{\partial z} dz\right)^2 ,
\end{equation}
which expansion (at lowest order) brings:
\begin{equation}
\frac{\delta L}{ \delta z}  \approx  \! \frac{\partial u_z}{\partial z} + \frac{1}{2} \left[ 
   \left(\frac{\partial u_r}{\partial z}  \right)^2 + \left(\frac{\partial u_\theta}{\partial z}  \right)^2 \right] \! . \label{stretchen}
\end{equation}
The first term in Eq. (\ref{stretchen}) leads to a {\it linear} term in the energetics, that causes a renormalization of the rest position for a given wave (a {\it static} term, see below); only the second one contributes to the total time-dependent potential energy with a {\it quadratic term}.
We write the elastic energy as ${\cal E}_\sigma + {\cal E}_0$:
\begin{eqnarray}
{\cal E}_0 & = & \sigma_0 \left[ \frac{\partial u_z}{\partial z} - \frac{1}{2}  \left(\frac{\partial u_z}{\partial z}  \right)^2  \right], \label{epszero} \\
{\cal E}_\sigma & = & \frac{1}{2} \sigma_0 \left[ \left(\frac{\partial u_z}{\partial z}  \right)^2 +   \left(\frac{\partial u_r}{\partial z}  \right)^2 + \left(\frac{\partial u_\theta}{\partial z}  \right)^2 \right] \! . \label{eqEsigma}
\end{eqnarray}
The truncation in the expansion Eq. (\ref{stretchen}) therefore neglects nonlinear effects, which are outside of the scope of this article.
${\cal E}_\sigma$ is used when defining the stored mechanical energy in the rod, see  Subsection \ref{energy}. The addendum ${\cal E}_0$ is discussed at the end of this Subsection; we shall show that this term can be reasonably neglected. \\

Eq. (\ref{eqEsigma}) leads to a body force $\vec{f}_\sigma$ that adds up to the right-hand-side of Eq. (\ref{newton}).
It reads:
\begin{equation}
\vec{f}_\sigma = \sigma_0 \frac{\partial^2 \vec{u}}{\partial z^2} ,
\end{equation}
and in our definitions $\sigma_0>0$ for tensile stress.
The presence of this force modifies the dynamic equilibrium, but it turns out that the analytic expressions used to describe the Pochhammer-Chree waves without in-built stress are still solutions of this new Eq. (\ref{newton}). 
This is actually the reason behind the splitting of the stretching energy into a term which is analytically solvable, and the addendum ${\cal E}_0$ which should be analysed separately (see below).
However, this comes at the cost of a modification of the dispersion laws Eqs. (\ref{alphaEq},\ref{betaEq}):
\begin{eqnarray}
\alpha_{\left\{\eta\right\}}^2 \pm k_{e,p}^2 \left[  1 + \frac{ (1-2\nu) (1+\nu)}{ (1-\nu)} \frac{\sigma_0}{E_Y}  \right] & = & \frac{\omega_{ \left\{\eta\right\}}^2}{c_l^2}  , \label{alphaEq2} \\
\beta_{\left\{\eta\right\}}^2 \pm k_{e,p}^2  \left[ 1 + 2 (1+\nu) \frac{\sigma_0}{E_Y}  \right] & = & \frac{\omega_{ \left\{\eta\right\}}^2}{c_t^2} ,  \label{betaEq2}
\end{eqnarray}
 introducing two correcting terms within the $k^2$ dependencies which are proportional to $\sigma_0/E_Y$ (a small parameter for realistic physical situations). $\pm$ signs again stand for traveling ($k_{p}$, with $+$) and evanescent ($k_{e}$, with $-$).

		\begin{figure}[t!]		 
			 \includegraphics[width=11cm]{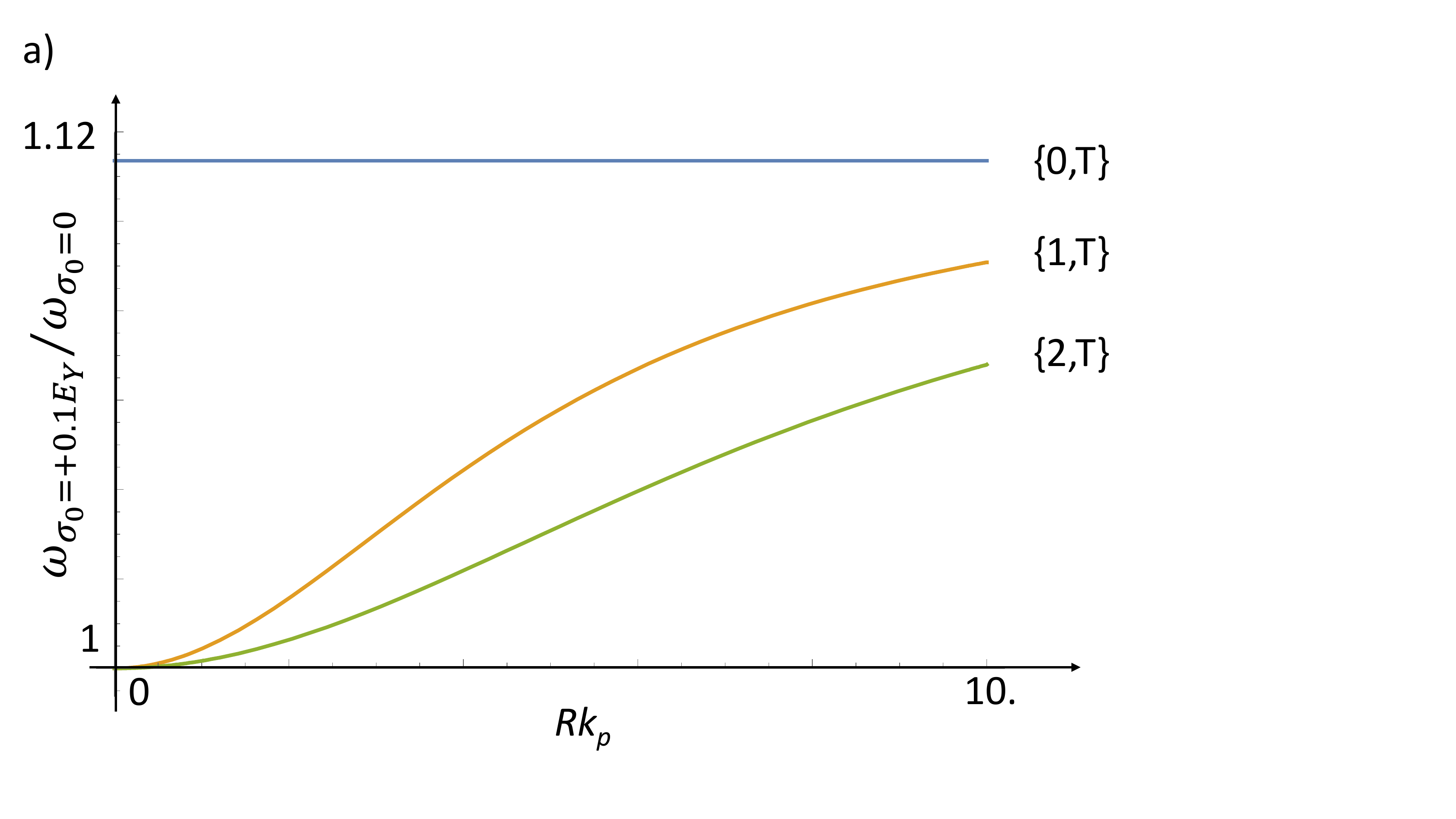}			 			 \includegraphics[width=11cm]{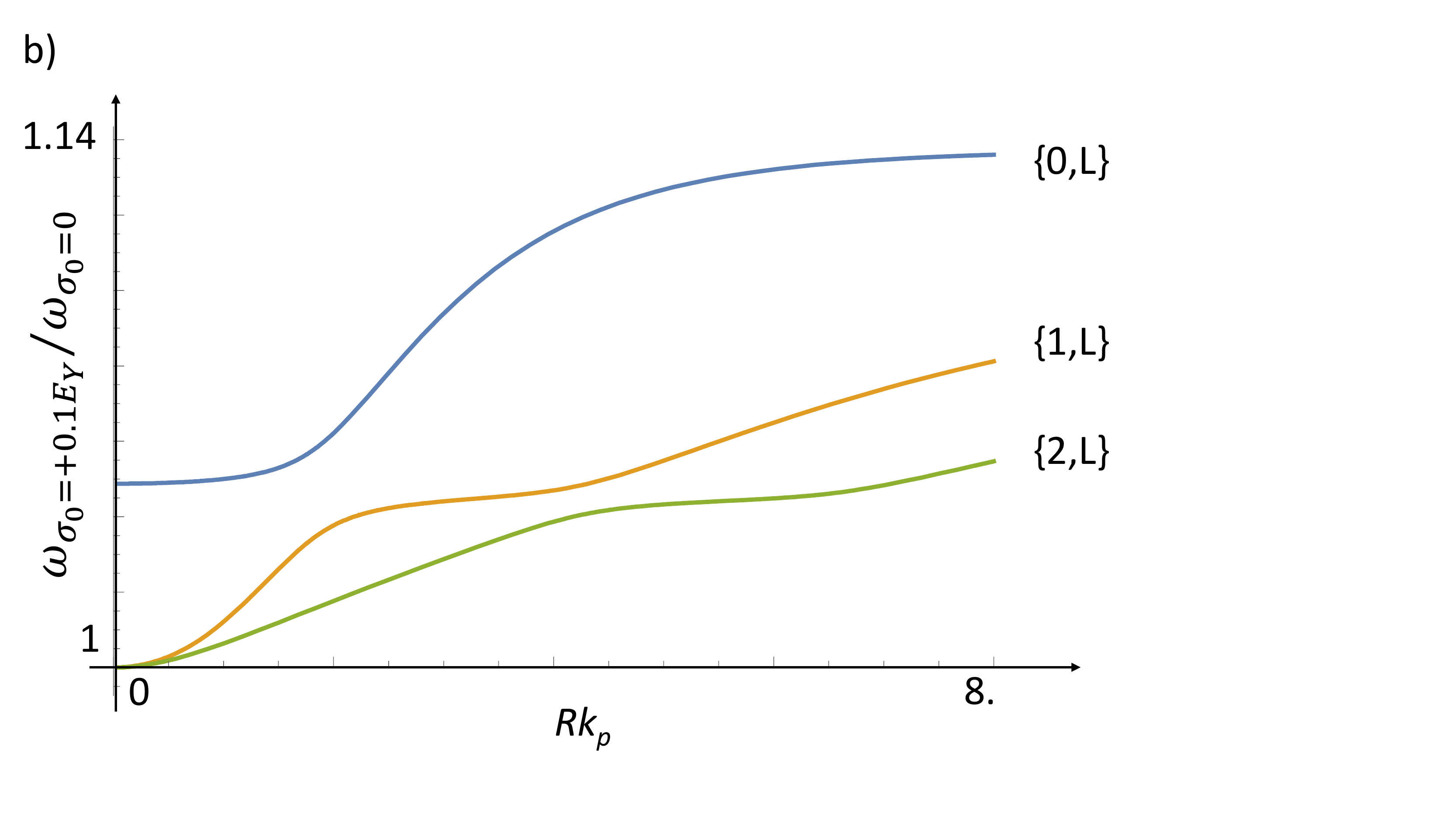} 	 \includegraphics[width=11cm]{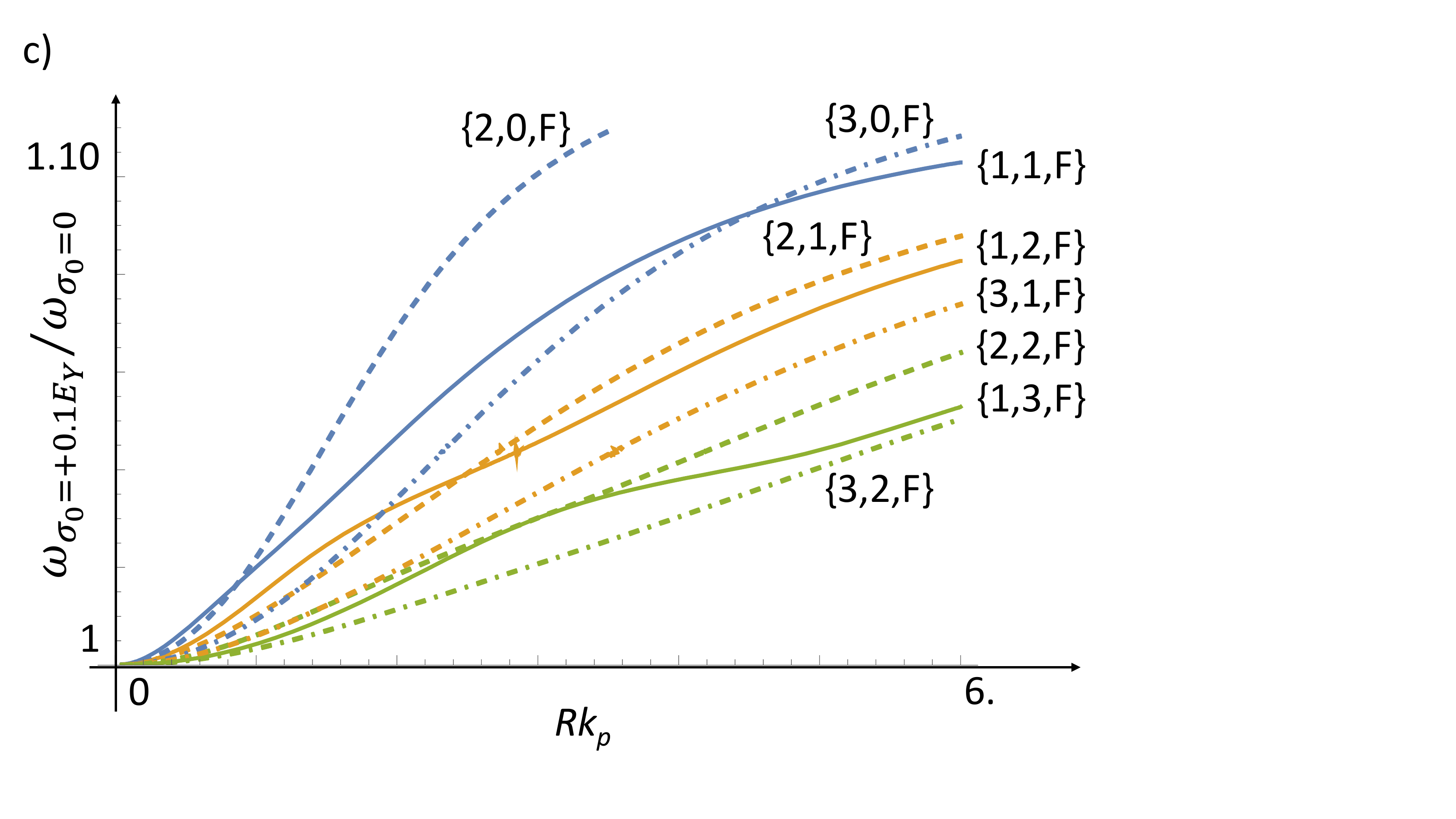}
			\caption{\small{
			Dispersion relations $\omega_{\left\{\eta\right\}}$ of the first traveling waves of each family, computed for $\sigma_0=+0.1 E_Y$, and normalized to the case $\sigma_0=0$. 
			a) Torsional family (analytic solution). 
			b) Longitudinal (numerical calculation).
			c) Flexural family (numerically computed; excluding the first flexure $\left\{1,0,\mbox{F}\right\}$).
			Plot realized with $\nu=0.2$, see text for details.
			}}
			\label{fig_11}
		\end{figure}
		
To our knowledge, the stress correction is not discussed in the literature devoted to waves in rods \cite{poche,pao1,pao2}.
For simplicity, All the illustrating graphs in this Appendix have been realized with $\sigma_0=0$. However, adding stress does not change significantly any of our conclusions, and it is therefore possible (but tedious) to take a finite $\sigma_0 \neq 0$ in practical numerical evaluations.  
For all traveling waves except the first flexural one, the stored axial stress just slightly modifies the dispersion relation $\omega_{\left\{\eta\right\}}$.
This is illustrated in Fig. \ref{fig_11} for the first traveling waves of each family, with a stored tensile stress $\sigma_0=+0.1 E_Y$ (beyond reach of conventional materials). With a compressive load, the graph is essentially mirror-imaged around the horizontal $y=1$ line (frequencies decrease, instead of increase). 
At small wavevector $k_p R \ll 1$, the correction is negligible for $m \neq 0$. At large wavevector $k_p R \gg 1$, the axial stress modifies the sound velocity (but does not affect the linear-in-$k_p$ dependence).
On the evanescent panel, the connecting points of the branches are slightly displaced by the added stress; the dispersion curve is thus slightly above (compressive) or below (tensile) the unstressed solution.

\begin{figure}[t!]
		\centering
	\includegraphics[width=12.1 cm]{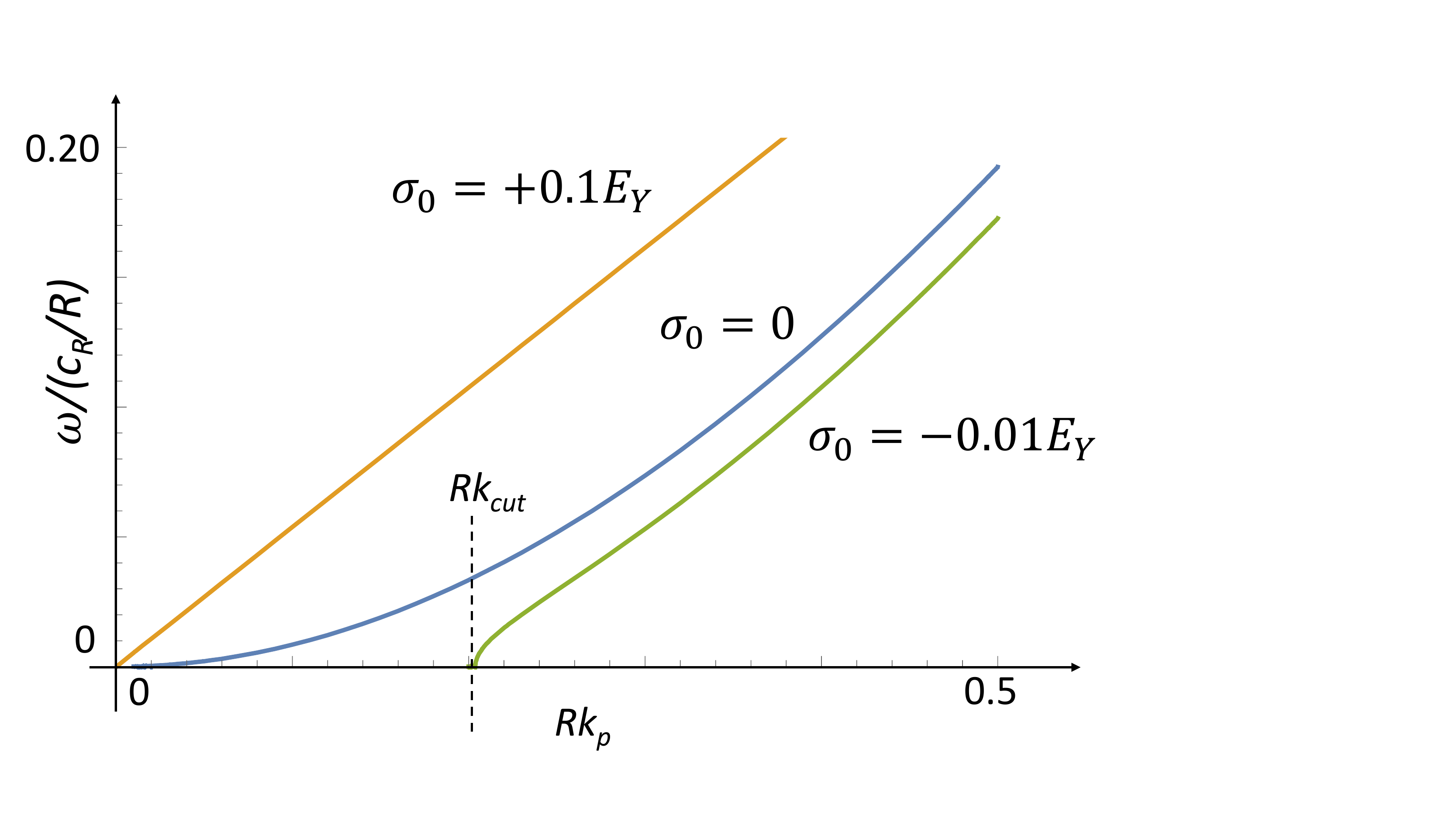}
			\caption{\small{ 
			Lowest traveling flexural solution $\left\{1,0,\mbox{F}\right\}$ dispersion relation calculated for $\sigma_0=0$, $\sigma_0=+0.1 E_Y$ (tensile) and for $\sigma_0=-0.01 E_Y$ (compressive). The law transforms from quadratic in $k_p^2$ into linear in $k_p$, when a large tensile stress is added. On the other hand, a compressive stress prevents small wavevectors to propagate (we used $\nu=0.2$, see text for details). }}
			\label{fig_12}
		\end{figure}

The main impact of the stress is to turn the quadratic dispersion law for the lowest traveling flexural wave $\left\{1,0,\mbox{F}\right\}$ at small wavevector $k_{p}$ into a linear one, for $\sigma_0 > 0$ (tensile). 
Reversely for a compressive load $\sigma_0 < 0$, a cutoff $k_{cut} R \approx 2 \sqrt{\left|\sigma_0\right|/E_Y}$ in $k_{p}$ appears and the smallest wavevectors cannot propagate anymore; this is shown in Fig. \ref{fig_12}. 
For a rod of finite length $L$, identifying  $k_{cut} =2 \pi/L$ produces the first {\it buckling} instability \cite{timoshenko}.
At large $k_p R \gg 1$, one obtains again a small renormalization of the sound velocity, as for the other waves (see Fig. \ref{fig_11}). 
On the evanescent side, a similar graph can be produced while {\it inverting} the role of tensile and compressive loads. Therefore, there is no $\left\{1,0,\mbox{F}\right\}$ evanescent wave available at small wavevectors $k_e$, for large  $ \sigma_0 >0 $. \\

\begin{figure}[t!]
		\centering
	\includegraphics[width=12.1 cm]{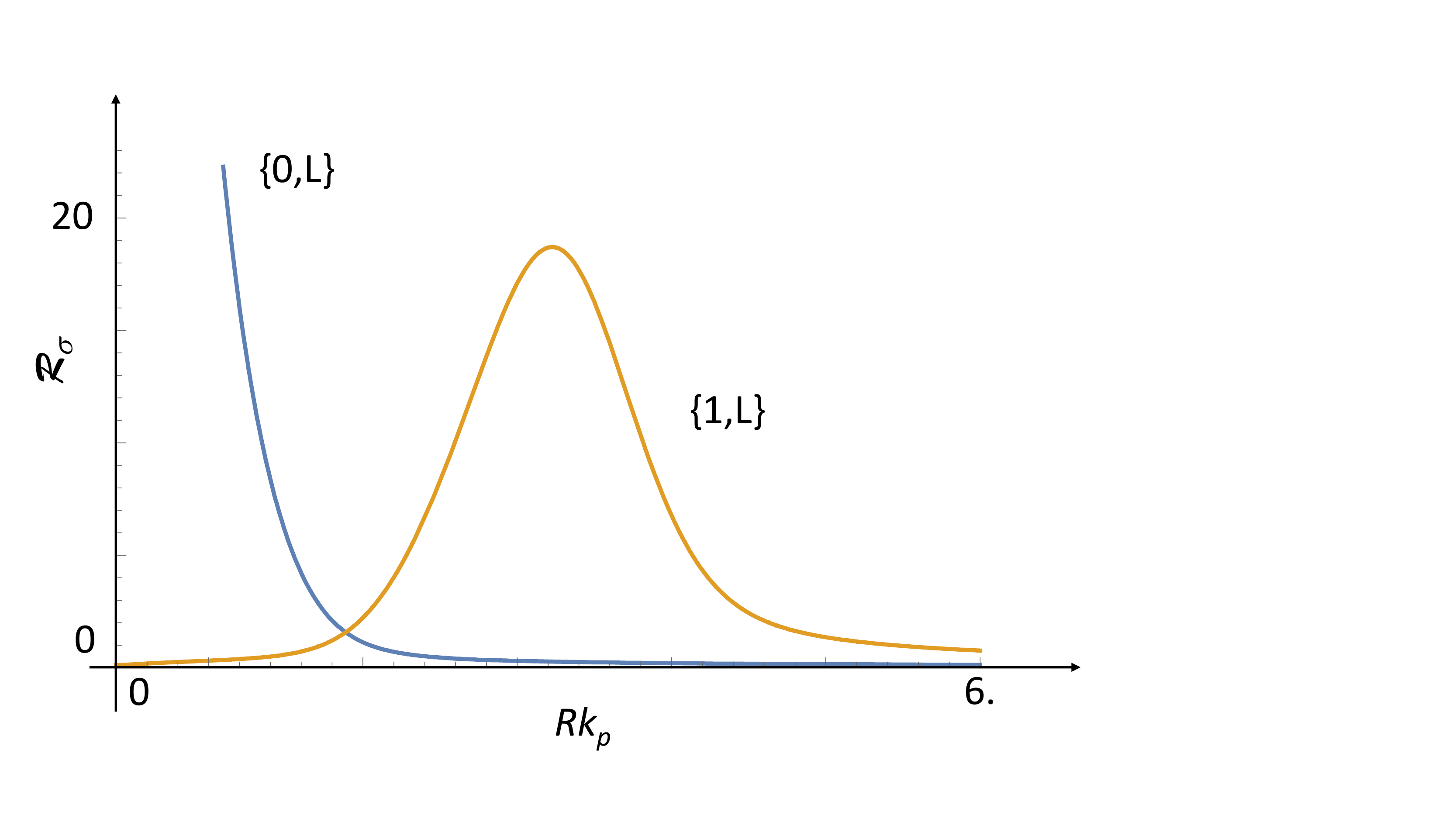}
			\caption{\small{ 
			${\cal R}_{\sigma}$ computed for the two first L waves ($\nu=0.2$, and $\sigma_0=+0.1 E_Y$). For the {\it strongly longitudinal} one ($m=0$), the analytic solution is getting worse as $k_p R \ll 1$; however for the $m=1$ solution, the two limits $k_p R \ll 1$ and $k_p R \gg 1$ verify ${\cal R}_{\sigma} \rightarrow 0$ (see text for discussion).  }}
			\label{fig_13}
		\end{figure}

Let us now come back to the addendum term in the energetics ${\cal E}_0 = \sigma_0 \partial u_z/\partial z - \frac{1}{2} \sigma_0 \left( \partial u_z/\partial z \right)^2$. We first consider the quadratic term, which has been introduced in order to preserve the validity of Pochhammer-Chree analytic solutions.
We define the quantity:
\begin{equation}
{\cal R}_{\sigma} =  \frac{\left( \partial u_z / \partial z \right)^2}{\left( \partial u_r / \partial z \right)^2+\left( \partial u_\theta / \partial z \right)^2} ,
\end{equation}
which quantifies the importance of the $u_z$ term with respect to the ``normal'' stretching contribution involving only $u_r$ and $u_\theta$.
For torsional (T) waves, ${\cal R}_{\sigma}$ is exactly zero: stating ${\cal E}_0=0$ is {\it exact}.
For all L and F waves, this expression tends to 0 for large $k_p R \gg 1$.
However, for $k_p R \ll 1$ (and $k_e R \ll 1$ as well) two situations have to be distinguished:
\begin{itemize}
\item for L waves, ${\cal R}_{\sigma}$ tends to 0 when $m$ is odd; it however grows as $k_p R \ll 1$ for $m$ even,
\item for F waves, ${\cal R}_{\sigma}$ tends to 0 when $m$ is even; it grows for $m$ odd. 
\end{itemize}
This is simply linked to the nature of the branch, as discussed in the previous Subsections: the correction is getting worse for waves which are {\it strongly longitudinal}, as could be expected.
This is illustrated in Fig. \ref{fig_13} with the computed ${\cal R}_{\sigma}$ for traveling solutions $\left\{0,\mbox{L}\right\}$ and $\left\{1,\mbox{L}\right\}$.
The point is that for all $m \neq 0$ waves, the correction at $k_p R \ll 1$ is {\it anyway negligible}; 
the fact that neglecting ${\cal E}_0$ is not always rigorous becomes thus irrelevant.
For the first flexure $\left\{1,0,\mbox{F}\right\}$ (which is $m$ even), when $k_{p,e} R \ll 1$ the modeling discarding ${\cal E}_0$ is correct; and for all waves when $k_p R \gg 1$, neglecting the quadratic term in ${\cal E}_0$ is a good assumption.   
The main limitation of this approach is that the model essentially {\it over-estimates} the effect of stress around $k_p R \sim 1$ for all modes,
and also for the first longitudinal mode $\left\{0,\mbox{L}\right\}$ at small wavevectors [the correction in Fig. \ref{fig_11} b) should physically tend to zero when $R k_p \rightarrow 0$, while it does not]. 
We shall not comment these facts any further, and will simply drop the ${\cal E}_0$ quadratic term. \\

We now come to the first term in the addendum ${\cal E}_0$, which is linear in displacement amplitude $u_z$. Consider a modification of our solution:
\begin{eqnarray}
u_{r \left\{\eta\right\}} & \rightarrow & u_{r \left\{\eta\right\}}  + \Phi_{r \left\{\eta\right\}} \left(r,\theta \right), \label{eq01} \\
u_{\theta \left\{\eta\right\}} & \rightarrow & u_{\theta \left\{\eta\right\}}  + \Phi_{\theta \left\{\eta\right\}} \left(r,\theta \right) , \\
u_{z \left\{\eta\right\}} & \rightarrow & u_{z \left\{\eta\right\}}  + \Phi_{z \left\{\eta\right\}} \left(r,\theta \right) \label{eq03} , 
\end{eqnarray}
where we introduced a $z$-and-$t$ independent contribution $\vec{\Phi}_{\left\{\eta\right\}}$.
As such, it does not impact the kinetic energy density Eq. (\ref{kin}), neither the stretching energy Eq. (\ref{eqEsigma}) [and ${\cal E}_0$] nor the dissipated power Eq. (\ref{dissipbulk}); it only modifies the bending energy ${\cal E}_{E_Y}$.
Injecting a linear superposition of such $\vec{u}_{\left\{\eta\right\}}$ solutions, Eq. (\ref{superpose}), into Eq. (\ref{pot}), one obtains:
\begin{eqnarray}
 {\cal E}_{E_Y}  & \rightarrow & {\cal E}_{E_Y} + {\cal E}_{\mbox{offset}} \left(r,\theta \right) \nonumber \\
\!\!\!\!&\!\!\!\!\!\!\!\!+ & \!\!\!\!\!\!\! \sum_{\left\{\eta\right\}} \left[  \frac{\partial U_{\left\{\eta\right\}}(z,t)}{\partial z} \left({\cal U}_{1 \left\{\eta\right\},\left\{\eta\right\}} + \sum_{\left\{\eta\right\} \neq \left\{\eta'\right\}} {\cal U}_{1 \left\{\eta\right\},\left\{\eta'\right\}} \right) \right. \nonumber \\
\!\!\!\!&\!\!\!\!\!\!\!\!+& \!\!\!\!\!\!\! \left. U_{\left\{\eta\right\}}(z,t) \left( {\cal U}_{0 \left\{\eta\right\},\left\{\eta\right\}} + \sum_{\left\{\eta\right\} \neq \left\{\eta'\right\}} {\cal U}_{0 \left\{\eta\right\},\left\{\eta'\right\}} \right) \right. \nonumber \\ 
\!\!\!\!&\!\!\!\!\!\!\!\!+& \!\!\!\!\!\!\! \left. \frac{ \partial^2  U_{\left\{\eta\right\}}(z,t)}{\partial z^2} \left( {\cal U}_{2 \left\{\eta\right\},\left\{\eta\right\}} + \sum_{\left\{\eta\right\} \neq \left\{\eta'\right\}} {\cal U}_{2 \left\{\eta\right\},\left\{\eta'\right\}} \right) \right] \!\! . \label{offst}
\end{eqnarray}
${\cal E}_{\mbox{offset}}$ depends only on the $\Phi_{r \left\{\eta\right\}}, \Phi_{\theta \left\{\eta\right\}},  \Phi_{z \left\{\eta\right\}}$ functions, while the other coefficients mix them with the $\phi_{r \left\{\eta\right\}}, \phi_{\theta \left\{\eta\right\}}, \phi_{z \left\{\eta\right\}}$ ones.
We spare the reader the explicit expressions of these terms.
Similarly, the expansion of the linear stretching term leads to:
\begin{equation}
\sigma_0 \frac{\partial u_z}{\partial z} = \sum_{\left\{\eta\right\}} \sigma_0  R \, \phi_{z \left\{\eta\right\}}(r,\theta)  \frac{ \partial^2  U_{\left\{\eta\right\}}(z,t)}{\partial z^2} . \label{firstorder}
\end{equation}
Comparing Eq. (\ref{offst}) with Eq. (\ref{firstorder}), we see that they can compensate each other if:
\begin{eqnarray}
{\cal U}_{0 \left\{\eta\right\},\left\{\eta\right\}}(r,\theta) & = & 0, \\
{\cal U}_{1 \left\{\eta\right\},\left\{\eta\right\}}(r,\theta) & = & 0, \\
{\cal U}_{2 \left\{\eta\right\},\left\{\eta\right\}}(r,\theta) & = & - \sigma_0  R \, \phi_{z \left\{\eta\right\}}(r,\theta) ,
\end{eqnarray}
for any $\left\{\eta\right\}$, provided the sums $\sum_{\left\{\eta\right\} \neq \left\{\eta'\right\}}$ vanish.
This set of 3 equations defines the 3 introduced $\vec{\Phi}_{\left\{\eta\right\}}$ functions.
The transformation Eqs. (\ref{eq01}-\ref{eq03}) therefore {\it removes} the linear term of ${\cal E}_0$ from the energy expansion, and replaces it with the energy density ${\cal E}_{\mbox{offset}}$.

For torsional (T) waves (or when we impose the situation $\sigma_0=0$ with L and F waves), one simply has $\vec{\Phi}_{\left\{\eta\right\}}=0$.
Physically, in all other situations the $\vec{\Phi}_{\left\{\eta\right\}} (r,\theta)$ vector corresponds to a new definition of the rest position for wave ${\left\{\eta\right\}}$, which {\it is not} homogeneously zero anymore for all components; it has a specific shape within the $r,\theta$ coordinates in the rod. This new rest position should derive from a (static) body force entered in Eq. (\ref{newton}). Its energetic consequence is the static energy density ${\cal E}_{\mbox{offset}} (r,\theta)$.
Note that the length $L$ (and similarly the radius $R$) has been chosen to be the rest dimension under the axial load $\sigma_0$. This means that the sum of all the static distortions should vanish, per definition: $\sum_{\left\{\eta\right\}} \vec{\Phi}_{\left\{\eta\right\}} = 0$.
The closure relations mentioned above for Eq. (\ref{offst}) with $\sum_{\left\{\eta\right\} \neq \left\{\eta'\right\}} \cdots =0$ should thus be a consequence of this property.
One can define a static stress $[\sigma_{0\,\left\{\eta\right\}\, ij}]$ associated to each mode by injecting the $\vec{\Phi}_{\left\{\eta\right\}}$ solutions into Eqs. (\ref{stress1}-\ref{stress6}).
By definition one has $\sum_{\left\{\eta\right\}} [\sigma_{0\,\left\{\eta\right\}\, ij}] = [\sigma_{0\,ij}]$, which has a single nonzero component $\sigma_{0\,zz}=\sigma_{0}$.
The explicit calculation of these terms and the demonstration of these rules are outside of the scope of the manuscript. 
It suffice to say for our purpose that the linear term cannot couple different waves, and it does not affect the $t$-dependent energetics.
In the core of the paper, the linear contribution of ${\cal E}_{0}$ (and the related $\vec{\Phi}_{\left\{\eta\right\}}$ functions) is thus neglected. 

\section{Modal coefficients, in-equilibrium}
\label{coefs}

In this Appendix we present the coefficients entering the energetics expansions 
obtained for ``string-like'' solutions. The derivation of the corresponding ``beam-like'' ones is much more complex, and shall not be addressed here.
We write:
\begin{eqnarray}
{\cal M}_{\left\{\eta\right\}} & = & \rho R^2 \left( a_{0\,\left\{\eta\right\}} + a_{1\,\left\{\eta\right\}} \left[k_p R \right]^2 \right), \\
{\cal S}_{\left\{\eta\right\}} & = & \sigma_0 \left( s_{1\,\left\{\eta\right\}} \left[k_p R \right]^2 + s_{2\,\left\{\eta\right\}} \left[k_p R \right]^4 \right), \\
{\cal B}_{\left\{\eta\right\}} & = & E_Y \left( p_{0\,\left\{\eta\right\}} + p_{1\,\left\{\eta\right\}} \left[k_p R \right]^2 \right. \nonumber \\
&+& \left.  p_{2\,\left\{\eta\right\}} \left[k_p R \right]^4 -2 p_{c\,\left\{\eta\right\}} \left[k_p R \right]^2 \right),
\end{eqnarray}
for the mass and spring constant per unit length coefficients, for torsional (T) and flexural (F) solutions.
The corresponding ``deviation terms'' introduced in Section \ref{ineq} are:
\begin{eqnarray}
\Delta {\cal M}_{\left\{\eta\right\}} & = & +\rho R^2\,  a_{1\,\left\{\eta\right\}} \left(k_p R \right)^2 , \\
\Delta {\cal S}_{\left\{\eta\right\}} & = & -\sigma_0\, s_{2\,\left\{\eta\right\}} \left(k_p R \right)^4  , \\
\Delta {\cal B}_{\left\{\eta\right\}} & = & -E_Y  \left( p_{0\,\left\{\eta\right\}}  \right. \nonumber \\
&+& \left.  p_{2\,\left\{\eta\right\}} \left[k_p R \right]^4 -2 p_{c\,\left\{\eta\right\}} \left[k_p R \right]^2 \right),
\end{eqnarray}
again written in the case of T and F modes. For Longitudinal (L) solutions, all the right-hand-sides should be divided by $( k_p R )^2$ because of our choice of amplitude definition, see Appendix \ref{pochhammer}.
The dimensionless coefficients appearing in the above expressions are explicitly listed in Tab. \ref{tabcoef}. They are defined from the $\phi_{i \left\{\eta\right\}}$ functions ($i=r,\theta,z$), which means that in general they also depend on $k_p$ (which is included in the mode index $\left\{\eta\right\}$) and the Poisson ratio $\nu$. Despite the similarities (see e.g. between some $a_{i\,\left\{\eta\right\}}$ and $s_{i\,\left\{\eta\right\}}$), we kept distinct notations for all terms in order to make the approach as clear as possible. From the (numerical) knowledge of the mode functions, all coefficients can be calculated (see thereafter).

Integrating the kinetic and potential energies over the length $L$ of the rod, we define respectively the effective mass ${\cal M}_{\left\{\eta\right\}} L/2$ and the effective spring constant ${\cal K}_{\left\{\eta\right\}}L/2$ that describe the dynamics of $U_{\omega_{\left\{\eta\right\}}}$ (see Section \ref{FDT}).
By construction, they verify $\omega_{\left\{\eta\right\}}^2 = (\frac{1}{2}{\cal K}_{\left\{\eta\right\}}L)/(\frac{1}{2}{\cal M}_{\left\{\eta\right\}}L)$.
For ``string-like'' solutions, $k_p = (\pi /L) \, q$ with $q >0$ an integer. On a given branch $i$, we see that this index runs up to about $L/a$, with $a$ the lattice spacing. This defines roughly the cutoffs $N_{cut\,i}$, which have to fulfill our ``Debye'' assumption. For two nearby modes separated by $\Delta k_p = (\pi /L) \, \Delta q$ (with $\Delta q \neq 0$ integer),
we have $\Delta k_p  R \ll 1$: at large $k_p$ on all branches, or for small $k_p$ on high-order branches (which are essentially flat near $k_p=0$), the mode parameters are almost equal for  $k_p$ and $k_p+ \Delta k_p$ wavelengths. This is especially true for the frequencies $\omega_{\left\{\eta\right\}}$.

\begin{widetext}

\begin{table}[h!]
\begin{center}
\hspace*{-1.3cm}
\begin{tabular}{|c|c|c|c|}    \hline
coefs.    & $a_{i\,\left\{\eta\right\}}$ &  $s_{i\,\left\{\eta\right\}}$  & $p_{i\,\left\{\eta\right\}}$      \\    \hline    \hline
 \shortstack{ $ $ \\ $i=0$ \\ $ $ }   & \shortstack{ $ $ \\ $\frac{1}{R^2}\int \!\!\! \int \left[ \phi_{r \left\{\eta\right\}}^2(r,\theta)+\phi_{\theta \left\{\eta\right\}}^2(r,\theta) \right] r dr d\theta $  \\ $ $ }   &  \shortstack{ $ $ \\  /   \\ $ $ }    & \shortstack{ $\int \!\!\! \int \frac{1 }{(1+\nu)}\left[\frac{1}{2} \left(\frac{\partial \phi_{\theta \left\{\eta\right\}}}{\partial r }(r,\theta)\right)^2 + \frac{(1-\nu)}{(1-2\nu)}\left(\frac{\partial \phi_{r \left\{\eta\right\}}}{\partial r }(r,\theta)\right)^2 \right.$ \\  $\left. + \frac{1}{r} \! \left( \frac{\partial \phi_{r \left\{\eta\right\}}}{\partial \theta }(r,\theta) - \phi_{\theta \left\{\eta\right\}}(r,\theta) \right) \! \frac{\partial \phi_{\theta \left\{\eta\right\}}}{\partial r }(r,\theta) +\frac{2\nu}{(1-2\nu) \, r} \! \left( \frac{\partial \phi_{\theta \left\{\eta\right\}}}{\partial \theta }(r,\theta) + \phi_{r \left\{\eta\right\}}(r,\theta) \right) \! \frac{\partial \phi_{r \left\{\eta\right\}}}{\partial r }(r,\theta) \right.$ \\ $\left. + \frac{1}{2 r^2} \! \left( \frac{\partial \phi_{r \left\{\eta\right\}}}{\partial \theta }(r,\theta) - \phi_{\theta \left\{\eta\right\}}(r,\theta) \right)^2 +\frac{(1-\nu)}{(1-2\nu)\, r^2} \! \left( \frac{\partial \phi_{\theta \left\{\eta\right\}}}{\partial \theta }(r,\theta) + \phi_{r \left\{\eta\right\}}(r,\theta) \right)^2  \right] r dr d\theta$  }                      \\    \hline
 \shortstack{ $ $ \\$i=1$ \\ $ $ }   & \shortstack{ $ $ \\  $\frac{1}{R^2} \int \!\!\! \int \left[ \phi_{z \left\{\eta\right\}}^2(r,\theta) \right] r dr d\theta $ \\ $ $ }    &  \shortstack{ $ $ \\ $a_{0\,\left\{\eta\right\}}$ \\ $ $ }  &  \shortstack{ $ \frac{1}{R^2}\int \!\!\! \int \frac{1 }{2(1+\nu)} \left[ \phi_{r \left\{\eta\right\}}^2(r,\theta)+\phi_{\theta \left\{\eta\right\}}^2(r,\theta) \right. $  \\  $+ \left. 2 \phi_{r \left\{\eta\right\}}(r,\theta)  R \frac{\partial \phi_{z \left\{\eta\right\}}}{\partial r }(r,\theta) +2 \phi_{\theta \left\{\eta\right\}}(r,\theta) \frac{R}{r}  \frac{\partial \phi_{z \left\{\eta\right\}}}{\partial \theta } (r,\theta)\right.$ \\ $\left.  +  \left(R \frac{\partial \phi_{z \left\{\eta\right\}}}{\partial r }(r,\theta)\right)^2 + \left(  \frac{R}{r}  \frac{\partial \phi_{z \left\{\eta\right\}}}{\partial \theta }(r,\theta)\right)^2   \right]    r dr d\theta $   }                 \\    \hline
 $i=2$    &  /          &  $a_{1\,\left\{\eta\right\}}$   &  $+\frac{(1-\nu) }{(1-2\nu)(1+\nu)} a_{1\,\left\{\eta\right\}}$                       \\    \hline
 $i=c$    &  /          &  /       &  $\frac{1}{R^2}\int \!\!\! \int \frac{+\nu }{(1-2\nu)(1+\nu)} \phi_{z \left\{\eta\right\}}(r,\theta) \left[ R \frac{\partial \phi_{r \left\{\eta\right\}}}{\partial r}(r,\theta) +\frac{R}{r} \left( \phi_{r \left\{\eta\right\}} (r,\theta)+ \frac{\partial \phi_{\theta \left\{\eta\right\}}}{\partial \theta}(r,\theta) \right) \right] r dr d\theta$     \\    \hline
\end{tabular}
\caption{ \label{tabcoef} Mass and spring (normalized) constants as a function of the modal functions $\phi_{i \left\{\eta\right\}}$. The $l_{i\,\left\{\eta\right\}}$ are obtained from the $p_{i\,\left\{\eta\right\}}$ with the replacement $\nu \rightarrow \nu_\lambda$ in the above expressions. For details, see text. }
\end{center}
\end{table}

\end{widetext}

The internal friction per unit length coefficient is given by:
\begin{eqnarray}
{\cal L}_{\left\{\eta\right\}} & = & \frac{\Lambda_{Y\,\left\{\eta\right\}}}{2 }  \left( l_{0\,\left\{\eta\right\}} + l_{1\,\left\{\eta\right\}} \left[k_p R \right]^2 \right. \nonumber \\
&+& \left.  l_{2\,\left\{\eta\right\}} \left[k_p R \right]^4 -2 l_{c\,\left\{\eta\right\}} \left[k_p R \right]^2 \right),
\end{eqnarray}
and similarly: 
\begin{eqnarray}
\Delta {\cal L}_{\left\{\eta\right\}} & = & -\frac{\Lambda_{Y\,\left\{\eta\right\}}}{2}  \left( l_{0\,\left\{\eta\right\}}  \right. \nonumber \\
&+& \left.  l_{2\,\left\{\eta\right\}} \left[k_p R \right]^4 -2 l_{c\,\left\{\eta\right\}} \left[k_p R \right]^2 \right),
\end{eqnarray}
for torsional (T) and flexural (F) solutions. As above, the longitudinal (L) case is obtained by dividing the right-hand-sides by $( k_p R )^2$.
From the choice we made in the parametrization of the internal dissipation mechanism, the $l_{i\,\left\{\eta\right\}}$ are identical to the $p_{i\,\left\{\eta\right\}}$ given in Tab. \ref{tabcoef} ($i=0,1,2,c$), provided one applies the modification $\nu \rightarrow \nu_\lambda$.

The last coefficients introduced in Section \ref{ineq} concern the clamping losses.
Because of the specificity of the stress profile at the anchoring point for each mode, each situation is somehow different. We shall discuss here only the ``string-like'' solutions, which verify:
\begin{eqnarray}
\frac{\partial U_{\left\{\eta\right\}}}{\partial z}(0,t) & = & k_p \, U_{\omega_{\left\{\eta\right\}}} (t)  , \\
\frac{\partial U_{\left\{\eta\right\}}}{\partial z}(L,t) & = & (-1)^q \, k_p \, U_{\omega_{\left\{\eta\right\}}} (t)  , \\
\frac{\partial^2 U_{\left\{\eta\right\}}}{\partial z^2}(0,t) & = & 0, \\
\frac{\partial^2 U_{\left\{\eta\right\}}}{\partial z^2}(L,t) & = & 0, 
\end{eqnarray}
for torsional (T) and flexural (F) modes, and:
\begin{eqnarray}
\tilde{U}_{\left\{\eta\right\}}(0,t)  & = & -R \frac{ U_{\omega_{\left\{\eta\right\}}} (t)}{k_p R}, 
\end{eqnarray}
\begin{eqnarray}
\tilde{U}_{\left\{\eta\right\}}(L,t)  & = & -(-1)^q \,R \frac{ U_{\omega_{\left\{\eta\right\}}} (t)}{k_p R}, \\
\frac{\partial U_{\left\{\eta\right\}}}{\partial z}(0,t) & = &k_p \, U_{\omega_{\left\{\eta\right\}}} (t), \\
\frac{\partial U_{\left\{\eta\right\}}}{\partial z}(L,t) & = &-(-1)^q \,k_p \, U_{\omega_{\left\{\eta\right\}}} (t),
\end{eqnarray}
for longitudinal (L) ones [following the replacement procedure of Eqs. (\ref{zeroFirst}-\ref{bound6})]. $q$ is the mode number related to the wavevector $k_p = q \, \pi/L$ that was introduced for ``string-like'' modes.
Replacing these within the power density expressions, and integrating over the clamp volumes leads at first order in $\varepsilon$ to:
\begin{eqnarray}
\int_{\substack{\mbox{clamps} \\ 0<z<\varepsilon \\ L<z<L+\varepsilon}}  \int_0^{2 \pi} \!\!\!  \int_0^R {\cal \dot{D}}_{clamp \, \left\{\eta\right\},\left\{\eta \right\} } \, r dr d\theta dz & = & \nonumber \\
&&\!\!\!\!\!\!\!\!\!\!\!\!\!\!\!\!\!\!\!\!\!  \!\!\!\!\!\!\!\!\!\!\!\!\!\!\!\!\!\!\!\!\!  \!\!\!\!\!\!\!\!\!\!\!\!\!\!\!\!\!\!\!\!\! \!\!\!\!\!\!\!\!\!\!\!\!\!\!\!\!\!\!\!\!\!  \left( \frac{1 + (-1)^{2q}}{2} \right) {\cal C}_{\left\{\eta\right\}} \varepsilon \left[ \omega_{\left\{\eta\right\}} \, U_{\omega_{\left\{\eta\right\}}}(t) \right]^2 \! , 
\end{eqnarray}
with:
\begin{eqnarray}
{\cal C}_{\left\{\eta\right\}} & = &  \frac{2\, E_Y^2}{\Lambda_{c\,\left\{\eta\right\}}}\frac{ (k_p R)^2}{\omega_{\left\{\eta\right\}}^2} c_{1\,\left\{\eta\right\}} , \label{clampTF}
\end{eqnarray}
for T and F modes, and:
\begin{eqnarray}
{\cal C}_{\left\{\eta\right\}} & = & \frac{2\, E_Y^2}{\Lambda_{c\,\left\{\eta\right\}}}  \frac{ (k_p R)^2}{\omega_{\left\{\eta\right\}}^2} \left( \frac{ c_{0\,\left\{\eta\right\}} }{\left( k_p R \right)^4} \right. \nonumber \\
%
%
%
&+& \left. c_{2\,\left\{\eta\right\}} -2 \frac{ c_{c\,\left\{\eta\right\}} }{\left( k_p R \right)^2} \right),
\end{eqnarray}
for L ones, where we deliberately kept as prefactor a term (equal to 1) reminding that the two ends are included in the calculation.
The $c_{i\,\left\{\eta\right\}}$ coefficients ($i=0,1,2,c$) depend on both $\nu$ and $\nu_c$; they are defined from the $\phi_{i \left\{\eta\right\}}$ functions ($i=r,\theta,z$), but contain also the first order expansion correction describing the clamping strain profile [functions $\varphi_{i \left\{\eta\right\}}^{(1)}(r,\theta)$ introduced in Eq. (\ref{phiexpand}), Subsection \ref{clamps}].
They are listed in Tab. \ref{tabcoef2}. As for the other terms of Tab. \ref{tabcoef}, they depend on $k_p$ through the index $q$ (which is part of the mode label $\left\{\eta\right\}$).
Note that the idealized clamping conditions chosen in Subsection \ref{modes} for our description remain rather generic: the actual strain profile within the anchors is by construction taken into account by the $\varphi_{i \left\{\eta\right\}}^{(n)}$ functions, $n>0$. How to calculate them is another matter, outside of the scope of this manuscript.
Another simplification made was to consider the damping parameters $\nu_\lambda, \nu_c$ independent of frequency. 
While there is no fundamental reason to impose this, it simplifies the writing of our coefficients in Tabs. \ref{tabcoef},\ref{tabcoef2}. Pragmatically, all frequency dependencies can be incorporated into the $\Lambda_{Y\,\left\{\eta\right\}},\Lambda_{c\,\left\{\eta\right\}}$ parameters without losing much generality.
Technically, the {\it perfect clamping}  is obtained from a geometrical argument $\varepsilon \rightarrow 0$ (vanishing clamp zone), while the material-and-geometry-dependent $\Lambda_{c\,\left\{\eta\right\}}$ shall remain finite (characteristic of the way rod-waves radiate into the bulk support; a complex problem outside of our scope). The actual meaning of the clamp modeling through $\varepsilon$ is further discussed in Section \ref{FDT}. 

\begin{widetext}

\begin{table}[h!]
\begin{center}
\begin{tabular}{|c|c|}    \hline
coefs.    & $c_{i\,\left\{\eta\right\}}$  \\    \hline    \hline
 $i=0$    &     \shortstack{ $\int \!\!\! \int \frac{1 }{(1+\nu)^2}\left[\frac{(1+\nu_c)}{2} \left(\frac{\partial \phi_{\theta \left\{\eta\right\}}}{\partial r }(r,\theta)\right)^2 + \frac{(1-2\nu+3 \nu^2)-2\nu(2-\nu)\,\nu_c}{(1-2\nu)^2}\left(\frac{\partial \phi_{r \left\{\eta\right\}}}{\partial r }(r,\theta)\right)^2 \right.$ \\  $\left. + \frac{(1+\nu_c)}{r} \! \left( \frac{\partial \phi_{r \left\{\eta\right\}}}{\partial \theta }(r,\theta) - \phi_{\theta \left\{\eta\right\}}(r,\theta) \right) \! \frac{\partial \phi_{\theta \left\{\eta\right\}}}{\partial r }(r,\theta) +\frac{2\nu(2-\nu)-2(1+2\nu^2)\, \nu_c}{(1-2\nu)^2 \, r} \! \left( \frac{\partial \phi_{\theta \left\{\eta\right\}}}{\partial \theta }(r,\theta) + \phi_{r \left\{\eta\right\}}(r,\theta) \right) \! \frac{\partial \phi_{r \left\{\eta\right\}}}{\partial r }(r,\theta) \right.$ \\ $\left. + \frac{(1+\nu_c)}{2 r^2} \! \left( \frac{\partial \phi_{r \left\{\eta\right\}}}{\partial \theta }(r,\theta) - \phi_{\theta \left\{\eta\right\}}(r,\theta) \right)^2 +\frac{(1-2\nu+3\nu^2)-2\nu(2-\nu)\,\nu_c}{(1-2\nu)^2\, r^2} \! \left( \frac{\partial \phi_{\theta \left\{\eta\right\}}}{\partial \theta }(r,\theta) + \phi_{r \left\{\eta\right\}}(r,\theta) \right)^2  \right.$ \\
$ \left. + \frac{(1+\nu_c)}{2} \! \left(\varphi_{r \left\{\eta\right\}}^{(1)}(r,\theta) \right)^2 +\frac{(1+\nu_c)}{2} \! \left( \varphi_{\theta \left\{\eta\right\}}^{(1)}(r,\theta)  \right)^2 \right] r dr d\theta$ }  
                        \\    \hline
 $i=1$    &      \shortstack{ $ \frac{1}{R^2}\int \!\!\! \int \frac{(1+\nu_c) }{2(1+\nu)^2} \left[ \phi_{r \left\{\eta\right\}}^2(r,\theta)+\phi_{\theta \left\{\eta\right\}}^2(r,\theta) \right. $  \\  $+ \left. 2 \phi_{r \left\{\eta\right\}}(r,\theta)  R \frac{\partial \phi_{z \left\{\eta\right\}}}{\partial r }(r,\theta) +2 \phi_{\theta \left\{\eta\right\}}(r,\theta) \frac{R}{r}  \frac{\partial \phi_{z \left\{\eta\right\}}}{\partial \theta } (r,\theta)\right.$ \\ $\left.  +  \left(R \frac{\partial \phi_{z \left\{\eta\right\}}}{\partial r }(r,\theta)\right)^2 + \left(  \frac{R}{r}  \frac{\partial \phi_{z \left\{\eta\right\}}}{\partial \theta }(r,\theta)\right)^2 \right. $ \\ $\left. + 2\frac{ ( 1-2\nu+3\nu^2)-2\nu ( 2-\nu ) \, \nu_c  }{(1+\nu_c)(1-2\nu)^2} \left( \varphi_{z \left\{\eta\right\}}^{(1)}(r,\theta) \right)^2 \right]    r dr d\theta $   }                             \\    \hline
 $i=2$    &     $+\frac{(1 - 2\nu + 3 \nu^2)  -2  \nu (2 - \nu) \, \nu_c  }{(1-2\nu)^2(1+\nu)^2} a_{1\,\left\{\eta\right\}}   $                    \\    \hline
 $i=c$    &   $\frac{1}{R^2}\int \!\!\! \int \frac{\nu (2-\nu) -(2 \nu^2+1) \, \nu_c }{(1-2\nu)^2(1+\nu)^2} \phi_{z \left\{\eta\right\}}(r,\theta) \left[ R \frac{\partial \phi_{r \left\{\eta\right\}}}{\partial r}(r,\theta) +\frac{R}{r} \left( \phi_{r \left\{\eta\right\}} (r,\theta)+ \frac{\partial \phi_{\theta \left\{\eta\right\}}}{\partial \theta}(r,\theta) \right) \right] r dr d\theta$                        \\    \hline
\end{tabular}
\caption{ \label{tabcoef2} Clamp dissipation (normalized) constants as a function of the modal functions $\phi_{i \left\{\eta\right\}}$ (and $\varphi_{i \left\{\eta\right\}}^{(1)}$, the first order correction within the clamping zone). They depend on {\it both} $\nu$ and $\nu_c$, by construction; for details, see text. }
\end{center}
\end{table}

\end{widetext}

To conclude this appendix, we shall give some straightforward examples illustrating Tabs. \ref{tabcoef} and \ref{tabcoef2}.
The case of torsional (T) modes $\left\{m,\mbox{T},k_p\right\}$ is rather enlightening, and fully analytic.
The mass and spring constant coefficients reduce to:
\begin{eqnarray}
a_{0\,\left\{\eta\right\}} & = & \Omega(m)\, \frac{\pi}{2} , \\
a_{1\,\left\{\eta\right\}} & = & 0, 
\end{eqnarray}
and:
\begin{eqnarray}
p_{0\,\left\{\eta\right\}} & = & \frac{\pi}{2 (1+\nu)} \left( \beta_{\left\{\eta\right\}} R\right)^2 , \label{p1} \\
p_{1\,\left\{\eta\right\}} & = & \Omega(m)\,\frac{\pi}{4 (1+\nu)} , \\
p_{2\,\left\{\eta\right\}} & = & 0, \\
p_{c\,\left\{\eta\right\}} & = & 0, \label{p4}
\end{eqnarray}
having defined $\Omega(m=0)=1$ and $\Omega(m \neq 0)=2$.
Consequently, $s_{1\,\left\{\eta\right\}} = a_{0\,\left\{\eta\right\}}$ and $s_{2\,\left\{\eta\right\}} = 0$.
Neglecting the correction $\varphi_{z \left\{\eta\right\}}^{(1)}$, the clamp dissipation constant writes:
\begin{equation}
c_{1\,\left\{\eta\right\}} = \Omega(m)\, \frac{\pi \, (1+\nu_c)}{4 (1+\nu)^2},
\end{equation}
which is the only one required for T modes [see Eq. (\ref{clampTF})]. The internal damping terms $l_{i\,\left\{\eta\right\}}$ ($i=0,1,2,c$) are obtained from Eqs. (\ref{p1}-\ref{p4}) with the replacement $\nu \rightarrow \nu_\lambda$. \\

		\begin{figure}[t!]		 
			 \includegraphics[width=11cm]{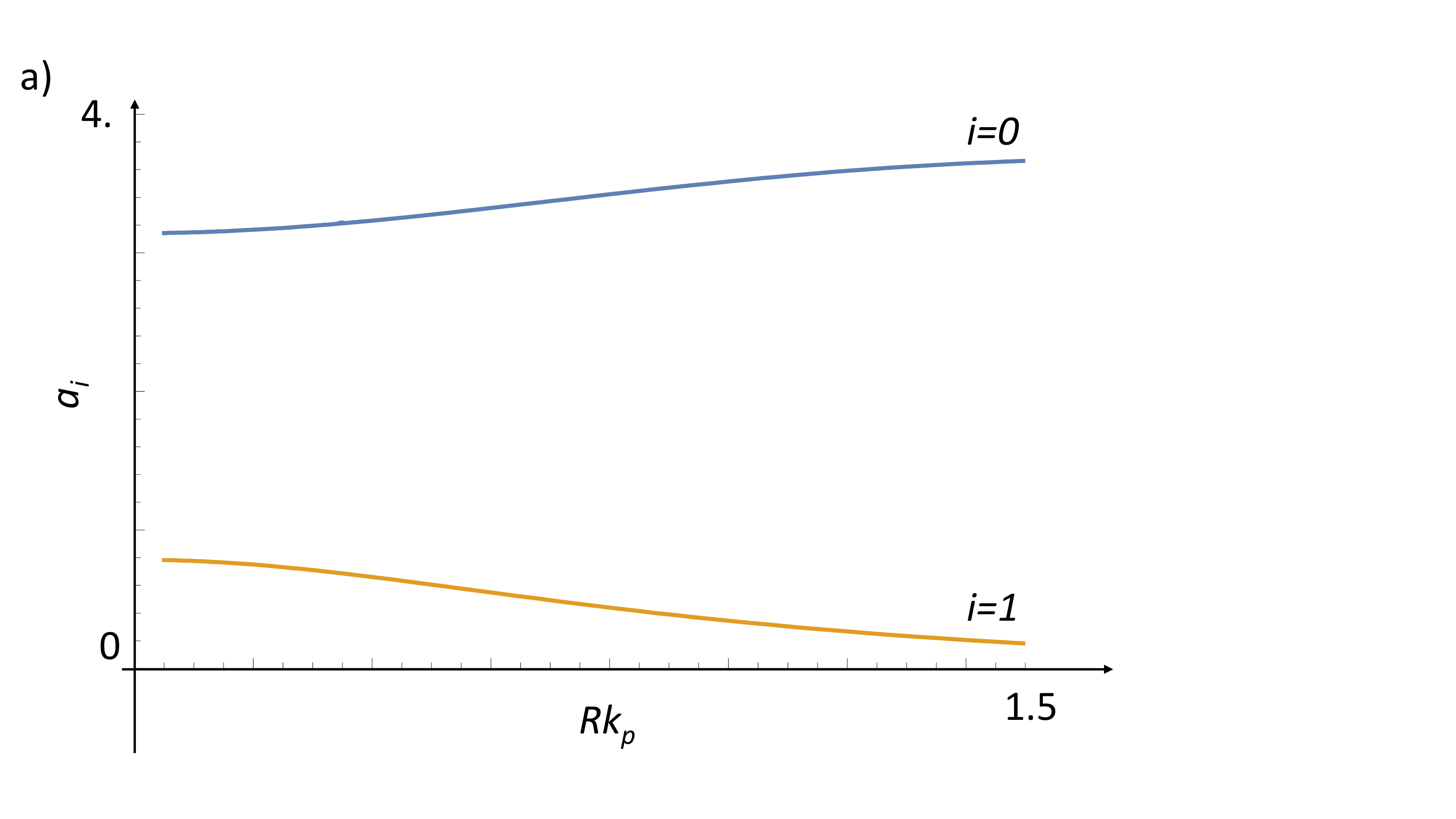}			 			 \includegraphics[width=11cm]{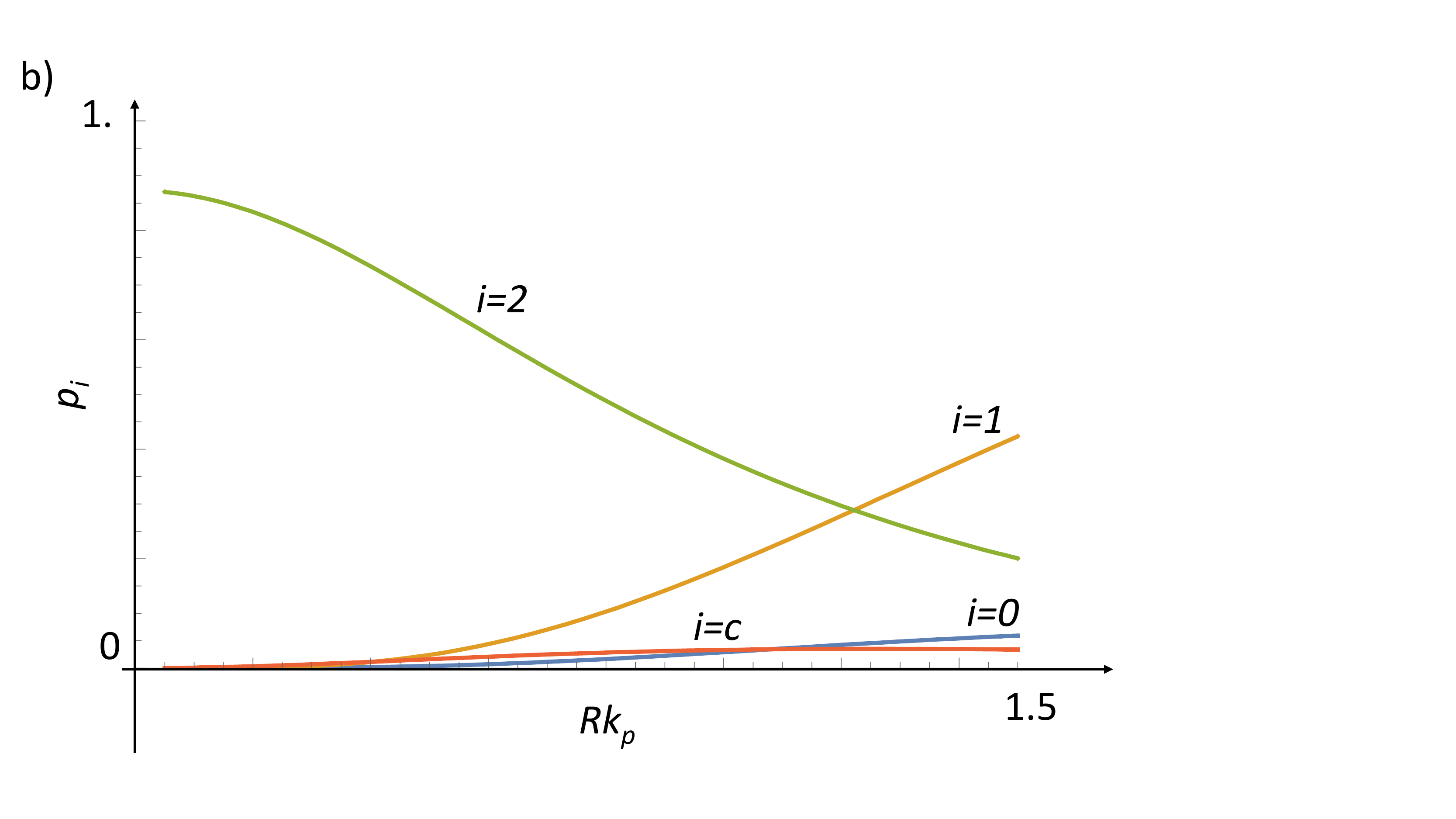}
			\caption{\small{
			Normalized modal parameters for flexural (F) modes $n=1$,$m=0$ at small wavevectors $k_p$. 
			a) Mass-related parameters $a_{i\,\left\{\eta\right\}}$. b) Bending-related parameters $p_{i\,\left\{\eta\right\}}$. 
			The stretching and internal damping terms can be obtained from these, following the rules described in the text. Plot realized with $\nu=0.2$ and no axial stress ($\sigma_0=0$). In the limit $R k_p \ll 1$, the terms $a_{0\,\left\{\eta\right\}}$ and $p_{2\,\left\{\eta\right\}}$ dominate, leading to the usual mass and spring constant expressions (see text).
			}}
			\label{fig_14}
		\end{figure}
		
The motion amplitude parameter $U_{\omega_{\left\{\eta\right\}}} $, which is defined as the tangential displacement of the rod's periphery, is directly linked to the usual angular motion parameter $\Theta_{\omega_{\left\{\eta\right\}}}$ by:
\begin{equation}
U_{\omega_{\left\{\eta\right\}}} (t) = R \, \Theta_{\omega_{\left\{\eta\right\}}} (t) .
\end{equation}
We then easily derive the {\it moment of inertia} $\rho \,\Omega(m) \frac{\pi}{2} R^4$ and {\it torsional rigidity} $\left[\frac{E_Y}{2(1+\nu)}+\sigma_0\right] \,\Omega(m) \frac{\pi}{2} R^4$, which match well-known definitions for the lowest $m=0$ branch without stored stress $\sigma_0=0$ (as they should) \cite{clelandBk}.
Note the presence of the extra term $\propto p_{0\,\left\{\eta\right\}}$ in the potential energy for T modes $m>0$.
The clamp friction Eq. (\ref{dissclamp}) can then be recast into ${\cal C}_{\left\{\eta\right\}} \varepsilon R^2 \, \dot{\Theta}_{\omega_{\left\{\eta\right\}}}^2$, 
which leads to an {\it effective friction torque} $\tau_{z\,\left\{\eta\right\}}$ acting onto each anchor:
\begin{eqnarray}
\tau_{z\,\left\{\eta\right\}} &= &\left[\frac{ (1+\nu_c) \,E_Y^2}{4(1+\nu)^2 R \,\Lambda_{c\,\left\{\eta\right\}} \omega_{\left\{\eta\right\}}^2} \right] \frac{\varepsilon }{R} \times \nonumber \\
&& \Omega(m)\,\frac{\pi}{2} R^4  \, (k_p R)^2\, \dot{\Theta}_{\omega_{\left\{\eta\right\}}} ,
\end{eqnarray}
with the brackets containing all material-dependent parameters. Looking into Eq. (\ref{dissipclamps}), this torque is actually due solely to a shear $\sigma_{c\,\theta z}$. This is the only clamp ``loss channel'' available to torsional waves within the first-order in $\varepsilon$ modeling. \\

For flexural (F) modes, the problem at hand is much more complex and non-analytic.
In Fig. \ref{fig_14} we show the coefficients $a_{i\,\left\{\eta\right\}}$ and $p_{i\,\left\{\eta\right\}}$ computed for modes $\left\{n=1,m=0,\mbox{F},k_p\right\}$, for small $k_p$ wavevectors and no axial stress applied.
When $R k_p \ll 1$, we derive a mode mass $\rho \, \pi R^2$ and a mode spring constant $E_Y \, ( 1 + 1.5\, \nu^2+0.3\, \nu^3 + 15.0\, \nu^4 +  35.0\, \nu^5 ) \frac{\pi}{4} R^4 \, k_p^4$ (both per unit length), which match usual Euler-Bernoulli definitions (in the limit $\nu \rightarrow 0$, as they should) \cite{clelandBk}.
The $\nu$-expansion of the spring constant coefficient is fit on numerical data within $1~\%$ in the range $-0.35 < \nu < +0.35$.
The clamp damping term $c_{1\,\left\{\eta\right\}}$ (neglecting the $\varphi_{z \left\{\eta\right\}}^{(1)}$ function) is plotted in Fig. \ref{fig_16} a). As for T modes, this is the only one relevant for a first-order in $\varepsilon$ modeling. When $R k_p \ll 1$, it behaves as $\propto (k_p R)^4$, which is actually due to both $\sigma_{c\,r z}$ and $\sigma_{c\,\theta z}$ components of the clamp's stress tensor. In Cartesian terms, these reduce to a stress tensor $\sigma_{c\,y z}$, and the clamp friction can be interpreted as an {\it effective friction torque} $\tau_{x\,\left\{\eta\right\}}$ acting upon each anchor (for a flexure in the $\vec{y}$ direction, Fig. \ref{fig_10}). This is the only clamping ``loss channel'' available to these modes, in the limit of small $\varepsilon$. \\

		\begin{figure}[t!]		 
			 \includegraphics[width=11cm]{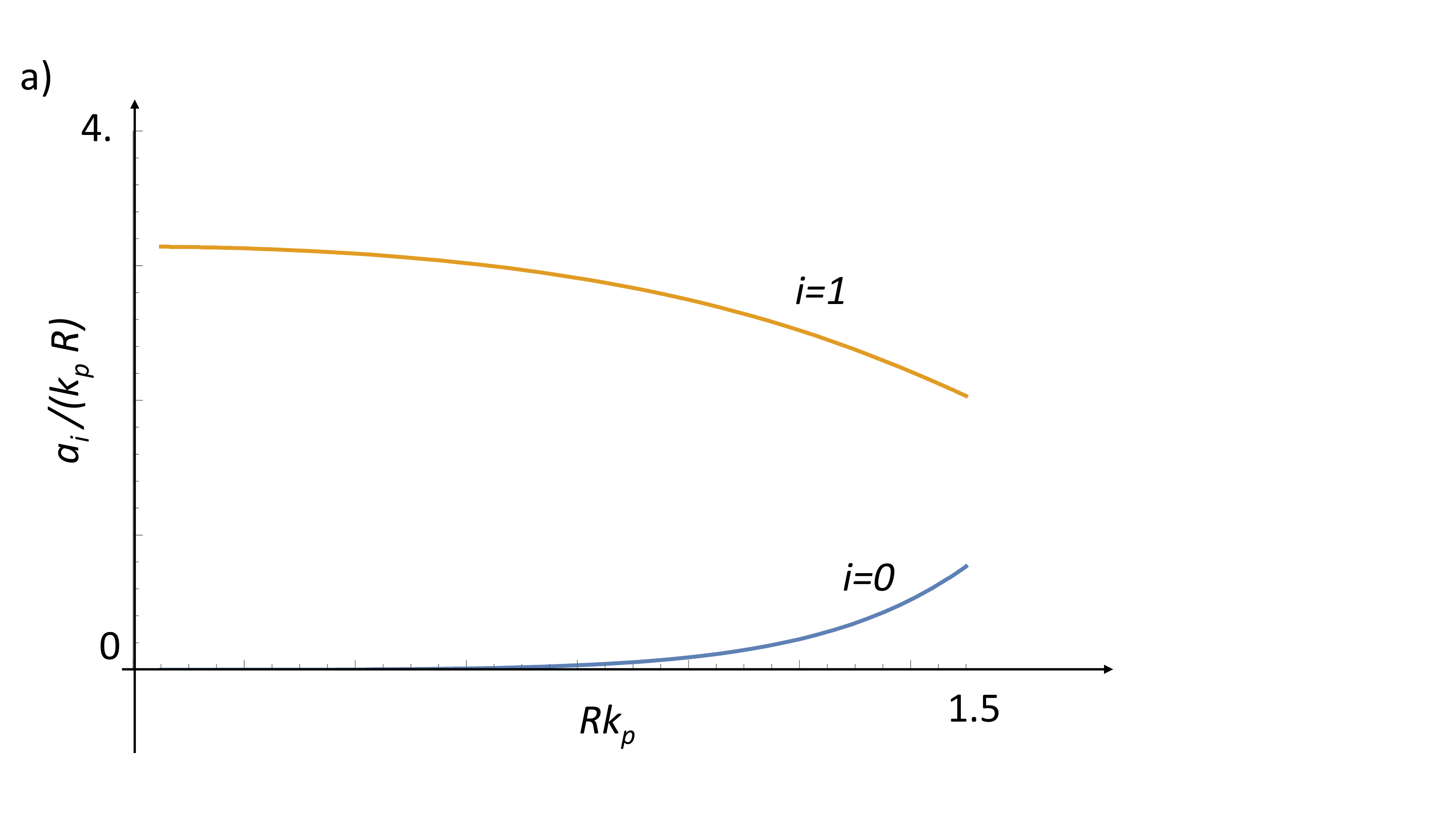}			 			 \includegraphics[width=11cm]{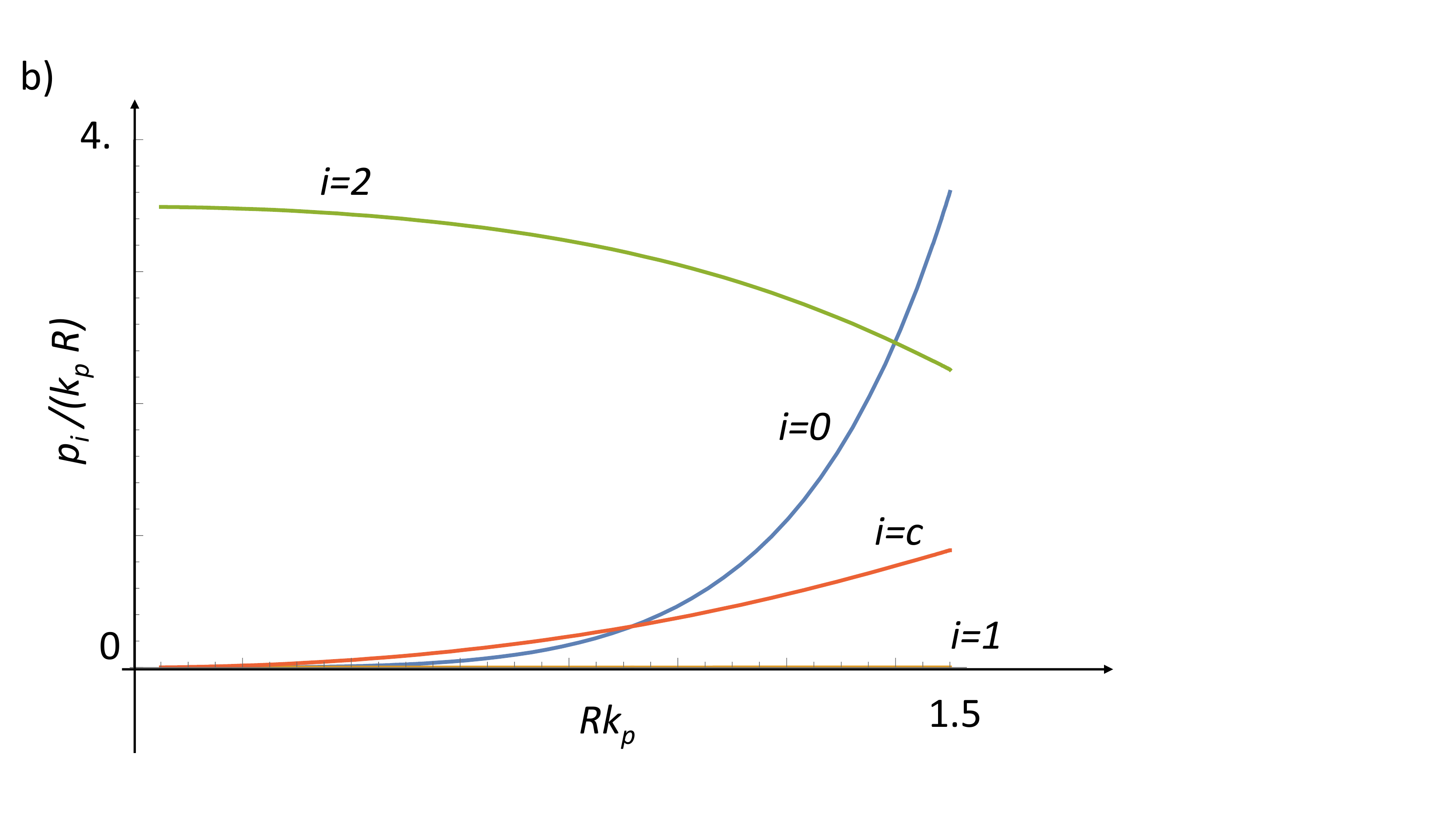}
			\caption{\small{
			Normalized modal parameters for longitudinal (L) modes $m=0$, at small wavevectors $k_p$. 
			a) Mass-related parameters $a_{i\,\left\{\eta\right\}}$. b) Bending-related parameters $p_{i\,\left\{\eta\right\}}$. The $1/(k_p R)$ normalisation comes from our definition of motion amplitudes, see Appendix \ref{pochhammer}.
			The stretching and internal damping terms can be obtained from these, following the rules described in the text. Plot realized with $\nu=0.2$ and no axial stress ($\sigma_0=0$). In the limit $R k_p \ll 1$, the terms $a_{1\,\left\{\eta\right\}}$ and $p_{2\,\left\{\eta\right\}}$ dominate, leading to the usual mass and spring constant expressions (see text).
			}}
			\label{fig_15}
		\end{figure}

In Fig. \ref{fig_15} we present the coefficients $a_{i\,\left\{\eta\right\}}$ and $p_{i\,\left\{\eta\right\}}$ computed for modes $\left\{m=0,\mbox{L},k_p\right\}$, at small $k_p$ wavevectors (and no axial load).
The solution is again found numerically; asymptotically at $R k_p \ll 1$,
the mass per unit length tends to $\rho \, \pi R^2 (k_p R)^2$ and the spring constant per unit length $E_Y \, ( 1 +1.5 \, \nu^2+ 1.0 \,\nu^3+15.0 \, \nu^4+30.0  \, \nu^5) (k_p R)^2 \pi R^2 \, k_p^2$.
The $(k_p R)^2$ term appearing in both expressions is due to our choice of motion amplitude, see Appendix \ref{pochhammer}.
As for F modes, the $\nu$ expansion has been fit on the calculated results, within $1~\%$ in the range $-0.35 < \nu < +0.35$.
Dividing mass and spring constants (per unit length) by $(k_p R)^2$, one recovers the usual beam theory expressions (in the limit $\nu \rightarrow 0$, as it should be) \cite{clelandBk}.
The clamp friction is more involved than in the previous cases, and requires 3 coefficients plotted in Fig. \ref{fig_16} b) [while assuming $\varphi_{i \left\{\eta\right\}}^{(1)} = 0$, $i=r,\theta$]. It is dominated by $c_{2\,\left\{\eta\right\}}$, which tends to $\approx \pi$ for $k_p R \ll 1$.
Its origin lies actually in the components $\sigma_{c\,ii}$ of the stress tensor ($i=r, \theta,z$), with a clear predominance of the 	$\sigma_{c\,zz}$ contribution for small $k_p$. One can interpret it as an {\it effective friction force} $F_z$ acting on each of the anchors, which represents the clamping ``loss channel'' available (at lowest order in $\varepsilon$) to longitudinal (L) modes.

\section{Modal coefficients, thermal gradient}
\label{coefII}

		\begin{figure}[t!]		 
			 \includegraphics[width=11cm]{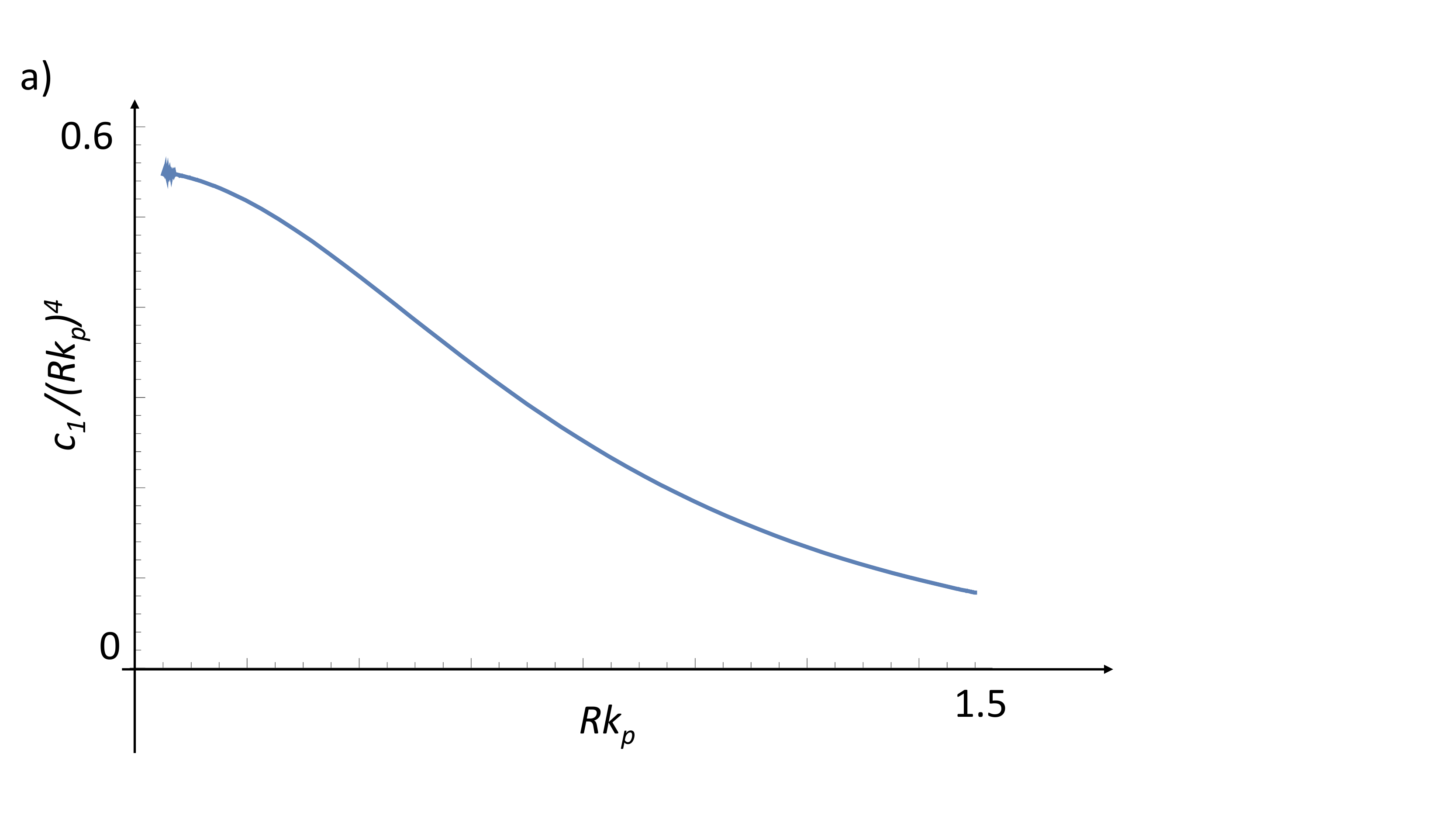}			 			 \includegraphics[width=11cm]{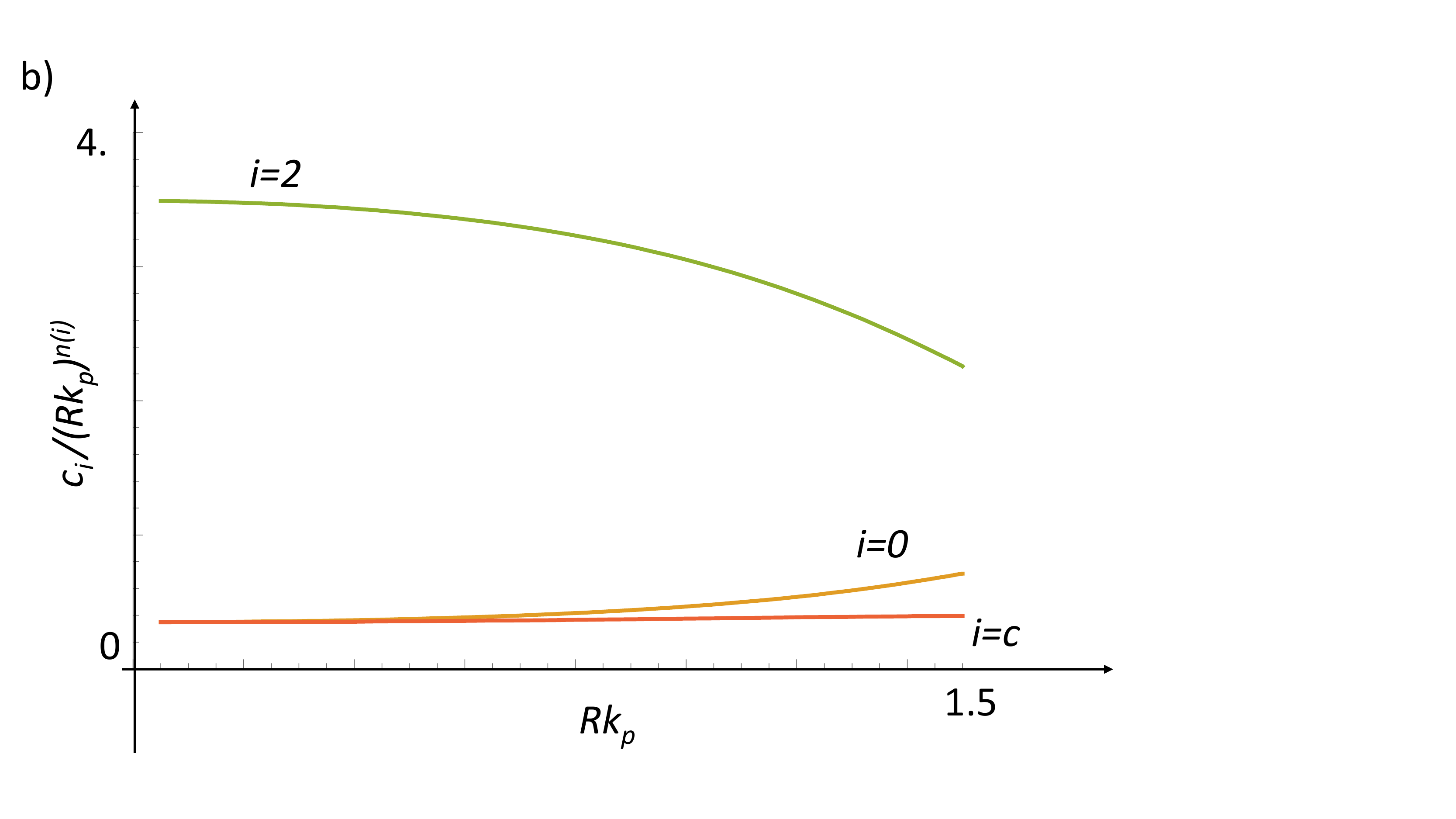}
			\caption{\small{
			Normalized modal parameters $c_{i\,\left\{\eta\right\}}$ defining clamp friction.
			a) For flexural (F) modes $n=1$,$m=0$ at small wavevectors $k_p$. b) For longitudinal (L) modes $m=0$, at small wavevectors $k_p$. Each coefficient is normalized to its leading order dependence, with $n(0)=4$, $n(2)=0$ and $n(c)=2$. Plot realized with $\nu=\nu_c=0.2$ and no axial stress ($\sigma_0=0$). The clamp correction functions $\varphi_{i \left\{\eta\right\}}^{(1)}$ have been set to zero (see text for details).
			}}
			\label{fig_16}
		\end{figure}

In this Appendix we extend the formalism of the previous one, in order to be able to deal with the situation where a thermal gradient is present.
For ``string-like'' solutions (the much more complex ``beam-like'' case shall not be discussed), we define the cross-terms in the energetics $\left\{\eta\right\} \neq \left\{\eta' \right\}$ to be:
\begin{eqnarray}
{\cal M}_{\left\{\eta\right\},\left\{\eta'\right\}} & = & \rho R^2 \left( a_{0\,\left\{\eta\right\},\left\{\eta'\right\}} + a_{1\,\left\{\eta\right\},\left\{\eta'\right\}} \left[k_p k_p' R^2 \right] \right), \\
{\cal S}_{\left\{\eta\right\},\left\{\eta'\right\}} & = & \sigma_0 \Big( s_{1\,\left\{\eta\right\},\left\{\eta'\right\}} \left[k_p k_p' R^2 \right]  \nonumber \\
&+&   s_{2\,\left\{\eta\right\},\left\{\eta'\right\}} \left[k_p k_p' R^2 \right]^2 \Big), \\
{\cal B}_{\left\{\eta\right\},\left\{\eta'\right\}} & = & E_Y \Big( p_{0\,\left\{\eta\right\},\left\{\eta'\right\}} + p_{1\,\left\{\eta\right\},\left\{\eta'\right\}} \left[k_p k_p' R^2 \right]  \nonumber \\
&+&     p_{2\,\left\{\eta\right\},\left\{\eta'\right\}} \left[k_p k_p' R^2 \right]^2   \nonumber \\
&-&   p_{c\,\left\{\eta\right\},\left\{\eta'\right\}} \left[k_p R \right]^2 - p_{c\,\left\{\eta'\right\},\left\{\eta\right\}} \left[k_p' R \right]^2 \! \Big),
\end{eqnarray}
for the mass and spring constant per unit length coefficients, for torsional (T) and flexural (F) solutions.
We write $k_p$ and $k_p'$ the wavevectors corresponding to modes $\left\{\eta\right\}$ and $\left\{\eta'\right\}$, respectively.
The corresponding ``deviation terms'' introduced in Section \ref{gradient} are then:
\begin{eqnarray}
\Delta {\cal M}_{\left\{\eta\right\},\left\{\eta'\right\}} & = & +\rho R^2\,  a_{1\,\left\{\eta\right\},\left\{\eta'\right\}} \left(k_p k_p' R^2 \right) , 
\end{eqnarray}
\begin{eqnarray}
\Delta {\cal S}_{\left\{\eta\right\},\left\{\eta'\right\}} & = & -\sigma_0\, s_{2\,\left\{\eta\right\},\left\{\eta'\right\}} \left(k_p k_p' R^2 \right)^2  , \\
\Delta {\cal B}_{\left\{\eta\right\},\left\{\eta'\right\}} & = & -E_Y  \Big( p_{0\,\left\{\eta\right\},\left\{\eta'\right\}}   \nonumber \\
&+&  p_{2\,\left\{\eta\right\},\left\{\eta'\right\}} \left[k_p k_p' R^2 \right]^2  \nonumber \\
& - &   p_{c\,\left\{\eta\right\},\left\{\eta'\right\}} \left[k_p R \right]^2 - p_{c\,\left\{\eta'\right\},\left\{\eta\right\}} \left[k_p' R \right]^2 \! \Big),
\end{eqnarray}
again written in the case of T and F modes. For Longitudinal (L) solutions, all the right-hand-sides should be divided by $ k_p k_p' R^2 $ because of our choice of amplitude definition, see Appendix \ref{pochhammer}.
By definition, we have:
\begin{eqnarray}
{\cal K}_{\left\{\eta\right\},\left\{\eta'\right\}} & = & {\cal S}_{\left\{\eta\right\},\left\{\eta'\right\}}  + {\cal B}_{\left\{\eta\right\},\left\{\eta'\right\}} , \\
\Delta {\cal K}_{\left\{\eta\right\},\left\{\eta'\right\}} & = & \Delta {\cal S}_{\left\{\eta\right\},\left\{\eta'\right\}}  + \Delta {\cal B}_{\left\{\eta\right\},\left\{\eta'\right\}} .
\end{eqnarray}
The adimensional coefficients appearing in the above are obtained from the mode functions $\phi_{i \left\{\eta\right\}}$ ($i=r,\theta,z$), and are given in Tab. \ref{tabcoefcross}. Note the symmetry by exchange $\left\{\eta\right\} \leftrightarrow \left\{\eta'\right\}$ in all these formulas.
Also, with this writing imposing $\left\{\eta\right\}=\left\{\eta'\right\}$ one recovers the expressions of the previous Appendix. \\

The dissipation expansions Eqs. (\ref{bulkD},\ref{bulkC}) also present $\left\{\eta\right\} \neq \left\{\eta'\right\}$ elements.
The internal friction per unit length cross-terms are thus defined as:
\begin{eqnarray}
{\cal L}_{\left\{\eta\right\},\left\{\eta'\right\}} & = & \frac{\Lambda_{Y\,\left\{\eta\right\}}}{2 }  \Big( l_{0\,\left\{\eta\right\},\left\{\eta'\right\}} + l_{1\,\left\{\eta\right\},\left\{\eta'\right\}} \left[k_p k_p'R^2 \right]  \nonumber \\
&+&    l_{2\,\left\{\eta\right\},\left\{\eta'\right\}} \left[k_p k_p' R^2 \right]^2  \nonumber \\
&-&    l_{c\,\left\{\eta\right\},\left\{\eta'\right\}} \left[k_p R \right]^2 - l_{c\,\left\{\eta'\right\},\left\{\eta\right\}} \left[k_p' R \right]^2 \Big), \label{dissipbulkcrossI}
\end{eqnarray}
and similarly: 
\begin{eqnarray}
\Delta {\cal L}_{\left\{\eta\right\},\left\{\eta'\right\}} & = & -\frac{\Lambda_{Y\,\left\{\eta\right\}}}{2}  \Big( l_{0\,\left\{\eta\right\},\left\{\eta'\right\}}  \nonumber \\
&+&    l_{2\,\left\{\eta\right\},\left\{\eta'\right\}} \left[k_p k_p' R^2 \right]^2 \nonumber \\
&-&  l_{c\,\left\{\eta\right\},\left\{\eta'\right\}} \left[k_p R \right]^2 - l_{c\,\left\{\eta'\right\},\left\{\eta\right\}} \left[k_p' R \right]^2 \Big),
\end{eqnarray}
for torsional (T) and flexural (F) solutions. As above, the longitudinal (L) case is obtained by dividing the right-hand-sides by $ k_p k_p' R^2$.
Similarly to Appendix \ref{coefs}, the $l_{i\,\left\{\eta\right\},\left\{\eta'\right\}}$ are identical to the $p_{i\,\left\{\eta\right\},\left\{\eta'\right\}}$ given in Tab. \ref{tabcoefcross} ($i=0,1,2,c$), provided one applies the modification $\nu \rightarrow \nu_\lambda$.

A similar derivation is performed with the clamping terms.
It leads for T and F modes to:
\begin{eqnarray}
\!\!\!\!\!\!\!\!\!\!\!\!\!\!\! {\cal C}_{\left\{\eta\right\},\left\{\eta'\right\}} &  \!\!\! = &   \!\!\!\! \frac{2 \, E_Y^2}{\Lambda_{c\,\left\{\eta\right\}}}\frac{ (k_p k_p' R^2)}{\omega_{\left\{\eta\right\}}\omega_{\left\{\eta'\right\}}} c_{1\,\left\{\eta\right\},\left\{\eta'\right\}} , \label{clampTFcross}
\end{eqnarray}
 and:
\begin{eqnarray}
{\cal C}_{\left\{\eta\right\},\left\{\eta'\right\}} & = &  \frac{2 \, E_Y^2}{\Lambda_{c\,\left\{\eta\right\}}}  \frac{ (k_p k_p' R^2)}{\omega_{\left\{\eta\right\}}\omega_{\left\{\eta'\right\}}} \left( \frac{ c_{0\,\left\{\eta\right\},\left\{\eta'\right\}} }{\left( k_p k_p' R^2 \right)^2} \right. \nonumber \\
&+& \left. c_{2\,\left\{\eta\right\},\left\{\eta'\right\}}- \frac{ c_{c\,\left\{\eta\right\},\left\{\eta'\right\}} + c_{c\,\left\{\eta'\right\},\left\{\eta\right\}} }{\left( k_p k_p' R^2 \right)} \right), \label{dissipclampI} 
\end{eqnarray}
for L ones. The $c_{i\,\left\{\eta\right\},\left\{\eta'\right\}}$ ($i=0,1,2, c$) are explicitly given in Tab. \ref{tabcoefcross2}. Imposing $\left\{\eta\right\} = \left\{\eta'\right\}$ in these friction coefficients, we reproduce the ones introduced in the preceding Appendix \ref{coefs}.
Note that Eqs. (\ref{dissipbulkcrossI}-\ref{dissipclampI}) are symmetric under the exchange $\left\{\eta\right\} \leftrightarrow \left\{\eta'\right\}$ provided $\Lambda_{Y\,\left\{\eta\right\}}=\Lambda_{Y\,\left\{\eta'\right\}}$ and $\Lambda_{c\,\left\{\eta\right\}}=\Lambda_{c\,\left\{\eta'\right\}}$, which might {\it not} be the case for arbitrary modes $\left\{\eta\right\} \neq\left\{\eta'\right\}$.
However, the only terms of practical interest in the development of Section \ref{gradient} are the ones with $\left\{\eta\right\}$ and $\left\{\eta'\right\}$
belonging to the same branch, with very close $k_p$ and $k_p'$ wavevectors.
It is therefore perfectly reasonable to assume $\Lambda_{Y\,\left\{\eta\right\}} \approx \Lambda_{Y\,\left\{\eta'\right\}}$ and $\Lambda_{c\,\left\{\eta\right\}}\approx \Lambda_{c\,\left\{\eta'\right\}}$ in any of our computations, restoring thus the symmetry with $ {\cal C}_{\left\{\eta\right\},\left\{\eta'\right\}} \approx {\cal C}_{\left\{\eta'\right\},\left\{\eta\right\}} \approx   {\cal C}_{\left\{\eta'\right\}} \approx  {\cal C}_{\left\{\eta\right\}}$. 

A parameter of particular importance in Section \ref{gradient} is ${\cal M}_{\left\{\eta\right\},\left\{\eta'\right\}} \, \omega_{\left\{\eta\right\}} \omega_{\left\{\eta'\right\}} / {\cal K}_{\left\{\eta\right\},\left\{\eta'\right\}}$. 
For a given branch, it can be expanded in terms of $\Delta k_p = k_p' - k_p$, the difference between wavevectors; the result is analytical for torsional (T) waves:
\begin{eqnarray}
&&\frac{{\cal M}_{\left\{\eta\right\},\left\{\eta'\right\}} \, \omega_{\left\{\eta\right\}} \omega_{\left\{\eta'\right\}} }{ {\cal K}_{\left\{\eta\right\},\left\{\eta'\right\}}} \approx \nonumber \\
&& \!\!\!\! \!\!\!\! \!\!\!\!\!\!\!\! 1  +  \frac{\left(\beta_{\left\{\eta\right\}}R\right)^2 \left[1 + 2 \,\sigma_0 (1+\nu)/E_Y \right]}{2 \left[\left(\beta_{\left\{\eta\right\}}R\right)^2 + k_p^2 \left(1 + 2 \, \sigma_0(1+\nu)/E_Y \right) \right]^2} \left(\Delta k_p R \right)^2 \! ,
\end{eqnarray}
at lowest order.
It turns out that this $\Delta k_p$ dependence is negligible for any situation of practical interest.
A similar result holds for L and F modes, but requires numerical evaluation to be tested.


\begin{widetext}

\begin{table}[h!]
\begin{center}
\hspace*{-1.3cm}
\begin{tabular}{|c|c|c|}    \hline
coefs.    & $a_{i\,\left\{\eta\right\},\left\{\eta'\right\}}$   & $p_{i\,\left\{\eta\right\},\left\{\eta'\right\}}$      \\    \hline    \hline
 \shortstack{ $ $ \\ $i=0$ \\ $ $ }   & \shortstack{ $ $ \\ $\frac{1}{R^2}\int \!\!\! \int \left[ \phi_{r \left\{\eta\right\}}\phi_{r \left\{\eta'\right\}}+\phi_{\theta \left\{\eta\right\}} \phi_{\theta \left\{\eta'\right\}} \right] r dr d\theta $  \\ $ $ }       & \shortstack{ $\int \!\!\! \int \frac{1 }{(1+\nu)}\left[\frac{1}{2}  \frac{\partial \phi_{\theta \left\{\eta\right\}}}{\partial r }\frac{\partial \phi_{\theta \left\{\eta'\right\}}}{\partial r } + \frac{(1-\nu)}{(1-2\nu)} \frac{\partial \phi_{r \left\{\eta\right\}}}{\partial r }\frac{\partial \phi_{r \left\{\eta'\right\}}}{\partial r } \right.$ \\  $\left. + \frac{1}{2r} \! \left( \frac{\partial \phi_{r \left\{\eta\right\}}}{\partial \theta }  - \phi_{\theta \left\{\eta\right\}}  \right) \! \frac{\partial \phi_{\theta \left\{\eta'\right\}}}{\partial r } + \frac{1}{2r} \! \left( \frac{\partial \phi_{r \left\{\eta'\right\}}}{\partial \theta }  - \phi_{\theta \left\{\eta'\right\}}  \right) \! \frac{\partial \phi_{\theta \left\{\eta\right\}}}{\partial r }  \right.$ \\ $\left. +\frac{\nu}{(1-2\nu) \, r} \! \left( \frac{\partial \phi_{\theta \left\{\eta\right\}}}{\partial \theta }  + \phi_{r \left\{\eta\right\}}  \right) \! \frac{\partial \phi_{r \left\{\eta'\right\}}}{\partial r } +\frac{\nu}{(1-2\nu) \, r} \! \left( \frac{\partial \phi_{\theta \left\{\eta'\right\}}}{\partial \theta }  + \phi_{r \left\{\eta'\right\}}  \right) \! \frac{\partial \phi_{r \left\{\eta\right\}}}{\partial r } \right. $ \\ $\left. + \frac{1}{2 r^2} \! \left( \frac{\partial \phi_{r \left\{\eta\right\}}}{\partial \theta }  - \phi_{\theta \left\{\eta\right\}}  \right) \left( \frac{\partial \phi_{r \left\{\eta'\right\}}}{\partial \theta }  - \phi_{\theta \left\{\eta'\right\}}  \right)   \right. $   \\ $\left. +\frac{(1-\nu)}{(1-2\nu)\, r^2} \! \left( \frac{\partial \phi_{\theta \left\{\eta\right\}}}{\partial \theta }  + \phi_{r \left\{\eta\right\}}  \right)\left( \frac{\partial \phi_{\theta \left\{\eta'\right\}}}{\partial \theta }  + \phi_{r \left\{\eta'\right\}}  \right)  \right] r dr d\theta$  }                           \\    \hline
 \shortstack{ $ $ \\$i=1$ \\ $ $ }   & \shortstack{ $ $ \\  $\frac{1}{R^2} \int \!\!\! \int \left[ \phi_{z \left\{\eta\right\}}\phi_{z \left\{\eta'\right\}} \right] r dr d\theta $ \\ $ $ }     &  \shortstack{ $ \frac{1}{R^2}\int \!\!\! \int \frac{1 }{2(1+\nu)} \left[ \phi_{r \left\{\eta\right\}}\phi_{r \left\{\eta'\right\}} +\phi_{\theta \left\{\eta\right\}} \phi_{\theta \left\{\eta'\right\}}  \right. $  \\  $+ \left.   \phi_{r \left\{\eta'\right\}}   R \frac{\partial \phi_{z \left\{\eta\right\}}}{\partial r }  +  \phi_{\theta \left\{\eta'\right\}}  \frac{R}{r}  \frac{\partial \phi_{z \left\{\eta\right\}}}{\partial \theta } +\phi_{r \left\{\eta \right\}}   R \frac{\partial \phi_{z \left\{\eta'\right\}}}{\partial r }  +  \phi_{\theta \left\{\eta \right\}}  \frac{R}{r}  \frac{\partial \phi_{z \left\{\eta'\right\}}}{\partial \theta }  \right.$ \\ $\left.  +   R^2 \frac{\partial \phi_{z \left\{\eta\right\}}}{\partial r }\frac{\partial \phi_{z \left\{\eta'\right\}}}{\partial r }  + \left(  \frac{R}{r} \right)^2 \frac{\partial \phi_{z \left\{\eta\right\}}}{\partial \theta } \frac{\partial \phi_{z \left\{\eta'\right\}}}{\partial \theta }    \right]    r dr d\theta $   }                 \\    \hline
 $i=2$    &  /            &  $+\frac{(1-\nu) }{(1-2\nu)(1+\nu)} a_{1\,\left\{\eta\right\},\left\{\eta'\right\}}$                       \\    \hline
 $i=c$    &  /            &  $\frac{1}{R^2}\int \!\!\! \int \frac{+\nu }{(1-2\nu)(1+\nu)} \phi_{z \left\{\eta'\right\}}  \left[ R \frac{\partial \phi_{r \left\{\eta\right\}}}{\partial r}  +\frac{R}{r} \left( \phi_{r \left\{\eta\right\}} + \frac{\partial \phi_{\theta \left\{\eta\right\}}}{\partial \theta}   \right) \right] r dr d\theta$     \\    \hline
\end{tabular}
\caption{ \label{tabcoefcross}  Mass and spring (normalized) cross-constants as a function of the modal functions $\phi_{i \left\{\eta\right\}}$. By definition  $s_{1\,\left\{\eta\right\},\left\{\eta'\right\}}=a_{0\,\left\{\eta\right\},\left\{\eta'\right\}}$ and $s_{2\,\left\{\eta\right\},\left\{\eta'\right\}}=a_{1\,\left\{\eta\right\},\left\{\eta'\right\}}$. The $l_{i\,\left\{\eta\right\},\left\{\eta'\right\}}$ are obtained from the $p_{i\,\left\{\eta\right\},\left\{\eta'\right\}}$ with the replacement $\nu \rightarrow \nu_\lambda$ in the above expressions. }
\end{center}
\end{table}

\begin{table}[h!]
\begin{center}
\begin{tabular}{|c|c|}    \hline
coefs.    & $c_{i\,\left\{\eta\right\},\left\{\eta'\right\}}$  \\    \hline    \hline
 $i=0$    &     \shortstack{ $\int \!\!\! \int \frac{1 }{(1+\nu)^2}\left[\frac{(1+\nu_c)}{2}  \frac{\partial \phi_{\theta \left\{\eta\right\}}}{\partial r }\frac{\partial \phi_{\theta \left\{\eta'\right\}}}{\partial r } + \frac{(1-2\nu+3 \nu^2)-2\nu(2-\nu)\,\nu_c}{(1-2\nu)^2} \frac{\partial \phi_{r \left\{\eta\right\}}}{\partial r } \frac{\partial \phi_{r \left\{\eta'\right\}}}{\partial r } \right.$ \\  $\left. + \frac{ (1+\nu_c)}{2 r} \! \left( \frac{\partial \phi_{r \left\{\eta\right\}}}{\partial \theta }  - \phi_{\theta \left\{\eta\right\}}  \right) \! \frac{\partial \phi_{\theta \left\{\eta'\right\}}}{\partial r } 
+ \frac{ (1+\nu_c)}{2 r} \! \left( \frac{\partial \phi_{r \left\{\eta'\right\}}}{\partial \theta }  - \phi_{\theta \left\{\eta'\right\}}  \right) \! \frac{\partial \phi_{\theta \left\{\eta \right\}}}{\partial r }\right. $ \\ 
$ \left.  +\frac{ \nu(2-\nu)- (1+2\nu^2)\, \nu_c}{(1-2\nu)^2 \, r} \! \left( \frac{\partial \phi_{\theta \left\{\eta\right\}}}{\partial \theta }  + \phi_{r \left\{\eta\right\}}  \right) \! \frac{\partial \phi_{r \left\{\eta'\right\}}}{\partial r } 
+\frac{ \nu(2-\nu)- (1+2\nu^2)\, \nu_c}{(1-2\nu)^2 \, r} \! \left( \frac{\partial \phi_{\theta \left\{\eta'\right\}}}{\partial \theta }  + \phi_{r \left\{\eta'\right\}}  \right) \! \frac{\partial \phi_{r \left\{\eta \right\}}}{\partial r }  \right.$ \\ $\left. + \frac{(1+\nu_c)}{2 r^2} \! \left( \frac{\partial \phi_{r \left\{\eta\right\}}}{\partial \theta }  - \phi_{\theta \left\{\eta\right\}}  \right)  \left( \frac{\partial \phi_{r \left\{\eta'\right\}}}{\partial \theta }  - \phi_{\theta \left\{\eta'\right\}}  \right)         \right.       $  \\   $  \left.    +\frac{(1-2\nu+3\nu^2)-2\nu(2-\nu)\,\nu_c}{(1-2\nu)^2\, r^2} \! \left( \frac{\partial \phi_{\theta \left\{\eta\right\}}}{\partial \theta }  + \phi_{r \left\{\eta\right\}} \right) \left( \frac{\partial \phi_{\theta \left\{\eta'\right\}}}{\partial \theta }  + \phi_{r \left\{\eta'\right\}} \right) \right.$ \\
$ \left. + \frac{(1+\nu_c)}{2} \!  \varphi_{r \left\{\eta\right\}}^{(1)}\varphi_{r \left\{\eta'\right\}}^{(1)}  +\frac{(1+\nu_c)}{2} \!  \varphi_{\theta \left\{\eta\right\}}^{(1)} \varphi_{\theta \left\{\eta'\right\}}^{(1)} \right] r dr d\theta$ }  
                        \\    \hline
 $i=1$    &      \shortstack{ $ \frac{1}{R^2}\int \!\!\! \int \frac{(1+\nu_c) }{2(1+\nu)^2} \left[ \phi_{r \left\{\eta\right\}} \phi_{r \left\{\eta'\right\}} +\phi_{\theta \left\{\eta\right\}} \phi_{\theta \left\{\eta'\right\}}  \right. $  \\  $+ \left.   \phi_{r \left\{\eta'\right\}}   R \frac{\partial \phi_{z \left\{\eta\right\}}}{\partial r }  +  \phi_{\theta \left\{\eta'\right\}}  \frac{R}{r}  \frac{\partial \phi_{z \left\{\eta\right\}}}{\partial \theta } +  \phi_{r \left\{\eta\right\}}   R \frac{\partial \phi_{z \left\{\eta'\right\}}}{\partial r }  +  \phi_{\theta \left\{\eta\right\}}  \frac{R}{r}  \frac{\partial \phi_{z \left\{\eta'\right\}}}{\partial \theta } \right.$ \\ $\left.  +   R^2 \frac{\partial \phi_{z \left\{\eta\right\}}}{\partial r }\frac{\partial \phi_{z \left\{\eta'\right\}}}{\partial r } + \left(  \frac{R}{r} \right)^2 \frac{\partial \phi_{z \left\{\eta\right\}}}{\partial \theta }\frac{\partial \phi_{z \left\{\eta'\right\}}}{\partial \theta }  \right. $ \\ $\left. + 2\frac{ ( 1-2\nu+3\nu^2)-2\nu ( 2-\nu ) \, \nu_c  }{(1+\nu_c)(1-2\nu)^2}   \varphi_{z \left\{\eta\right\}}^{(1)} \varphi_{z \left\{\eta'\right\}}^{(1)}   \right]    r dr d\theta $   }                             \\    \hline
 $i=2$    &     $+\frac{(1 - 2\nu + 3 \nu^2)  -2  \nu (2 - \nu) \, \nu_c  }{(1-2\nu)^2(1+\nu)^2} a_{1\,\left\{\eta\right\},\left\{\eta'\right\}}   $                    \\    \hline
 $i=c$    &   $\frac{1}{R^2}\int \!\!\! \int \frac{\nu (2-\nu) -(2 \nu^2+1) \, \nu_c }{(1-2\nu)^2(1+\nu)^2} \phi_{z \left\{\eta'\right\}}  \left[ R \frac{\partial \phi_{r \left\{\eta\right\}}}{\partial r}  +\frac{R}{r} \left( \phi_{r \left\{\eta\right\}}  + \frac{\partial \phi_{\theta \left\{\eta\right\}}}{\partial \theta}  \right) \right] r dr d\theta$                        \\    \hline
\end{tabular}
\caption{ \label{tabcoefcross2} Clamp dissipation (normalized) cross-constants as a function of the modal functions $\phi_{i \left\{\eta\right\}}$ (and $\varphi_{i \left\{\eta\right\}}^{(1)}$, the first order correction within the clamping zone). }
\end{center}
\end{table}

\end{widetext}

Finally, the heat flow is calculated from Eq. (\ref{flow}) with the above expressions of the required cross-coefficients.
It leads to the definition of $g_{\left\{\eta\right\},\left\{\eta'\right\}}$ in Eq. (\ref{kappadef}), which reads:
\begin{eqnarray}
&& g_{\left\{\eta\right\},\left\{\eta'\right\}} = \left(k_p k_p' R^2 \right) \times \nonumber \\
&&   c_{1\,\left\{\eta\right\},\left\{\eta'\right\}}   / 
 \left[\sum_{\substack{i=0,1,2}} p_{i\,\left\{\eta\right\},\left\{\eta'\right\}}\left(k_p k_p' R^2 \right)^i \right. \nonumber \\
&& \left. -\left(p_{c\,\left\{\eta\right\},\left\{\eta'\right\}}\left[k_p R\right]^2+ p_{c\,\left\{\eta'\right\},\left\{\eta\right\}}\left[k_p' R\right]^2\right) \right. \nonumber \\
&& \left. +\frac{\sigma_0}{E_Y}\sum_{\substack{i=1,2}} s_{i\,\left\{\eta\right\},\left\{\eta'\right\}} \left(k_p k_p' R^2 \right)^i \right], \label{lambdacI}
\end{eqnarray}
for T and F modes, and:
\begin{eqnarray}
&& g_{\left\{\eta\right\},\left\{\eta'\right\}} = \left(k_p k_p' R^2 \right) \times \nonumber \\
&& \left[\frac{ c_{0\,\left\{\eta\right\},\left\{\eta'\right\}} }{\left( k_p k_p' R^2 \right)^2}  
+  c_{2\,\left\{\eta\right\},\left\{\eta'\right\}} - \frac{   c_{c\,\left\{\eta\right\},\left\{\eta'\right\}} + c_{c\,\left\{\eta'\right\},\left\{\eta\right\}} }{ \left( k_p k_p' R^2 \right)} \right] \times \nonumber \\
&& \left( k_p k_p' R^2 \right) / 
 \left[\sum_{\substack{i=0,1,2}} p_{i\,\left\{\eta\right\},\left\{\eta'\right\}}\left(k_p k_p' R^2 \right)^i \right. \nonumber \\
&& \left. -\left(p_{c\,\left\{\eta\right\},\left\{\eta'\right\}}\left[k_p R\right]^2+ p_{c\,\left\{\eta'\right\},\left\{\eta\right\}}\left[k_p' R\right]^2\right) \right. \nonumber \\
&& \left. +\frac{\sigma_0}{E_Y}\sum_{\substack{i=1,2}} s_{i\,\left\{\eta\right\},\left\{\eta'\right\}} \left(k_p k_p' R^2 \right)^i \right], \label{lambdacII}
\end{eqnarray}
for L ones. With modes of relevance being nearby, we simplify again the problem into $g_{\left\{\eta\right\},\left\{\eta'\right\}} \approx g_{\left\{\eta\right\}}$ by replacing in the above all cross-terms by their corresponding modal counterpart, i.e. $p_{i\,\left\{\eta\right\},\left\{\eta'\right\}} \approx p_{i\,\left\{\eta\right\}}$ and so on. 
This lead in Section \ref{FDT} to the definition of the average clamping loss parameter $\bar{\Lambda}_{c} $, Eq. (\ref{deflambdabar}):
\begin{equation}
\frac{1}{\bar{\Lambda}_{c} } = \frac{1}{3N} \!\!\!\!\!\!\!\! \sum_{\substack{\mbox{all branches $i$} \\ \mbox{and wavevectors $q$}\\ \mbox{i.e. all   modes  $\left\{\eta\right\}$}}} \frac{ g_{\left\{\eta\right\}} }{\Lambda_{c\,\left\{\eta\right\} }}, 
\end{equation}
that we repeat here for convenience. 
For our model to be meaningful, this sum should be {\it finite}. 
Experimentally, $\Lambda_{c\,\left\{\eta\right\}} $ is found to grow with increasing mode number (see e.g. Ref. \cite{martialPHD} for a study on NEMS beams).
We can therefore verify the convergence of the sum by checking the asymptotic behaviour of the coefficients appearing in Eqs. (\ref{lambdacI},\ref{lambdacII}).
For torsional (T) modes, one easily shows that $g_{\left\{\eta\right\}} \rightarrow \mbox{Cste}$ as $k_p \rightarrow + \infty$ (for any mode index $m$; for $m=0$ it is strictly a constant). For F and L modes, the problem has again to be solved numerically (see Figs. \ref{fig_14}, \ref{fig_15}, and \ref{fig_16} for an example of parameter sets).

\section{Integrated generalized spectra}
\label{integrals}

The properties of the generalised stochastic Langevin forces acting onto the modes are derived from the integrals:
\begin{eqnarray}
&& \int_{0}^{+\infty}  \chi_{\left\{\eta\right\}}(\omega) \, \chi_{\left\{\eta'\right\}}(\omega)^{*}  d\omega  \nonumber \\
&& =  \left[\int_{-\infty}^{0}\chi_{\left\{\eta\right\}}(\omega) \, \chi_{\left\{\eta'\right\}}(\omega)^{*}  d\omega \right]^* , \\
&& \int_{0}^{+\infty} \omega^2 \, \chi_{\left\{\eta\right\}}(\omega) \, \chi_{\left\{\eta'\right\}}(\omega)^{*}  d\omega  \nonumber \\
&& =  \left[\int_{-\infty}^{0} \omega^2 \, \chi_{\left\{\eta\right\}}(\omega) \, \chi_{\left\{\eta'\right\}}(\omega)^{*}  d\omega \right]^* , 
\end{eqnarray}
which we will study here numerically. 
Two situations have to be considered: either the mode frequencies $\omega_{\left\{\eta\right\}}, \omega_{\left\{\eta'\right\}}$ are much closer than their 
relaxation rates $\Gamma_{\left\{\eta\right\}}, \Gamma_{\left\{\eta'\right\}}$, or they are far apart. \\

		\begin{figure}[t!]
		\centering
	\includegraphics[width=13 cm]{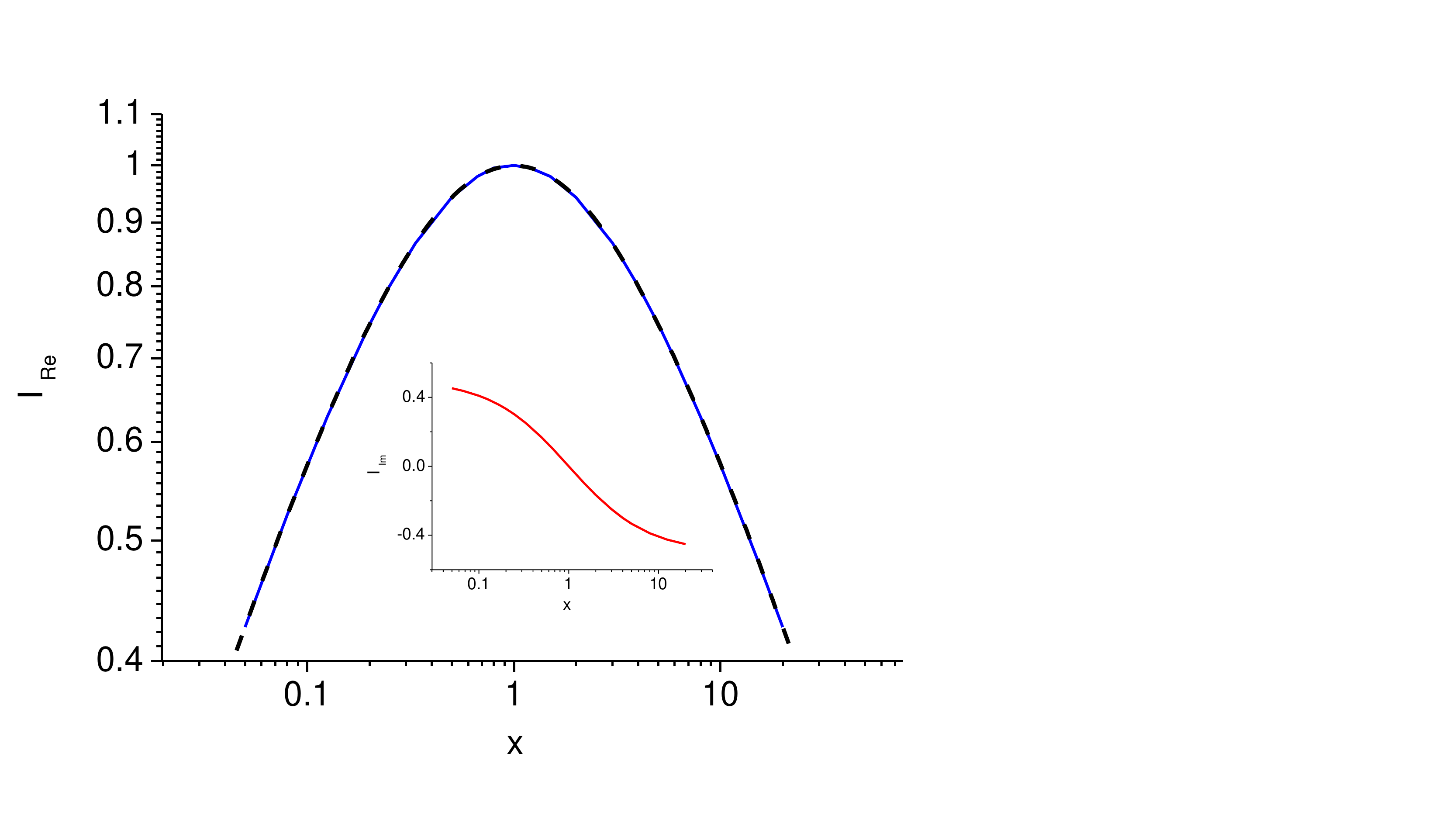}
			\caption{\small{ 
			Functions $ \mbox{I}_{\mbox{Re}}(x)$ and $ \mbox{I}_{\mbox{Im}}(x)$ that define the spectrum integrals when $\omega_{\left\{\eta\right\}} \approx \omega_{\left\{\eta'\right\}}$. The dashed line is a fit  (see text). }}
			\label{fig_17}
		\end{figure}
		
Consider first the case $\omega_{\left\{\eta\right\}} \approx \omega_{\left\{\eta'\right\}}$, which is the most common one. 
Typically, this is a good approximation (within a few $\%$) as soon as $\left|\omega_{\left\{\eta\right\}}- \omega_{\left\{\eta'\right\}}\right|< \Gamma_{\left\{\eta\right\}}/10, \Gamma_{\left\{\eta'\right\}}/10$.
Then we write, in the high-Q limit:
\begin{eqnarray}
&& \int_{0}^{+\infty}  \chi_{\left\{\eta\right\}}(\omega) \, \chi_{\left\{\eta'\right\}}(\omega)^{*}  d\omega  \nonumber \\
&& \!\!\!\!\!\! = \frac{\pi}{2} \omega_{\left\{\eta\right\}} \sqrt{Q_{\left\{\eta\right\}} Q_{\left\{\eta'\right\}}} \left[  \mbox{I}_{\mbox{Re}}\!\! \left( \frac{Q_{\left\{\eta\right\}}}{Q_{\left\{\eta'\right\}}}\right) -\mathrm{i} \frac{ \mbox{I}_{\mbox{Im}}\!\!\left( \frac{Q_{\left\{\eta\right\}}}{Q_{\left\{\eta'\right\}}}\right)}{  \sqrt{Q_{\left\{\eta\right\}} Q_{\left\{\eta'\right\}}}   }\right] \! \! , \label{integrIm} \\
&& \int_{0}^{+\infty}  \omega^2 \, \chi_{\left\{\eta\right\}}(\omega) \, \chi_{\left\{\eta'\right\}}(\omega)^{*}  d\omega  \nonumber \\
&& = \omega_{\left\{\eta\right\}}^2 \int_{0}^{+\infty}  \chi_{\left\{\eta\right\}}(\omega) \, \chi_{\left\{\eta'\right\}}(\omega)^{*}  d\omega \, .
\end{eqnarray}
The functions $ \mbox{I}_{\mbox{Re}}$ and $ \mbox{I}_{\mbox{Im}}$ are plotted in Fig. \ref{fig_17}. They verify the symmetries:
\begin{eqnarray}
\mbox{I}_{\mbox{Re}}(x) & = & \mbox{I}_{\mbox{Re}}(1/x) , \\
\mbox{I}_{\mbox{Im}}(x) & = & - \mbox{I}_{\mbox{Im}}(1/x) .
\end{eqnarray}
By construction, we chose $\mbox{I}_{\mbox{Re}}(1)=1$ which reproduces the conventional case where the mode labels $\left\{\eta\right\},\left\{\eta'\right\}$ refer to the same one; besides, $\mbox{I}_{\mbox{Im}}$ is bound and remains  $\left|\mbox{I}_{\mbox{Im}}\right| < 0.5$. 
Because of the Q-dependence of the imaginary component of Eq. (\ref{integrIm}), we see that this contribution will disappear in the high-Q limit. This is what justifies keeping only the real part of the force noise spectrum in Section \ref{FDT}, ensuring thus {\it time-reversal invariance} for the correlation functions. 
We fit $\mbox{I}_{\mbox{Re}}(x)= 2/(x^{0.5}+1/x^{0.5})$ on the numerical data of Fig. \ref{fig_17} (dashed line).\\

		\begin{figure}[t!]
		\centering
	\includegraphics[width=13 cm]{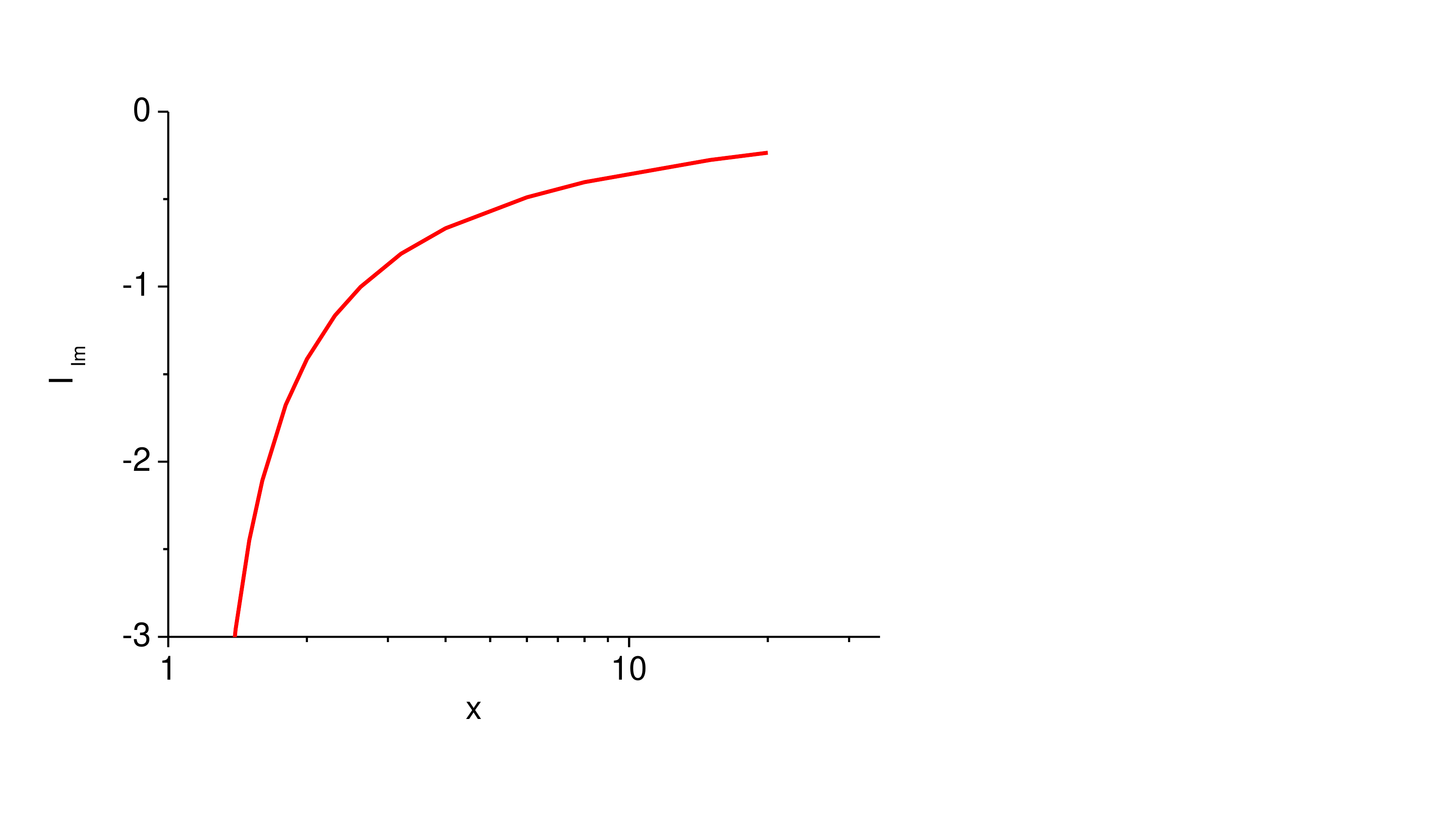}
			\caption{\small{ 
			Function $ \mbox{I}_{\mbox{Im}}(x)$ [equivalently $ \mbox{J}_{\mbox{Im}}$] that defines the spectrum integrals when $\omega_{\left\{\eta\right\}} \neq \omega_{\left\{\eta'\right\}}$ (see text). }}
			\label{fig_18}
		\end{figure}
		
Next, let us assume that the two frequencies $\omega_{\left\{\eta\right\}} \neq \omega_{\left\{\eta'\right\}}$ are clearly separated in the spectrum, meaning that the peaks generated by $\chi_{\left\{\eta\right\}}$ and $\chi_{\left\{\eta'\right\}}$, which have a half-height-width of $\Gamma_{\left\{\eta\right\}}, \Gamma_{\left\{\eta'\right\}}$ respectively, are clearly identified.
Numerical integration leads then to, in the high-Q limit:
\begin{eqnarray}
&& \int_{0}^{+\infty}  \chi_{\left\{\eta\right\}}(\omega) \, \chi_{\left\{\eta'\right\}}(\omega)^{*}  d\omega  \nonumber \\
&&  = \frac{\pi}{2} \sqrt{\omega_{\left\{\eta\right\}}  \omega_{\left\{\eta'\right\}} } \left[  \frac{  \mbox{I}_{\mbox{Re}} }{ \sqrt{Q_{\left\{\eta\right\}} Q_{\left\{\eta'\right\}}}   } -\mathrm{i} \, \mbox{I}_{\mbox{Im}}  \right]  , \label{integrIm2} \\
&& \int_{0}^{+\infty}  \omega^2 \, \chi_{\left\{\eta\right\}}(\omega) \, \chi_{\left\{\eta'\right\}}(\omega)^{*}  d\omega  \nonumber \\
&& = \omega_{\left\{\eta\right\}} \omega_{\left\{\eta'\right\}} \, \frac{\pi}{2} \sqrt{\omega_{\left\{\eta\right\}}  \omega_{\left\{\eta'\right\}} } \left[  \frac{  \mbox{J}_{\mbox{Re}} }{ \sqrt{Q_{\left\{\eta\right\}} Q_{\left\{\eta'\right\}}}   } -\mathrm{i} \, \mbox{J}_{\mbox{Im}}  \right]  , 
\end{eqnarray}
where now the functions $\mbox{I}_{\mbox{Re}} ,\mbox{I}_{\mbox{Im}} $ and $\mbox{J}_{\mbox{Re}} ,\mbox{J}_{\mbox{Im}} $ depend in principle on both $ \omega_{\left\{\eta\right\}}/ \omega_{\left\{\eta'\right\}} $ and $Q_{\left\{\eta\right\}}/ Q_{\left\{\eta'\right\}}$.
It turns out that $\mbox{I}_{\mbox{Im}} = \mbox{J}_{\mbox{Im}}$, which depend on the ratio  $ \omega_{\left\{\eta\right\}}/ \omega_{\left\{\eta'\right\}} $ only. We plot them in Fig. \ref{fig_18}.
They verify the symmetry property:
\begin{eqnarray}
\mbox{I}_{\mbox{Im}}(x) & = & - \mbox{I}_{\mbox{Im}}(1/x) ,
\end{eqnarray}
as in the previous case. We chose arbitrarily the normalization value $\mbox{I}_{\mbox{Im}}(2.61) \approx -1$ such that the numerical prefactor appearing in Eq. (\ref{integrIm2}) is the same as in Eq. (\ref{integrIm}).
Note that the apparent divergence in Fig. \ref{fig_18} when $x \rightarrow 1$ simply marks the transition towards the previous situation $\omega_{\left\{\eta\right\}}\approx \omega_{\left\{\eta'\right\}}$.
Within the high-Q limit, we shall neglect $\mbox{I}_{\mbox{Re}}, \mbox{J}_{\mbox{Re}}$ when computing the noise spectra in Section \ref{FDT}.  
This is fairly symmetric to the treatment performed when the modes are close in frequency, but the implications are drastically different: now, the correlation functions for the Langevin forces are {\it anti-symmetric} by time-reversal. 
This counter-intuitive feature would require to be confirmed experimentally, and is explicitly discussed in the core of the manuscript.
		
\section{From standing wave modes to mixing} 
\label{beatings}
%

In this last Appendix, we shall discuss the physical meaning of the correlations we introduced.
Heat is fed into the rod from the clamp: nonlinear processes within the region of size $\varepsilon$ are creating the $ \left\langle  U_{\omega_{\left\{\eta\right\}}\!}(t)\, U_{\omega_{\left\{\eta'\right\}}\!}(t) \right\rangle \neq 0 $, as stated by Eq. (\ref{dissclampcross}). 

Let us come back to Eqs. (\ref{equr1}-\ref{equz1}). The stochastic amplitude $U_{\omega_{\left\{\eta\right\}}\!}(t)$ of one mode can be rewritten:
\begin{equation}
U_{\omega_{\left\{\eta\right\}}\!}(t) = U_{\left\{\eta\right\},0}(t) \cos(\omega_{\left\{\eta\right\}} t)  +  U_{\left\{\eta\right\},\pi/2}(t) \sin(\omega_{\left\{\eta\right\}} t)  , \label{transform}
\end{equation}
with both $U_{\left\{\eta\right\},0}$ and $U_{\left\{\eta\right\},\pi/2}$ two random (uncorrelated and centered) variables.
Equivalently, one can define $U_{\left\{\eta\right\},0}=-2U_{\left\{\eta\right\},A}\sin\phi_{\left\{\eta\right\}}$ and $U_{\left\{\eta\right\},\pi/2}=+2U_{\left\{\eta\right\},A}\cos\phi_{\left\{\eta\right\}}$. Eq. (\ref{transform}) was already introduced in Section \ref{modes} for well-defined amplitudes; we apply it here to stochastic ones.
These are just standard Rotating Frame Transformations (RWT), at a frame frequency $\omega_{\left\{\eta\right\}}$.
The definition of $U_{\left\{\eta\right\},0}$, $U_{\left\{\eta\right\},\pi/2}$ is of practical importance: they correspond to what is measured when using demodulation techniques.
Furthermore, they imply that the standing wave $\left\{\eta\right\}$ is the result of the superposition of two traveling ones:
\begin{eqnarray}
 U_{\left\{\eta\right\}}(z,t) &=& U_{\left\{\eta\right\},A}(t) \cos[\omega_{\left\{\eta\right\}} t -\phi_{\left\{\eta\right\}}(t) -k_{\left\{\eta\right\}} z] \nonumber \\
 &-& U_{\left\{\eta\right\},A}(t) \cos[\omega_{\left\{\eta\right\}} t -\phi_{\left\{\eta\right\}}(t) + k_{\left\{\eta\right\}} z] , \label{reflex}
\end{eqnarray}
which can be understood as the reflection of the wave traveling towards positive $z$ onto the end clamp, and coming backwards with a $\pi$ change of phase (and vice-versa). Similar expressions stand for Eqs. (\ref{equr2}-\ref{equz2}), but arising from a reflection with no phase change [a $+$ instead of a $-$ in Eq. (\ref{reflex})]. As a result, one should perform the replacement $- \sin \phi_{\left\{\eta\right\}} \rightarrow \cos \phi_{\left\{\eta\right\}} $, $\cos \phi_{\left\{\eta\right\}} \rightarrow \sin \phi_{\left\{\eta\right\}}$ in the definitions of 
$U_{\left\{\eta\right\},0}, U_{\left\{\eta\right\},\pi/2}$ (Section \ref{modes}).
In order to be of any use, the RWT comes along with the Rotating Wave Approximation (RWA): the new fluctuating variables defined in the rotating frame are assumed to be infinitely slow, even though they depend explicitly on time $t$. This means that in calculations, their time-derivatives will be neglected.

From the above, we see that correlations in the Rotating Frame verify:
\begin{eqnarray}
&& \left\langle  U_{\omega_{\left\{\eta\right\}}\!} \, U_{\omega_{\left\{\eta'\right\}}\!}  \right\rangle  =  \\
&&\!\!\!\!\!\!\!\!\!\!\!\!\!\!\!  +\left[\frac{\left\langle  U_{\omega_{\left\{\eta\right\}},0} \, U_{\omega_{\left\{\eta'\right\}},0}  \right\rangle + \left\langle  U_{\omega_{\left\{\eta\right\}},\pi/2} \, U_{\omega_{\left\{\eta'\right\}},\pi/2}  \right\rangle}{2} \right] \cos(\omega_{\left\{\eta\right\}} - \omega_{\left\{\eta'\right\}})\, t \nonumber \\
&&\!\!\!\!\!\!\!\!\!\!\!\!\!\!\!  +\left[\frac{\left\langle  U_{\omega_{\left\{\eta\right\}},\pi/2} \, U_{\omega_{\left\{\eta'\right\}},0}  \right\rangle - \left\langle  U_{\omega_{\left\{\eta\right\}},0} \, U_{\omega_{\left\{\eta'\right\}},\pi/2}  \right\rangle}{2} \right] \sin(\omega_{\left\{\eta\right\}} - \omega_{\left\{\eta'\right\}})\, t \nonumber \\
&&\!\!\!\!\!\!\!\!\!\!\!\!\!\!\!  +\left[\frac{\left\langle  U_{\omega_{\left\{\eta\right\}},0} \, U_{\omega_{\left\{\eta'\right\}},0}  \right\rangle - \left\langle  U_{\omega_{\left\{\eta\right\}},\pi/2} \, U_{\omega_{\left\{\eta'\right\}},\pi/2}  \right\rangle}{2} \right] \cos(\omega_{\left\{\eta\right\}} + \omega_{\left\{\eta'\right\}})\, t \nonumber \\
&&\!\!\!\!\!\!\!\!\!\!\!\!\!\!\!  +\left[\frac{\left\langle  U_{\omega_{\left\{\eta\right\}},\pi/2} \, U_{\omega_{\left\{\eta'\right\}},0}  \right\rangle + \left\langle  U_{\omega_{\left\{\eta\right\}},0} \, U_{\omega_{\left\{\eta'\right\}},\pi/2}  \right\rangle}{2} \right] \sin(\omega_{\left\{\eta\right\}} + \omega_{\left\{\eta'\right\}})\, t , \nonumber \\ 
&& \frac{\left\langle  \dot{U}_{\omega_{\left\{\eta\right\}}\!} \, \dot{U}_{\omega_{\left\{\eta'\right\}}\!}  \right\rangle}{\omega_{\left\{\eta\right\}}\omega_{\left\{\eta'\right\}}}  \approx  \\
&&\!\!\!\!\!\!\!\!\!\!\!\!\!\!\!  +\left[\frac{\left\langle  U_{\omega_{\left\{\eta\right\}},0} \, U_{\omega_{\left\{\eta'\right\}},0}  \right\rangle + \left\langle  U_{\omega_{\left\{\eta\right\}},\pi/2} \, U_{\omega_{\left\{\eta'\right\}},\pi/2}  \right\rangle}{2} \right] \cos(\omega_{\left\{\eta\right\}} - \omega_{\left\{\eta'\right\}})\, t \nonumber \\
&&\!\!\!\!\!\!\!\!\!\!\!\!\!\!\!  +\left[\frac{\left\langle  U_{\omega_{\left\{\eta\right\}},\pi/2} \, U_{\omega_{\left\{\eta'\right\}},0}  \right\rangle - \left\langle  U_{\omega_{\left\{\eta\right\}},0} \, U_{\omega_{\left\{\eta'\right\}},\pi/2}  \right\rangle}{2} \right] \sin(\omega_{\left\{\eta\right\}} - \omega_{\left\{\eta'\right\}})\, t \nonumber \\
&&\!\!\!\!\!\!\!\!\!\!\!\!\!\!\!  -\left[\frac{\left\langle  U_{\omega_{\left\{\eta\right\}},0} \, U_{\omega_{\left\{\eta'\right\}},0}  \right\rangle - \left\langle  U_{\omega_{\left\{\eta\right\}},\pi/2} \, U_{\omega_{\left\{\eta'\right\}},\pi/2}  \right\rangle}{2} \right] \cos(\omega_{\left\{\eta\right\}} + \omega_{\left\{\eta'\right\}})\, t \nonumber \\
&&\!\!\!\!\!\!\!\!\!\!\!\!\!\!\!  -\left[\frac{\left\langle  U_{\omega_{\left\{\eta\right\}},\pi/2} \, U_{\omega_{\left\{\eta'\right\}},0}  \right\rangle + \left\langle  U_{\omega_{\left\{\eta\right\}},0} \, U_{\omega_{\left\{\eta'\right\}},\pi/2}  \right\rangle}{2} \right] \sin(\omega_{\left\{\eta\right\}} + \omega_{\left\{\eta'\right\}})\, t . \nonumber
\end{eqnarray}
The oscillations at $\omega_{\left\{\eta\right\}}+\omega_{\left\{\eta'\right\}}$ must vanish; therefore, the Rotating Frame correlation functions satisfy:
\begin{eqnarray}
\left\langle  U_{\omega_{\left\{\eta\right\}},0} \, U_{\omega_{\left\{\eta'\right\}},0}  \right\rangle & = & + \left\langle  U_{\omega_{\left\{\eta\right\}},\pi/2} \, U_{\omega_{\left\{\eta'\right\}},\pi/2}  \right\rangle , \label{rule1} \\
\left\langle  U_{\omega_{\left\{\eta\right\}},\pi/2} \, U_{\omega_{\left\{\eta'\right\}},0}  \right\rangle & = & - \left\langle  U_{\omega_{\left\{\eta\right\}},0} \, U_{\omega_{\left\{\eta'\right\}},\pi/2}  \right\rangle , \label{rule2}
\end{eqnarray}
from which we recover the result $\left\langle  \dot{U}_{\omega_{\left\{\eta\right\}}\!} \, \dot{U}_{\omega_{\left\{\eta'\right\}}\!}  \right\rangle=\omega_{\left\{\eta\right\}}\omega_{\left\{\eta'\right\}} $ $ \left\langle  U_{\omega_{\left\{\eta\right\}}\!} \, U_{\omega_{\left\{\eta'\right\}}\!}  \right\rangle$.
Since $\left\langle  U_{\omega_{\left\{\eta\right\}}\!} \, U_{\omega_{\left\{\eta'\right\}}\!}  \right\rangle$ is time-independent by construction, we finally obtain:
\begin{eqnarray}
&& \left\langle  U_{\omega_{\left\{\eta\right\}},0} \, U_{\omega_{\left\{\eta'\right\}},0}  \right\rangle  = \label{corr1} \\
&& \left\langle  U_{\omega_{\left\{\eta\right\}}\,} \, U_{\omega_{\left\{\eta'\right\}}\,}  \right\rangle \cos(\omega_{\left\{\eta\right\}} - \omega_{\left\{\eta'\right\}})\, t, \nonumber \\
&& \left\langle  U_{\omega_{\left\{\eta\right\}},\pi/2} \, U_{\omega_{\left\{\eta'\right\}},0}  \right\rangle  = \label{corr2} \\
&& \left\langle  U_{\omega_{\left\{\eta\right\}} \,} \, U_{\omega_{\left\{\eta'\right\}}\,}  \right\rangle \sin(\omega_{\left\{\eta\right\}} - \omega_{\left\{\eta'\right\}})\, t . \nonumber
\end{eqnarray}
The amplitude correlations in the Rotating Frame have therefore the same magnitude as the initial one in the laboratory frame, Eq. (\ref{equiparticross}), and they oscillate at the beating frequency $\omega_{\left\{\eta\right\}}-\omega_{\left\{\eta'\right\}}$. Note that the above results trivially match the conventional situation $\left\{\eta\right\}=\left\{\eta'\right\}$. Eqs. (\ref{corr1},\ref{corr2}) are typically properties that could be probed in an experiment.

Consider now Eqs. (\ref{modekin2},\ref{modebend2}). Injecting explicitly the traveling wave solutions in them, and regrouping immediately potential and kinetic terms by taking $\omega_{\left\{\eta\right\}}   \omega_{\left\{\eta' \right\}} {\cal M}_{\left\{\eta\right\},\left\{\eta' \right\} }/{\cal K}_{\left\{\eta\right\},\left\{\eta' \right\} } \approx 1$, they lead to the new writing for the energy density:
\begin{widetext}
\begin{eqnarray}
   \frac{d E_{tot} (z)}{dz}   \approx  \sum_{\left\{\eta\right\},\left\{\eta' \right\}}\frac{1}{2}  {\cal K}_{\left\{\eta\right\},\left\{\eta' \right\} } \!\!\! \!\!\! &&\Bigg[ +   \left(\frac{\left\langle  U_{\omega_{\left\{\eta\right\}},0} \, U_{\omega_{\left\{\eta'\right\}},0}  \right\rangle + \left\langle  U_{\omega_{\left\{\eta\right\}},\pi/2} \, U_{\omega_{\left\{\eta'\right\}},\pi/2}  \right\rangle}{2} \right)  \frac{1}{2} \cos\left[(\omega_{\left\{\eta\right\}}-\omega_{\left\{\eta'\right\}}) t -(k_{\left\{\eta\right\}}-k_{\left\{\eta'\right\}}) z \right] \nonumber \\
&& \,\,\,\, \, + \left(\frac{\left\langle  U_{\omega_{\left\{\eta\right\}},0} \, U_{\omega_{\left\{\eta'\right\}},0}  \right\rangle + \left\langle  U_{\omega_{\left\{\eta\right\}},\pi/2} \, U_{\omega_{\left\{\eta'\right\}},\pi/2}  \right\rangle}{2} \right)  \frac{1}{2} \cos\left[(\omega_{\left\{\eta\right\}}-\omega_{\left\{\eta'\right\}}) t + (k_{\left\{\eta\right\}}-k_{\left\{\eta'\right\}}) z \right]    \nonumber \\
&& \,\,\,\, \, + \left(\frac{\left\langle  U_{\omega_{\left\{\eta\right\}},\pi/2} \, U_{\omega_{\left\{\eta'\right\}},0}  \right\rangle - \left\langle  U_{\omega_{\left\{\eta\right\}},0} \, U_{\omega_{\left\{\eta'\right\}},\pi/2}  \right\rangle}{2} \right)  \frac{1}{2} \sin\left[(\omega_{\left\{\eta\right\}}-\omega_{\left\{\eta'\right\}}) t - (k_{\left\{\eta\right\}}-k_{\left\{\eta'\right\}}) z \right]    \nonumber \\
&& \,\,\,\, \, + \left(\frac{\left\langle  U_{\omega_{\left\{\eta\right\}},\pi/2} \, U_{\omega_{\left\{\eta'\right\}},0}  \right\rangle - \left\langle  U_{\omega_{\left\{\eta\right\}},0} \, U_{\omega_{\left\{\eta'\right\}},\pi/2}  \right\rangle}{2} \right)  \frac{1}{2} \sin\left[(\omega_{\left\{\eta\right\}}-\omega_{\left\{\eta'\right\}}) t + (k_{\left\{\eta\right\}}-k_{\left\{\eta'\right\}}) z \right]    \nonumber \\
&& \,\,\,\, \, -\left(\frac{\left\langle  U_{\omega_{\left\{\eta\right\}},0} \, U_{\omega_{\left\{\eta'\right\}},0}  \right\rangle - \left\langle  U_{\omega_{\left\{\eta\right\}},\pi/2} \, U_{\omega_{\left\{\eta'\right\}},\pi/2}  \right\rangle}{2} \right)  \frac{1}{2}  \cos\left[(\omega_{\left\{\eta\right\}}+\omega_{\left\{\eta'\right\}}) t -(k_{\left\{\eta\right\}}+k_{\left\{\eta'\right\}}) z \right]  \nonumber \\
&& \,\,\,\, \, -\left(\frac{\left\langle  U_{\omega_{\left\{\eta\right\}},0} \, U_{\omega_{\left\{\eta'\right\}},0}  \right\rangle - \left\langle  U_{\omega_{\left\{\eta\right\}},\pi/2} \, U_{\omega_{\left\{\eta'\right\}},\pi/2}  \right\rangle}{2} \right)  \frac{1}{2} \cos\left[(\omega_{\left\{\eta\right\}}+\omega_{\left\{\eta'\right\}}) t +(k_{\left\{\eta\right\}}+k_{\left\{\eta'\right\}}) z \right]  \nonumber \\
&& \,\,\,\, \,  - \left(\frac{\left\langle  U_{\omega_{\left\{\eta\right\}},\pi/2} \, U_{\omega_{\left\{\eta'\right\}},0}  \right\rangle + \left\langle  U_{\omega_{\left\{\eta\right\}},0} \, U_{\omega_{\left\{\eta'\right\}},\pi/2}  \right\rangle}{2} \right)  \frac{1}{2} \sin\left[(\omega_{\left\{\eta\right\}}+\omega_{\left\{\eta'\right\}}) t -(k_{\left\{\eta\right\}}+k_{\left\{\eta'\right\}}) z \right]  \nonumber \\
&&\,\,\,\, \, - \left(\frac{\left\langle  U_{\omega_{\left\{\eta\right\}},\pi/2} \, U_{\omega_{\left\{\eta'\right\}},0}  \right\rangle + \left\langle  U_{\omega_{\left\{\eta\right\}},0} \, U_{\omega_{\left\{\eta'\right\}},\pi/2}  \right\rangle}{2} \right)  \frac{1}{2} \sin\left[(\omega_{\left\{\eta\right\}}+\omega_{\left\{\eta'\right\}}) t +(k_{\left\{\eta\right\}}+k_{\left\{\eta'\right\}}) z \right]\, \Bigg] \nonumber \\
&& \!\!\!\!\!\!\!\!\!\!\!\!\!\!\!\!\! +0 \, ,\label{eqbeat} 
\end{eqnarray}
\end{widetext}
in which we kept only the terms that do not involve $\Delta {\cal K}_{\left\{\eta\right\},\left\{\eta' \right\} }$ and $\Delta {\cal M}_{\left\{\eta\right\},\left\{\eta' \right\} }$ as they vanish in Eq. (\ref{grad2}), which is reminded with a $0$ in the above equation.
We clearly see in Eq. (\ref{eqbeat}) that the standing wave pattern responsible for the temperature profile $T(z)$ is actually due to {\it mixing} 
at frequencies $\omega_{\left\{\eta\right\}} \pm \omega_{\left\{\eta'\right\}}$, with wavevectors $k_{\left\{\eta\right\}} \pm k_{\left\{\eta'\right\}}$.
From the rules Eqs. (\ref{rule1},\ref{rule2}), only the mixing 
 at $\omega_{\left\{\eta\right\}} - \omega_{\left\{\eta'\right\}}$, with   $k_{\left\{\eta\right\}} - k_{\left\{\eta'\right\}}$ are nonzero.
These correspond to beatings that store energy in the rod; on the other hand, the terms with $\omega_{\left\{\eta\right\}} + \omega_{\left\{\eta'\right\}}$ and   $k_{\left\{\eta\right\}} + k_{\left\{\eta'\right\}}$ can only propagate (and do not store energy). As such, they should be involved only in heat transport: these should be considered as the true ``propagation channels'' within the rod.

\bibliographystyle{ieeetran}

\begin{thebibliography}{9}

\bibitem{volker91} V. Heine, I.J. Robertson, M.C. Payne, Phil. Trans. Soc. Lond. A {\bf 334}, 393 (1991).
\bibitem{clelandBk} Andrew N. Cleland, {\it Foundations of nanomechanics}, Springer (2003).
\bibitem{ziman} J.M. Ziman {\it Electrons and Phonons: The Theory of Transport Phenomena in Solids}, Oxford university press (2001).
\bibitem{siliconMD} Sebastian G. Volz and Gang Chen, Phys. Rev. B {\bf 61}, 2651 (2000).
\bibitem{siliconMD2} Mark D. Kluge and John R. Ray, J. Chem. Phys. {\bf 85}, 4028 (1986).
\bibitem{OptoTh} Francesco Fogliano, Benjamin Besga, Antoine Reigue, Laure Mercier de L\'epinay, Philip Heringlake,
Clement Gouriou, Eric Eyraud, Wolfgang Wernsdorfer, Benjamin Pigeau and Olivier Arcizet, Nat. Comm. {\bf 12}, 4124 (2021).
\bibitem{VinantePRL} A. Vinante, M. Bahrami, A. Bassi, O. Usenko, G. Wijts, and T. H. Oosterkamp, Phys. Rev. Lett. {\bf 116}, 090402 (2016).
\bibitem{sansa} Marc Sansa, Eric Sage, Elizabeth C. Bullard, Marc G\'ely, Thomas Alava, Eric Colinet,
Akshay K. Naik, Luis Guillermo Villanueva, Laurent Duraffourg, Michael L. Roukes, Guillaume Jourdan and S\'ebastien Hentz, Nat. Nanotech. {\bf 11}, 552 (2016).
\bibitem{regal2008} C. Regal, J.D. Teufel, K.W. Lehnert, Nat. Phys. {\bf 4}, 555 (2008).
\bibitem{statbook} B. Diu, C. Guthmann, D. Lederer, B. Roulet, {\it Physique statistique}, Hermann (1997). 
\bibitem{bellon1} Mickael Geitner, Felipe Aguilar Sandoval, Eric Bertin, and Ludovic Bellon, Phys. Rev. E {\bf 95}, 032138 (2017).
\bibitem{bellon2} Felipe Aguilar Sandoval, Mickael Geitner, Eric Bertin,  and Ludovic Bellon, J. of Appl. Phys. {\bf 117}, 234503 (2015).
\bibitem{bellon3} Alex Fontana, Richard Pedurand, Vincent Dolique, Ghaouti Hansali, Ludovic Bellon, 
Phys. Rev. E {\bf 103}, 062125 (2021). 
\bibitem{poche} R.N. Thurston, J. of Sound and Vibration {\bf 159}(3), 441 (1992).

\bibitem{hunklinger} S. Hunklinger, J. de Physique Colloques, {\bf 43} (C9), pp. C9-461-C9-474 (1982).

\bibitem{ashcroft} N. W. Ashcroft and N. D. Mermin, {\it Solid State Physics}, Saunders College Philadelphia (1976).

\bibitem{SiPh1} Audrey Valentin, Johann S\'ee, Sylvie Galdin-Retailleau and Philippe Dollfus, J. Phys.: Condens. Matter {\bf 20}, 145213 (2008). 
\bibitem{SiPh2} Sanghamitra Neogi and Davide Donadioa, Eur. Phys. J. B {\bf 88}: 73 (2015). 
\bibitem{SiPh3} A. Sparavigna, Phys. Rev. B {\bf 67}, 144305 (2003).

\bibitem{abe} Haruka Abe, Hideyuki Kato, and Tetsuya Baba, Jpn. J. Appl. Phys. {\bf 50}, 11RG01 (2011). 

\bibitem{kappaSi} T. Rufa, R.W. Henn, M. Asen-Palmer, E. Gmelin, M. Cardona, H.-J. Pohl,
G.G. Devyatych, P.G. Sennikov, Solid State Communications {\bf 115}, 243 (2000).

\bibitem{TLSus} D. Cattiaux, I. Golokolenov, S. Kumar, M. Sillanp\"a\"a, L. Mercier de L\'epinay, R. R. Gazizulin, X. Zhou, A. D. Armour, O. Bourgeois, A. Fefferman and E. Collin, Nat. Comm. {\bf 12}, 6182 (2021).
\bibitem{ustinov} Alexander Bilmes, Sebastian Zanker, Andreas Heimes, Michael Marthaler, Gerd Sch\"on, Georg Weiss, Alexey V. Ustinov, and J\"urgen Lisenfeld, Phys. Rev. B {\bf 96}, 064504 (2017).

\bibitem{phillips} W. A. Phillips, Rep. Prog. Phys. {\bf 50}, pp. 1657-1708 (1987). 

\bibitem{bondarenk} A. A. Bondarenko, European Women in Mathematics, Proceedings of the 13th General Meeting, p. 103, World Scientific (2010).

\bibitem{onoe} Morio Onoe, H.D. McNiven, R.D. Mindlin, J. of Appl. Mech. {\bf 29}(4), 729 (1962).
\bibitem{pao1} Yih-Hsing Pao and R.D. Mindlin, J. of Appl. Mech. {\bf 27}(3), 513 (1960).
\bibitem{pao2} Yih-Hsing Pao, J. of Appl. Mech. {\bf 29}(1), 61 (1962).

\bibitem{ini1} L. Pochhammer, J. f\"ur reine und angewandte Mathematik (Crelle) {\bf 81}, 324 (1876).
\bibitem{ini2} C. Chree, Transactions of the Cambridge Philosophical Society {\bf 14}, 250 (1889).

\bibitem{kotthaus} Quirin P. Unterreithmeier, Thomas Faust, and J\"org P. Kotthaus, Phys. Rev. Lett. {\bf 105}, 027205 (2010).

\bibitem{ilya_nanolett} I. Golokolenov, S. Kumar, B. Alperin, B. Fernandez, A. Fefferman and E. Collin, 
J. of Appl. Phys. {\bf 133}, 124302 (2023).

\bibitem{olivePRB} Olivier Maillet, Dylan Cattiaux, Xin Zhou, Rasul R. Gazizulin, Olivier Bourgeois, Andrew D. Fefferman, Eddy Collin, Phys. Rev. B {\bf 107}, 064104 (2023).

\bibitem{timoshenko} Stephen P. Timoshenko, James M. Gere, {\it Theory of Elastic Stability}, Second Ed., Dover Publications Inc., Mineola, New York (2009).

\bibitem{judge1} Douglas M. Photiadis and John A. Judge, Appl. Phys. Lett. {\bf 85}, 482 (2004).
\bibitem{judge2} John A. Judge, Douglas M. Photiadis, Joseph F. Vignola, Brian H. Houston and Jacek Jarzynski, J. of Appl. Phys. {\bf 101}, 013521 (2007).
\bibitem{cross} M. C. Cross and Ron Lifshitz, Phys. Rev. B {\bf 64}, 085324 (2001).

\bibitem{Ignacio} I. Wilson-Rae Phys. Rev. B {\bf 77}, 245418 (2008).

\bibitem{regal} P.-L. Yu, K. Cicak, N. S. Kampel, Y. Tsaturyan, T. P. Purdy, R. W. Simmonds, and C. A. Regal, Appl. Phys. Lett.  {\bf 104}, 023510 (2014).
\bibitem{kipp} Amir Hossein Ghadimi, Dalziel Joseph Wilson, and Tobias J. Kippenberg, Nano Lett. {\bf 17}, 3501 (2017).

\bibitem{suhel} A. Suhel, B. D. Hauer, T. S. Biswas, K. S. D. Beach, and J. P. Davis, Appl. Phys. Lett. {\bf 100}, 173111 (2012).
\bibitem{biswas} T.S. Biswas, A. Suhel, B.D. Hauer, A. Palomino, K.S.D. Beach and J.P. Davis, Appl. Phys. Lett. {\bf 101}, 093105 (2012).

\bibitem{schliesser} Y. Tsaturyan, A. Barg, E. S. Polzik and A. Schliesser, Nat. Nanotech. {\bf 12}, 776 (2017).

\bibitem{antisym_corr} A. V. Balatsky and Elihu Abrahams, Phys. Rev. Lett. {\bf 74}, 1004 (1995).

\bibitem{meAQS} E. Collin, AVS Quantum Sci. {\bf 4}, 020501 (2022).

\bibitem{Jukka} Jukka P. Pekola and Bayan Karimi, Rev. of Mod. Phys. {\bf 93}, 041001 (2021). 
\bibitem{martialPHD} M. Defoort, {Nonlinear dynamics in nano-electromechanical systems at low temperatures}, P.h.D. thesis, Grenoble University (2014).


\end{thebibliography}

\end{document}